\documentclass[conference,letterpaper]{IEEEtran}

\usepackage[utf8]{inputenc} 
\usepackage[T1]{fontenc}
\usepackage{url}
\usepackage{ifthen}
\usepackage{cite}
\usepackage[cmex10]{amsmath} 

\usepackage{amsmath,amssymb,amsfonts,mathrsfs,multicol,ulem,bm}

\usepackage[amsmath,thmmarks]{ntheorem}
\usepackage{graphicx}
\usepackage[rightcaption]{sidecap}
\usepackage{caption}
\usepackage{soul}
\usepackage{pdfpages}
\usepackage{verbatim}

\theoremstyle{plain}
\theorembodyfont{\upshape}
\newtheorem{theorem}{Theorem}

\newtheorem{definition}[theorem]{Definition}
\newtheorem{lemma}[theorem]{Lemma}

\newtheorem{claim}[theorem]{Claim}
\theoremsymbol{\ensuremath{\square}}

\newcommand{\F}{\mathbb{F}}

\newcommand{\Z}{\mathbb{Z}}

\newcommand{\I}{\mathbb{I}}

\newcommand{\cL}{\mathcal{L}}

\newcommand{\ra}{\rangle}
\newcommand{\la}{\langle}

\newcommand{\bE}{\mathbf{E}}

\newcommand{\bu}{\mathbf{u}}
\newcommand{\bv}{\mathbf{v}}
\newcommand{\bw}{\mathbf{w}}
\newcommand{\bc}{\mathbf{c}}

\newcommand{\bx}{\mathbf{x}}

\newcommand{\bl}{\boldsymbol\ell}

\newcommand{\wt}{\text{wt}}
\newcommand{\dist}{\text{dist}}
\newcommand{\diag}{\text{diag}}
\newcommand{\SWAP}{\text{SWAP}}
\newcommand{\Aut}{\text{Aut}}
\newcommand{\SL}{\text{SL}}
\newcommand{\Sp}{\text{Sp}}

\newcommand{\GA}{\text{GA}}
\newcommand{\RM}{\text{RM}}
\newcommand{\QRM}{\text{QRM}}
\newcommand{\PQRM}{\text{PQRM}}
\newcommand{\oRM}{\overline{\text{RM}}}

\newcommand{\rel}{\text{rel}}
\newcommand{\CNOT}{\text{CNOT}}
\newcommand{\single}{\text{single}}
\newcommand{\SPAM}{\text{SPAM}}
\newcommand{\corr}{\text{corr}}

\newcommand{\hbc}{\hat{\mathbf{c}}}

\newcommand{\tbc}{\tilde{\mathbf{c}}}

\renewcommand{\epsilon}{\ensuremath\varepsilon}

\usepackage[linkcolor=black,colorlinks=true,citecolor=black,filecolor=black]{hyperref}

\usepackage{array} 
\usepackage{mathtools}
\usepackage{pgfplots}
\usepgfplotslibrary{fillbetween}
\pgfplotsset{compat=newest}
\usepgfplotslibrary{groupplots}
\usetikzlibrary{matrix}
\usetikzlibrary{quantikz2}
\usepgfplotslibrary{dateplot}
\newcounter{plot}[figure]

\usepackage{stmaryrd}
\usepackage{tikz}

\usepackage{cleveref}
\crefname{plot}{}{plots}
\Crefname{plot}{}{Plots}
\definecolor{blueviolet}{RGB}{138,43,226}
\definecolor{darkgray176}{RGB}{176,176,176}
\definecolor{lightgray211}{RGB}{211,211,211}
\definecolor{darkorange25512714}{RGB}{255,127,14}
\definecolor{dodgerblue}{RGB}{30,144,255}
\definecolor{gold}{RGB}{255,215,0}
\definecolor{red}{RGB}{255,0,0}
\definecolor{orange}{RGB}{255,93,0}
\definecolor{limegreen}{RGB}{50,205,50}
\definecolor{steelblue31119180}{RGB}{31,119,180}
\definecolor{teal}{RGB}{0,128,128}
\definecolor{olivegreen}{RGB}{55,126,34}
\definecolor{neonpurple}{RGB}{176,38,255}
\definecolor{coral}{RGB}{254,125,106}
\definecolor{lemonade}{RGB}{252,186,203}
\definecolor{lemon}{RGB}{255,247,0}
\definecolor{amber}{RGB}{255,191,0}
\definecolor{palegreen}{RGB}{180,238,180}
\definecolor{lightteal}{RGB}{64,224,208}
\definecolor{lavender}{RGB}{227,159,246}
\definecolor{darkviolet}{RGB}{153,50,204}
\definecolor{darkcyan}{RGB}{0,139,139}
\definecolor{darkgreen}{RGB}{0,100,0}
\definecolor{lightskyblue}{RGB}{136,205,250}

\usepackage{algorithm}
\usepackage[noend]{algpseudocode}
\algrenewcommand{\algorithmiccomment}[1]{\hskip3em // #1}

\makeatletter

\renewcommand*{\arraystretch}{0.95}
\newcommand{\sddots}{\raisebox{0pt}{$\scalebox{.5}{$\ddots$}$}}

\usepackage{braket}

\graphicspath{ {./imgs/} }

\begin{document}
\title{Computation with Quantum Reed-Muller Codes and their Mapping onto 2D Atom Arrays} 

\author{%
  \IEEEauthorblockN{Anqi Gong and
                    Joseph M. Renes}
  \IEEEauthorblockA{ETH Z\"urich, Switzerland,
                    gonga@student.ethz.ch}
  \\[-5.0ex]
}

\maketitle

\begin{abstract}
We give a fault tolerant construction for error correction and computation using two punctured quantum Reed-Muller (PQRM) codes. In particular, we consider the $\llbracket 127,1,15\rrbracket$ self-dual doubly-even code that has transversal Clifford gates (CNOT, H, S) and the triply-even $\llbracket 127,1,7\rrbracket$ code that has transversal T and CNOT gates. We show that code switching between these codes can be accomplished using Steane error correction. For fault-tolerant ancilla preparation we utilize the low-depth hypercube encoding circuit along with different code automorphism permutations in different ancilla blocks, while decoding is handled by the high-performance classical successive cancellation list decoder. In this way, every logical operation in this universal gate set is amenable to extended rectangle analysis. The CNOT exRec has a failure rate approaching $10^{-9}$ at $10^{-3}$ circuit-level depolarizing noise.

Furthermore, we map the PQRM codes to a 2D layout suitable for implementation in arrays of trapped atoms and try to reduce the circuit depth of parallel atom movements in state preparation.
The resulting protocol is strictly fault-tolerant for the $\llbracket 127,1,7\rrbracket$ code and practically fault-tolerant for the $\llbracket 127,1,15\rrbracket$ code. Moreover, each patch requires a permutation consisting of $7$ sub-hypercube swaps only. These are swaps of rectangular grids in our 2D hypercube layout and can be naturally created with acousto-optic deflectors (AODs).

Lastly, we show for the family of $\llbracket 2^{2r},{2r\choose r},2^r\rrbracket$ QRM codes that the entire logical Clifford group can be achieved using only permutations, transversal gates, and fold-transversal gates.
\end{abstract}

\section{Introduction}
In this paper we explore the role of code automorphisms in fault tolerance, in particular for Quantum Reed-Muller (QRM) codes\cite{SteaneRM}. 
These codes are appealing because classical RM codes \cite{RM-Reed, RM-Muller} are known for their large automorphism group \cite{theoryEC} while having good distance versus rate trade-off. 
Furthermore, these codes can be easily encoded by hypercube encoding circuits \cite{polarvsRM, polar}. When not too large, RM codes can also be decoded reasonably well using the successive cancellation list decoder \cite{Dumer3, SCL-CRC}. As CSS codes \cite{CSS-CS, CSS-Steane}, QRM codes inherit all these great properties from their classical counterparts.

QRM codes are of particular interest in the realm of universal quantum computation, since they and their punctured versions have very flexible gate set properties \cite{preskill, sublogarithmic}, and some are well-known for admitting transversal gates outside the Clifford group \cite{knill1996, sublogarithmic, Haah2018, barg2024RM}.
However, use of these codes also faces a few challenges. 
The first is how to fault-tolerantly prepare them on a quantum computer, where operations such as gates and measurements are noisy. 

A scheme based on bare-ancilla stabilizer measurement will likely not work because QRM codes are non-degenerate, i.e., the weight of the stabilizers is at least the code distance. 
Direct encoding of ancillae for Steane error-correction based on the low-depth hypercube encoding circuit is not fault-tolerant (FT) either, but there are methods \cite{Steane_golay,golay,cross2009comparative} to make it so.

Here we build upon the proposal of \cite{golay} in order to minimize resource overheads. The idea is to copy the potential faults onto identically encoded ancillary blocks which are subject to different automorphism permutations. Subsequently, the ancillary blocks are measured transversally, and only when the measurement results come out to be a stabilizer is the preparation deemed successful.
With suitably chosen permutations, faults in the entire circuit up to a certain order rarely conspire to add up to a stabilizer. 
In this way, the occurrence of malignant events can be signaled.
We apply this idea to punctured QRM codes of blocklength $127$ and find strictly fault-tolerant protocols that are resilient to three or four faults.

The second challenge is to fault-tolerantly implement a universal set of gates. Especially if we want to implement each gate transversally, we have to resort to using different codes \cite{Eastin_2009}. There are multiple proposals to get RM codes around this restriction. Chamberland \emph{et al.} \cite{105-1-9} suggest concatenating the $\llbracket 7,1,3\rrbracket$ Steane code and the $\llbracket 15,1,3\rrbracket$ code \cite{knill1996}. However, despite this $105$ qubit code having distance $9$ for logical CNOT gate, the effective distance for the Hadamard and T logical gates is only $3$ \cite{105-1-9, Campbell_2017}.
Conversion between the Steane code and the $\llbracket 15,1,3\rrbracket$ code has been suggested in \cite{Anderson2014}. This can be seen as the smallest instance of code-switching between 2D and 3D color codes \cite{GaugeColorCodes}. However, conversion by measuring the absent stabilizers using bare-ancilla substantially reduces noise resilience \cite{Beverland2021}. A recent work \cite{heußen2024codeswitching} addresses this issue by fault-tolerantly preparing an ancilla encoded in the target code and teleporting the logical state onto it using transversal CNOT gates followed by transversal measurement and correction.

Here we consider a scheme of code conversion between two codes of the same blocklength. 
In this case, the transversal CNOT, followed by measurement and correction is essentially Steane error correction (EC) \cite{SteaneEC}. 
Our proposal is inspired by \cite{15-1-3}, which suggested implementing the logical Hadamard gate as a transversal Hadamard gate followed by Steane EC for $\llbracket 15,1,3\rrbracket$. 
We observe that \cite{15-1-3} can be more generally interpreted as code-switching between codes that satisfy certain stabilizer containment relationships. We apply this idea to the $\llbracket 127,1,15\rrbracket$ code \cite{preskill} admitting transversal $H$ and $S$ gates and $\llbracket 127,1,7\rrbracket$ code \cite{sublogarithmic} admitting transversal $T$ gate. Combined with our automorphism-based FT ancilla preparation protocols tailored for these two codes, we perform extensive circuit-level simulations of extended rectangles (exRec) \cite{AGP} for each logical gate. Using the $d=15$ code alone for Clifford gates, the largest exRec (logical CNOT) has an error rate approaching $10^{-9}$ at physical depolarizing noise $10^{-3}$. The logical $T$ exRec implemented with code-switching between the $d=15$ and $d=7$ also shows suppression at this noise level. This code-switching $T$ gate on the $d=15$ code could also be used to distill higher fidelity $T$ state of the same code (and thus distance) \cite[Fig.~8]{Beverland2021}, \cite{Haah2018}, enabling a $T$ gate of fidelity similar to Clifford operations \cite{MSD}.

Beyond the construction of fault-tolerant gates for these codes, we have endeavored to tailor the required operations to be implementable on the neutral atom platform. 
We give an explicit 2D layout of qubits such that the required gates can be performed by using acousto-optic deflectors (AODs) manipulating the movement of a \emph{rectangular grid} of atoms. More specifically, we only need \emph{translations} of atom array, though AODs can do more general operations such as stretches and compressions \cite{Bluvstein_2023}. 

Additionally, since we are performing a near brute-force search in the automorphism space in order to find suitable permutations for protocols robust to three or four faults, the calculations necessary to test fault-tolerance need to be extremely fast. 
To speed up our tests, we borrow a technique from the meet-in-the-middle attack in classical cryptography.

Finally, we consider high-rate codes. It is well-known that QRM processes several high-rate families of codes, and that their automorphisms can be used to enact logical gates \cite{Grassl_2013}. The automorphism group of QRM codes, though enormous, is not enough to achieve all CNOT-type gates. However, we prove that this can be be achieved for a certain family of QRM codes by using the technique from \cite{Grassl_2013}, which is to interleave automorphism permutations with transversal CNOT gates between the current block and an ancillary block employing the \emph{same} code. Together with the transversal Hadamard and fold-transversal phase gate \cite{moussa-fold,fold_transversal, eberhardt2024logical}, we are able to complete the possible logical actions to the full Clifford group.

The remainder of the manuscript is organized as follows. Section~\ref{sec:preliminaries} provides some background knowledge in classical and quantum coding theory.
We summarize two important descriptions of classical RM codes in Section~\ref{sec:intro_RM}.
These descriptions are extensively used in the subsequent Section~\ref{sec:qrm} which gives the details of (punctured) quantum RM codes including automorphisms, transversal gates, and encoding circuits. 
Section~\ref{sec:logicalgates} describes how a universal set of quantum gates can be performed using transversal operations and code switching via Steane EC. 
We present the fault-tolerant encoding circuits required for Steane EC based on a layout of qubits in a 2D grid and parallel gates or movement between subgrids in Section~\ref{sec:FT_prep}. 
There, we also describe a meet-in-the-middle method of checking fault-tolerance of the automorphism-based encoding scheme.  
In Section~\ref{sec:simulation} and Appendix~\ref{sec:simulation_detail} we describe the performance of our scheme, based on numerical simulation of exRecs. 
Section~\ref{sec:high_rate} and Appendices~\ref{sec:CNOT_type} \& \ref{sec:phase_type} are dedicated to the Clifford gates of the high-rate QRM codes. The source code accompanying this work is available online\footnote{\href{https://github.com/gongaa/RM127}{https://github.com/gongaa/RM127}}.

\section{Preliminaries}
\label{sec:preliminaries}
\subsection{Classical codes}
An $[n,k]$ \emph{classical error correction code} encodes each sequence of $k$ \emph{information symbols} $\bu=u_1u_2\dots u_k$ into a length-$n$ \emph{codeword} $\bc=c_1c_2\dots c_n$. 
In this work, we will only deal with \emph{binary} codes, so $u_i,x_i$ take values in $\{0,1\}$. 
We call $n$ the \emph{blocklength}, and $k$ the \emph{dimension} of the code. A \emph{linear code} $C$ is such that if $\bc_1, \bc_2\in C$, then $\bc_1+\bc_2\in C$.

The \emph{Hamming weight} of a vector $\bx=x_1\dots x_t$, denoted $\wt(\bx)$, is the number of $x_i$ equal to $1$. Given two binary vectors $\bu, \bv$, we denote their sum over the binary field $\F_2$ (bitwise-XOR) by $\bu+\bv$. Further, define their \emph{overlap} (bitwise-AND) to be $\bu\wedge \bv$. One can check that 
\begin{equation}
\wt(\bu+\bv)=\wt(\bu)+\wt(\bv)-2\cdot\wt(\bu\wedge\bv).
    \label{eq:wt_sum}
\end{equation}
This equation implies that, if all the $k$ codeword generators specified for a linear code have even weight, then so does every codeword from this linear code. We call such a code an \emph{even} code.
The \emph{dual code} $C^{\perp}$ of a linear code $C$ contains all the $\bv$'s that are \emph{orthogonal} to every $\bu\in C$, i.e., $\wt(\bu\wedge \bv)$ is even.

Define the \emph{Hamming distance} $\dist(\bu,\bv)$ of two vectors $\bu, \bv$ to be the number of places where they differ; clearly $\dist(\bu,\bv)=\wt(\bu-\bv)$. For binary codes, $\bu-\bv=\bu+\bv$. The \emph{(minimum) distance of a code} is the minimum Hamming distance between its codewords: $d=\min\wt(\bu-\bv), \forall \bu,\bv\in C, \bu\neq\bv$. For a linear code, the distance is just the minimum weight of any nonzero codeword. 
As is standard, we refer to a $k$-dimensional blocklength-$n$ code of distance $d$ as an $[n,k,d]$ code. 

We will make use of the following operations on codes~\cite[Chapter~1]{theoryEC}.
\emph{Puncturing} a code refers to deleting some coordinates (positions) of each codeword, decreasing the blocklength. It is the inverse process of \emph{extending} a code by appending some parity-check results. For example, for each codeword, one can append its overall parity to the end, increasing the blocklength by $1$.

Meanwhile, \emph{expurgating} means discarding some codewords. For example, suppose $C$ is an $[n,k,d]$ binary code containing codewords of both odd and even weight. Then precisely half the codewords have even weight and half have odd weight. We can expurgate $C$ by throwing away the codewords of odd weight to get an $[n,k-1,d']$ code. Often $d'>d$ (for instance if $d$ is odd).
\emph{Shortening} is puncturing followed by expurgation. In other words, only a subcode of the original code is kept where some bits of the original codewords are restricted to a fixed value, e.g., zero.


Any permutation of the coordinate positions in a code $C$ creates an \emph{equivalent} code $C'$. That is, $C$ and $C'$ have the same minimum weight, weight distribution, etc. 
An \emph{automorphism} of $C$ is a permutation such that $C'=C$. The set of all automorphisms forms the \emph{automorphism group} of $C$, denoted $\Aut(C)$. 
A useful fact is that, if $C$ is linear, then $\Aut(C)=\Aut(C^{\perp})$. 

\subsection{CSS quantum codes}
\label{sec:CSS}
Quantum Calderbank-Shor-Steane (CSS) \cite{CSS-CS, CSS-Steane} codes are constructed from two classical linear codes $C_1$ and $C_2$ subject to the requirement that one is contained in the dual of the other, e.g., $C_2^\perp\subseteq C_1$. 
This implies $(C_2^{\perp})^{\perp}=C_2\supseteq C_1^{\perp}$. 
The codewords from $C_1^{\perp}$ and $C_2^\perp$ form the $X$ and $Z$-type stabilizers respectively. The $X$-type and $Z$-type logical operators are $C_2\backslash C_1^\perp$ and $C_1\backslash C_2^\perp$. 
If $C_1$ is an $[N,K_X]$ code, and $C_2$ is an $[N,K_Z]$ code, then the resulting quantum code is an $\llbracket N,K_X+K_Z-N\rrbracket$ code, i.e., encoding $K=K_X+K_Z-N$ logical qubits.

In this work, we use a notation equivalent to the above but more convenient for showing the gate set properties. We will specify $X$ and $Z$-type stabilizers as $C_X$ and $C_Z$, and logical operators as $L_X$ and $L_Z$. To relate to the above definition, set $C_X=C_1^{\perp}$ and $L_X=C_2\backslash C_1^\perp$. 
A row from $C_X$ is translated to an $X$-type stabilizer as, e.g. 1001 to $XIIX$, and a row from $C_Z$ to $Z$-type stabilizer as 1001 to $ZIIZ$.
It is sufficient to specify only $C_X$ and $L_X$, since $C_Z$ and $L_Z$ can be determined using Gaussian elimination.

The $X/Z$ distance of a quantum code is the minimum weight of any \emph{non-trivial} logical $X/Z$ operator. To be more explicit, the $Z$ distance is the minimum weight among coset codes $\bl+C_Z$ led by any $\mathbf{0}\neq \bl\in L_Z$.

In the following, we will mostly be dealing with odd-blocklength codes encoding $K=1$ qubit in which the $X$ and $Z$ logical are the all-one vector $\mathbf{1}$. It is thus beneficial to make things more explicit for this case here. The logical zero state is $|0\ra_L=\sum_{\bc\in C_X}|\bc\ra$ (normalization is ignored) and the logical one state is $|1\ra_L=\sum_{\bc\in C_X}|\mathbf{1}+\bc\ra$. One observes that any $\pi\in \Aut(C_X)$ preserves $|0\ra_L$, which is a superposition of all codewords from $C_X$. $\pi$ preserves $|1\ra_L$ as well because $\mathbf{1}\pi=\mathbf{1}$ for any permutation $\pi$. Therefore, $\pi$ preserves arbitrary state $\alpha |0\ra+\beta |1\ra$, i.e., an automorphism for $C_X$ is an automorphism for this specific CSS code we are considering.

For the $K>1$ case, the effect of classical component codes' automorphisms on the CSS code is more complicated \cite{Grassl_2013}. The high-rate RM codes we treat in Section~\ref{sec:high_rate} are special (thus easy) in the sense that $C_X=C_Z$ and hence $L_X=C_X^{\perp}\backslash C_X$. We know that $\Aut(C_X)=\Aut(C_X^{\perp})$, so a $\pi\in \Aut(C_X)$ will map a stabilizer $\in C_X=C_Z$ to a stabilizer, and map a non-trivial logical $\in C_X^{\perp}\backslash C_X$ to (possibly another) non-trivial logical operator. Hence $\pi$ effectively implements a logical CNOT-type gate. 

\subsection{Logical gates}
\label{sec:logical_gates}

An important set of universal gates are the Clifford gates augmented with $T$ gates. Besides CNOT, Clifford gates include Hadamard $H=\frac{1}{\sqrt{2}}\begin{psmallmatrix}1&1\\1&-1\end{psmallmatrix}$ and $S=\diag(1,e^{i\pi/2})=\sqrt{Z}$. 
We only consider CSS codes, which naturally admit transversal CNOT gate\cite{Steane_1999}. The $T$ gate is $T=\sqrt[\leftroot{-2}\uproot{2} 4]{Z}=\sqrt{S}=\diag(1,e^{i\pi/4})$.

Let us focus on the odd $N$ and $K=1$ case, where $X$-type and $Z$-type logical operators are both $\mathbf{1}$. Doubly-even self-dual codes are known for admitting transversal Clifford gates. 
Doubly-even means every codeword in $C_X$ (the $X$-type stabilizer) has weight divisible by $4$.
Self-dual means that expurgating $\mathbf{1}$ (and any odd-weight codewords) from $C_X^{\perp}$ obtains $C_X$. This implies $C_X=C_Z$, and hence such a code admits transversal Hadamard gate.  

The reason such codes have transversal $S$ gates is similar to the reason why odd $N$, $K=1$ codes with a triply-even $C_X$ (every codeword has weight divisible by $8$ \cite{triplyeven}) admit transversal $T$ gates \cite{Steane_1999}.
We explain this via the $\llbracket 15,1,3\rrbracket$ punctured QRM code \cite{knill1996}.
Since, for bare qubits, $T|0\ra=|0\ra$ and $T|1\ra=e^{i\pi/4}|1\ra$, it follows that $T^{\otimes 15}|\bc\ra=e^{i\cdot \wt(\bc)\pi/4}|c\ra$. If $8\vert\wt(\bc)$, then $e^{i\cdot \wt(\bc)\pi/4}=1$. Since $C_X$ is triply-even and $\wt(\mathbf{1})=15$, thus $T^{\otimes 15} |0\ra_L = T^{\otimes 15} \sum_{\bc\in C_X}|\bc\ra=|0\ra_L$ and $T^{\otimes 15} |1\ra_L = T^{\otimes 15} \sum_{\bc\in C_X}|\mathbf{1}+\bc\ra=e^{i\pi\cdot 7/4}|1\ra_L$. One sees that the transversal $T^{\otimes 15}$ gate implements a logical $T^{\dagger}$ gate on this code. Similarly, $T^{\dagger \otimes 15}$ implements a logical $T$ gate.
Of course, one can also let $C_Z$ be triply even, then the resulting code has transversal $TX=\sqrt[\leftroot{-2}\uproot{2} 4]{X}=HTH$ gate.

A generator matrix specifies a triply-even code if and only if\footnote{If 1.\ and 2.\ are relaxed to divisible by two, the code is called tri-orthogonal \cite{MSDlowoverhead}. One can then implement a logical $T$ via transversal $T$ and some diagonal correction such as $S$ or CZ.} \cite{triplyeven}: 1.\ Each row has weight divisible by eight; 2.\ The overlap between any two rows has weight divisible by four; 3.\ The overlap between any three rows has weight divisible by two.
Necessity of these conditions can be seen from Eq.~\ref{eq:wt_sum} and
\begin{align}
    \wt(\bu&+\bv+\bw) = \wt(\bu)+\wt(\bv)+\wt(\bw)\nonumber\\
    &-2\cdot \wt(\bu\wedge \bv) - 2\cdot \wt(\bu\wedge \bw) - 2\cdot \wt(\bv\wedge \bw)\nonumber \\&+ 4\cdot \wt(\bu\wedge\bv\wedge\bw).
    \label{eq:wt_three_terms}
\end{align}
The conditions are sufficient because, as one can prove, if a codeword is the sum of $x$ generators ($x\geq 4$), then the coefficient in front of an overlap of $x$ terms is $2^{x-1}$. With this observation, one can also generalize the above conditions to codes where every codeword has weight divisible by $2^{\nu}$.

We will show that the $C_X$ of $\llbracket 127,1,15\rrbracket$ \cite[Chapter~7]{preskill} is doubly-even and the $C_X$ of $\llbracket 127,1,7\rrbracket$ \cite{sublogarithmic} is triply-even. 
These conclusions will be more apparent using the polynomial formalism we introduce in the next section.

\section{Classical Reed-Muller codes}
\label{sec:intro_RM}
In this section we summarize two formalisms for describing classical Reed-Muller codes, shown in Figure~\ref{fig:polyeval}. 
The first \cite{polarvsRM} is based on a particular hypercube encoding circuit. 
The circuit itself is a reversible transformation of $n=2^m$ qubits and is identical for all codes in the (quantum) RM family with fixed $m$. 
The only difference for different codes is the particular inputs to the circuit. 
This picture also makes it simple to determine stabilizer containment relations and thus code-switching in Section \ref{sec:code_switch}. 
The second approach is the polynomial formalism \cite{theoryEC}, which allows one to immediately see the automorphism group and the gate set properties of QRM codes. It also provides a concise description of the logical operators and stabilizers, which we will employ for discussing high-rate codes in Section~\ref{sec:high_rate}.

\subsection{Hypercube encoding circuit and matrix}
\begin{figure}
    \centering
    \includegraphics[width=1.0\linewidth]{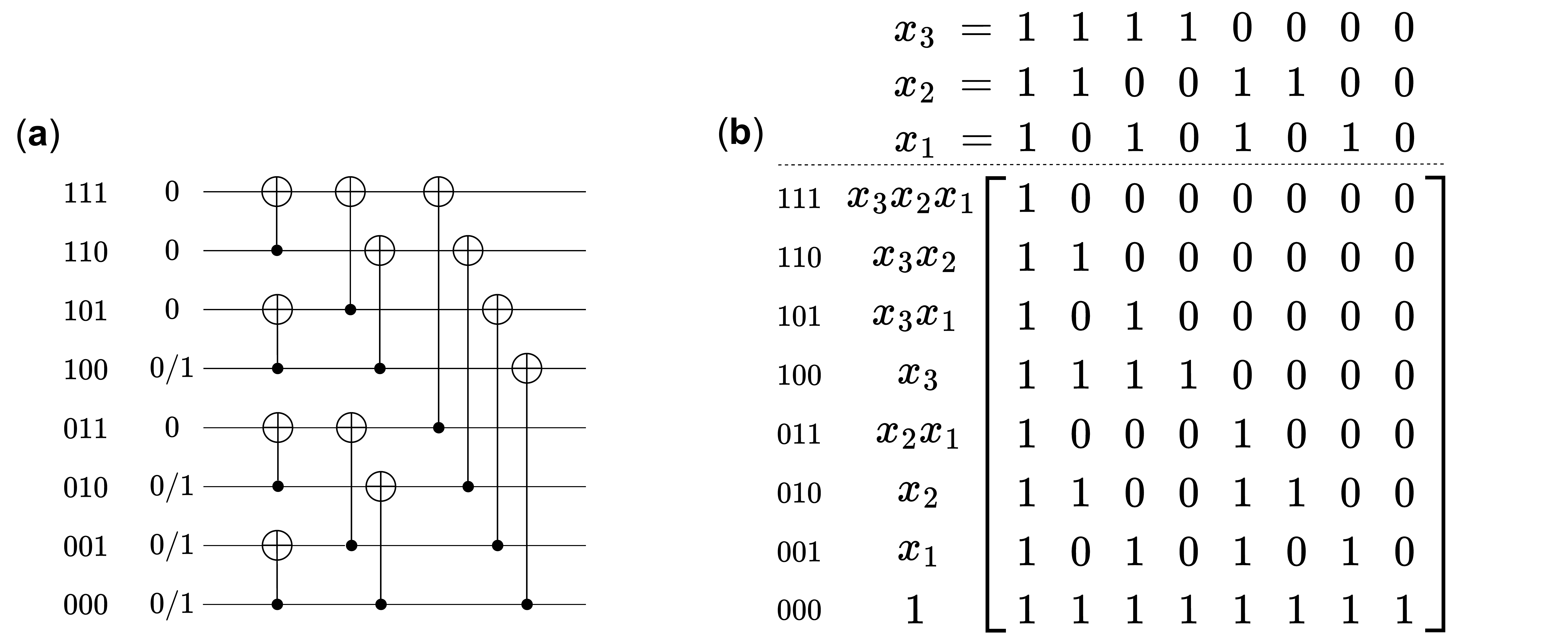}
    \caption{Two equivalent ways of understanding the classical Reed-Muller code. (\textbf{a}) via the hypercube encoding circuit. (\textbf{b}) polynomial evaluation. The information position (marked with $0$/$1$) on the left corresponds to which rows (evaluation vectors) are chosen from the encoding matrix.}
    \label{fig:polyeval}
\end{figure}

Fig~\ref{fig:polyeval}(a) shows the hypercube encoding circuit for $n=8$ code. 
Larger circuits can be constructed recursively: To construct an encoding circuit for $2^{k+1}$ bits, use two copies of encoding circuits for $2^k$ bits followed by transversal CNOT gates between the two blocks, all CNOT gates in the circuit oriented in the same direction. Labeling the bits from bottom to top using the binary expansion of $0$ to $n-1$, at time step $t$ a CNOT acts on each pair of bits that only differ at the $t^{th}$ bit. In other words, a bit interacts with one of its $m=\log(n)$ neighbors on the hypercube at a time.

The kernel of this circuit is $\begin{psmallmatrix}1&0\\1&1\end{psmallmatrix}$ as it represents the action of a single CNOT gate pointing in this upward direction. For the recursively-constructed hypercube encoding circuit on $2^m$ bits, the encoding matrix associated with it is $\bE=\begin{psmallmatrix}1&0\\1&1\end{psmallmatrix}^{\otimes m}$, e.g., the matrix in Fig.\ \ref{fig:polyeval}(b) is the encoding matrix for the circuit in Fig.\ \ref{fig:polyeval}(a). At the input side (left) of Fig.~\ref{fig:polyeval}(a), if one assigns a single one at a certain row while feeding zeros to the other rows, the result on the output side (right) will be the corresponding row of the encoding matrix.

The $2^m$ rows of $\bE$ are linearly independent since it is a lower triangular matrix with all-one diagonal. The $r^{th}$ order Reed-Muller code, denoted as $\RM(r,m)$, chooses the rows with labels containing $\leq r$ ones from $\bE$ to form its codeword generators \cite{polarvsRM}.

For example, Fig.~\ref{fig:polyeval} shows the $\RM(1,3)$ code. The input to the rows with labels containing $>1$ ones are always zero, indicating these rows are not selected as its codeword generators. We call this input bit a \emph{frozen} bit, otherwise an \emph{information} bit can be either $0$ or $1$.

From this row selection point of view, it is easy to see that the dimension of the $\RM(r,m)$ code is $k=1+{m\choose 1}+\dots+{m\choose r}$, and it is also obvious that $\RM(r',m)$ is contained in $\RM(r,m)$ if $r'<r$. One can show by induction that the minimum distance of $\RM(r,m)$ is $2^{m-r}$. 

To create the punctured Reed-Muller code $\RM(r,m)^*$ for $r<m$\footnote{The punctured $\RM(m,m)$ code is just the entire vector space $\mathbb F_{2^m-1}$.}, it is most convenient to delete the last (right-most) column from $\bE$ (otherwise we need to make the punctured column contain a single one through row operations). One can see that only the top row of $\bE$ does not have even weight; this row does not appear in $\RM(r,m)$ for $r<m$. Since only the last row has a 1 in the last column, deleting the last column only changes the weight of the last row. Denote the even-weight subcode of $\RM(r,m)^*$, i.e.\ the shortened  $\RM(r,m)$ code, by $\oRM(r,m)$. Its codewords generators are all the original ones from $\RM(r,m)$ except the last row. Contrary to $\oRM(r,m)$ being a linear code, the odd-weight subcode $\mathbf{1}+\oRM(r,m)$ is a coset code.

\subsection{Polynomial evaluation and automorphism}
\label{sec:polyeval}
In the polynomial formalism \cite[Chapter~13]{theoryEC}, $\RM(r,m)$ contains all evaluation vectors of polynomials in $x_1,\dots,x_m$ with degree $\leq r$. Let us explain this using Fig.~\ref{fig:polyeval}(b), where monomials are shown. The corresponding row of $\bE$ is the evaluation vector of the monomial labeled on the left. Taking the row $x_3 x_2$ as an example, 
this monomial is evaluated to one at two coordinates: $x_3=1,x_2=1,x_1=1$ and $x_3=1,x_2=1,x_1=0$. The coordinates (above the dotted line) are labeled vertically by the binary expansion of $0$ to $2^m-1$ from right to left. Later, we abbreviate the coordinates as, e.g., $x_3x_2x_1=111$.

We say that $\RM(r,m)$ is generated by monomials up to degree $r$. It is important to note that addition and multiplication are in the quotient ring $\F_2[x_1,\dots,x_m]/\la x_1^2=x_1,\dots, x_m^2=x_m\ra$, meaning that $2$ times any polynomial leads to zero, and $x_i^2=x_i$.

There are two important tricks in the polynomial formalism that we use throughout this work. The first is that the overlap of two polynomials (bitwise-AND of their evaluation vectors) is just their product. The second is that, if there are $i$ variables absent in a polynomial, then its weight is divisible by $2^i$. Here the weight counts the number of coordinates in which the polynomial evaluates to one. Once the value for the present variables are fixed, no matter how the absent variables vary (among the $2^i$ possible ways), the evaluation result of the polynomial is the same.

Two important properties of RM codes can be explained using the polynomial formalism. First, a row in $\bE$ with a label containing $s$ ones corresponds to a monomial containing $s$ variables and thus has weight $2^{m-s}$. Second, $\RM(r,m)$ is the dual code of $\RM(m-r-1,m)$. 
To see this, let $a$ and $b$ be monomials from each code, so that $a$ contains $\leq r$ variables and $b$ contains $\leq m-r-1$ variables. Then their overlap $ab$ must have an absent variable and therefore its weight is even.

Let $A=(a_{ij})$ be an invertible $m\times m$ binary matrix and $b$ be a binary length $m$ vector. The affine transform
\begin{equation}
    T: \text{replace } \begin{pmatrix}x_1\\ \vdots\\x_m\end{pmatrix} \text{ by } A\begin{pmatrix}x_1\\ \vdots\\x_m\end{pmatrix}+b
\label{eq:affine}
\end{equation}
preserves the degree of polynomials $f(x_1,\dots,x_m)$ and is thus an automorphism of the RM codes. The set of all such transformations $T$ forms the \emph{general affine group}, denoted by $\GA(m,\F_2)$. They form the automorphism group of the RM codes \emph{regardless of the order} $r$.

It is clear that the affine transform maps the codeword $f(x_1,\dots,x_m)$ to the codeword $f(\sum_j a_{1j}x_j+b_1,\dots, \sum_j a_{mj}x_j+b_j)$. Let us see how this is linked to the permutation of coordinates. As in Fig.~\ref{fig:polyeval}(b) we index the columns from right to left with the binary expansion of $0$ to $2^m-1$, represented as a length-$m$ vector, the coordinate permutation is precisely mapping the coordinate $x$ to $Ax+b$.

For example, take $m=2$ and an arbitrary polynomial in $x_1,x_2$, say $f(x_1,x_2)=x_1x_2+x_2+1$, its evaluation vector at $x_2x_1=11,10,01,00$ is $1011$. Take $A=\begin{psmallmatrix}1&1\\1&0\end{psmallmatrix}$ and $b=\begin{psmallmatrix}0\\0\end{psmallmatrix}$ which maps $x_1\mapsto x_1+x_2$ and $x_2\mapsto x_1$. Then the polynomial is mapped to $f(x_1+x_2,x_1)=g(x_1,x_2)=(x_1+x_2)x_1+x_1+1=x_1x_2+1$ (recall that $x_i^2=x_i$ and $x_i+x_i=0$), the evaluation vector at $x_2x_1=11,10,01,00$ is thus $0111$. With $A,b$ transforming the coordinates, instead of evaluating $g$ at $x_2x_1=11,10,01,00$, one can evaluate $f$ at $x_2x_1=10,01,11,00$ to obtain the same vector.

The affine transform is clearly an automorphism of RM codes, but it is less trivial to see that it forms \emph{all} the automorphism of these codes. For this, we refer the readers to the following theorem from \cite{theoryEC}.

\begin{theorem}
\label{thm:automorphsim}
    \textbf{(Thm. 24 of \cite[Chapter~13]{theoryEC})}. For $1\leq r\leq m-2$, $\Aut(\RM(r,m))=\GA(m,\F_2)$, $\Aut(\RM(r,m)^*)=\SL(m,\F_2)$.
\end{theorem}

The punctured Reed-Muller code $\RM(r,m)^*$ is obtained by deleting the last column of $\bE$, i.e., deleting the coordinate corresponding to $x_1=\dots=x_m=0$ in Fig.~\ref{fig:polyeval}(b). Therefore, to have an automorphism, the affine transform has to fix this coordinate, implying that $b$ must be zero. The automorphism group of punctured RM codes is thus the special linear group $\SL(m,\F_2)$ formed by all binary $m\times m$ invertible matrices $A$.


\section{(Punctured) Quantum Reed-Muller codes}
\label{sec:qrm}
Now we turn to quantum versions of Reed-Muller codes.

\subsection{QRM codes}
A CSS quantum RM code can be constructed from $\RM(r_x,m)$ and $\RM(r_z,m)$ as long as $r_x+r_z<m$ since they are orthogonal to each other. The orthogonality follows from $\RM(r_x,m)$ being the dual code of $\RM(m-r_x-1,m)\supset\RM(r_z,m)$ since $r_z\leq m-r_x-1$.
We denote the resulting code $\QRM(r_x,r_z,m)$.

\begin{definition}
\label{def:QRM}
    The blocklength $2^m$ quantum Reed-Muller code $\QRM(r_x,r_z,m)$ has $X$-type stabilizers $\RM(r_x,m)$, and $Z$-type stabilizers $\RM(r_z,m)$.
\end{definition}

From the discussion above, it follows that $\QRM(r_x,r_z,m)$ encodes $\sum_{i=r_x+1}^{m-r_z-1}{m\choose i}$ qubits. 
Its $X$- and $Z$-type logical operators are $\RM(m-r_z-1,m)\backslash\RM(r_x,m)$ and $\RM(m-r_x-1,m)\backslash\RM(r_z,m)$, respectively. 
This implies that the $X$ distance is $2^{r_z+1}$ and $Z$ distance is $2^{r_x+1}$.

To name some examples, the $\llbracket 8,3,2\rrbracket$  code implemented on the neutral atom platform \cite{Bluvstein_2023} and more generally, the $\llbracket 2^D,D,2\rrbracket$ codes \cite{FT-IQP} are $\QRM(0,D-2,D)$ codes. The $\llbracket 16,6,4\rrbracket$ tesseract code \cite{tesseract} is the $\QRM(1,1,4)$ code.

The hypercube circuit can also be used to encode QRM codes \cite{qpc_renes}. 
Now the inputs are not $0$s or $1$s, but particular quantum states. Again we label the rows from bottom to top using the binary expansion of $0$ to $2^m-1$.

\begin{lemma}
\label{lemma:initialization}
    To encode $\QRM(r_x,r_z,m)$, $|0\ra$ is input on rows with labels containing $\geq m-r_z$ ones, $|+\ra$ on rows with labels containing $\leq r_x$ ones, and an arbitrary quantum state to be encoded on the remaining rows.
\end{lemma}

\begin{figure}
    \centering
    \includegraphics[width=1.0\linewidth]{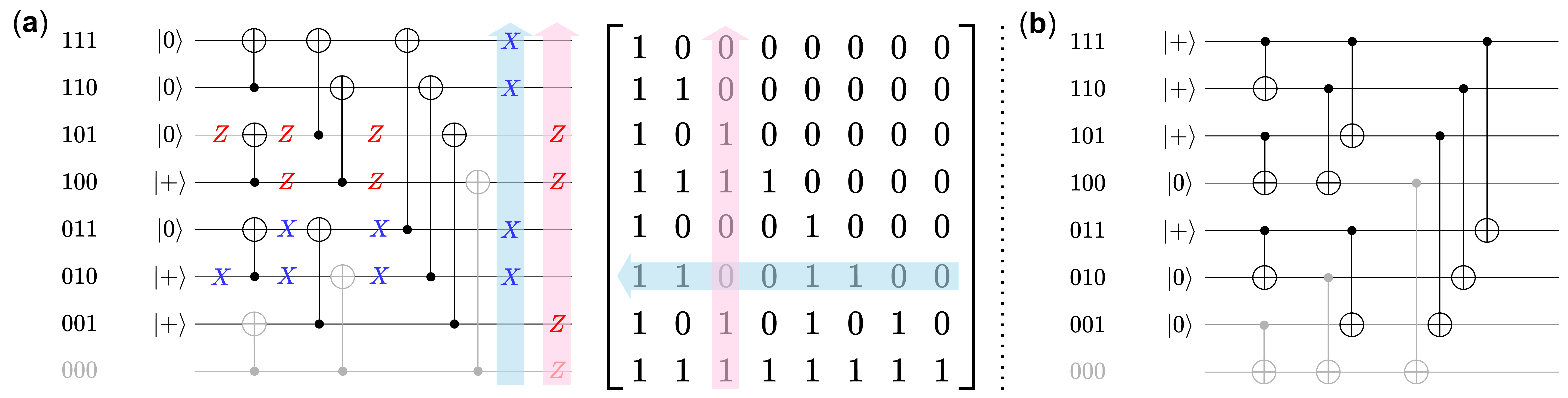}
    \caption{The QRM encoding circuit is the same as the one for classical RM codes, just the input are quantum states. (\textbf{a}) Showing the $X$ and $Z$ stabilizers of quantum RM codes obtained by propagating the stabilizers from the input (acting on a single qubit) to the output. To create the logical zero state $|0\ra_L$ of punctured QRM code, remove the bottom-most wire and all CNOT gates involving it. (\textbf{b}) To create the logical plus state $|+\ra_L$ with the correct chirality, one reverses the direction of all the CNOT gates, as well as bit-reverse and the input states. It is still the bottom-most wire that is removed.}
    \label{fig:pQRM}
\end{figure}

To see why this input assignment gives the stated stabilizers, consider a $|+\ra$ state at the input, e.g., 
on row $010$ of Fig.~\ref{fig:pQRM}.
It is stabilized by an $X$ operator. Propagating this stabilizer to the output side, we obtain a stabilizer for the state after encoding. That stabilizer corresponds to the same row of the encoding matrix $\bE$ because $X$ propagates like $1$ through classical CNOT gates. Therefore, it is obvious to see that by putting $|+\ra$ at the input side on rows with labels containing $\leq r_x$ ones, the resulting $X$-type stabilizers are just the $\RM(r_x,m)$ code. 

The $Z$-type stabilizers are not quite as straightforward to see. $Z$ propagates in the reversed way and hence it sees the kernel of $\bE$ as $\begin{psmallmatrix}1&1\\0&1\end{psmallmatrix}$. 
Therefore the propagated $Z$ stabilizer is the corresponding row of $\begin{psmallmatrix}1&1\\0&1\end{psmallmatrix}^{\otimes m}=\bE^T$, which is the same as the corresponding column of $\bE$. Noticing that $\bE$ is symmetric with respect to the skew diagonal, one may argue that the resulting $Z$-type stabilizers form the bit-reversed $\RM(r_z,m)$ code, instead of $\RM(r_z,m)$. But luckily, they are the same code because RM codes enjoy \emph{bit-reversal symmetry}, as can be seen taking $A$ to be the identity matrix and $b=\mathbf{1}$ in Eq.~\ref{eq:affine}.

In terms of the stabilizer commutation relationship, one notices that if two stabilizers commute at the beginning, then after propagating through a unitary encoding circuit, they still commute. Indeed at the input side, all stabilizers commute because no qubit is simultaneously frozen in $|0\ra$ and $|+\ra$.

\subsection{PQRM codes}
Let us now proceed to punctured QRM (PQRM) codes. They can be defined by starting from QRM codes and tracing out qubits, but for our purposes it is more direct to simply state their stabilizer generators. 
\begin{definition}
\label{def:PQRM}
    The punctured by one QRM code $\PQRM(r_x,r_z,m)$ has blocklength $2^m-1$, $X$-type stabilizers $\oRM(r_x,m)$, i.e., the even-weight subcode of $\RM(r_x,m)^*$, and $Z$-type stabilizers $\oRM(r_z,m)$. 
\end{definition}

The $\PQRM(r_x,r_z,m)$ code has one fewer $X$-type stabilizer and one fewer $Z$-type stabilizer than $\QRM(r_x,r_z,m)$, but also one fewer qubit, and therefore encodes $1+\sum_{i=r_x+1}^{m-r_z-1}{m\choose i}$ logical qubits.

We specialize to the case $r_x+r_z+1=m$ from now on (except in Section \ref{sec:high_rate} where we consider high-rate codes). 
These codes encode a single logical qubit, whose $X$- and $Z$-type logical operators are $\otimes_{i=1}^{2^m-1}X_i$ and $\otimes_{i=1}^{2^m-1}Z_i$, respectively.

Importantly, encoding the logical state $|0\ra_L$ can be performed by the hypercube circuit. 
\begin{lemma}\label{lemma:encode0}
To encode the $\ket0_L$ state of $\PQRM(r_x,m-r_x-1,m)$, reuse the hypercube encoding circuit for $\QRM(r_x,m-r_x-1,m)$, but remove the bottom-most qubit and all CNOT gates involving it.
\end{lemma}
This is depicted in Fig.~\ref{fig:pQRM}(a), and can be verified in the stabilizer propagation picture. The claim is more straightforward for the $X$-type stabilizers. By puncturing the bottom-most qubit, we remove the all-$X$ stabilizer since the input state in the associated $\QRM$ code is $\ket{+}$. Moreover, the punctured coordinate is always zero for stabilizers propagated from other qubits, so these stabilizers remain even weight after puncturing. 

For $Z$-type stabilizers, one needs to keep in mind the difference between stabilizers for a \emph{state} and stabilizers for a \emph{code}. 
The $Z$-type stabilizers for the PQRM code are $\oRM(r_z,m)$, while the $Z$-type stabilizers for $|{0}\ra_L$ state are the superset $\RM(r_z,m)^*$.
In Fig.~\ref{fig:pQRM}(a), the $Z$ stabilizer on the topmost qubit at the input side propagates to the all-$Z$ stabilizer for the $|0\ra_L$ state, while being a logical operator for the code. Recall that RM codes have bit-reversal symmetry, hence before puncturing, the $Z$-stabilizers for the code are $\RM(r_z,m)$; after puncturing, the $Z$-type stabilizers for the $|0\ra_L$ state is $\RM(r_z,m)$ with the bottom-most coordinate being removed, i.e., $\RM(r_z,m)^*$ is obtained. 

The logical state $|+\ra_L$ can be encoded in a very similar way, by removing the topmost wire in Fig.~\ref{fig:pQRM}(a). 
However, we will instead use Fig.~\ref{fig:pQRM}(b) that removes the bottom qubit.
\begin{lemma}
\label{lemma:PQRM_plus}
    To encode the $|+\ra_L$ state of $\PQRM(r_x,r_z=m-r_x-1,m)$, reverse all the CNOT gates and the input state assignments. $|0\ra$ is now input on rows with labels containing $\leq r_z$ ones, $|+\ra$ on rows with labels containing $\geq m-r_x$ ones.
\end{lemma}
This will be convenient later when we consider the qubit layout for implementation in atom arrays in Table~\ref{tab:hypercube_layout}, as it will be desirable to have the punctured qubit in the same corner for both $|0\ra_L$ and $|+\ra_L$, e.g. both in the bottom right corner.
The same argument as above ensures that the circuit in Fig.~\ref{fig:pQRM}(b) indeed prepares $|+\ra_L$, since by bit reversal symmetry it follows that the $X$-type stabilizers of the $|+\ra_L$ state are $\RM(r_x,m)^*$ and $Z$-type stabilizers are $\oRM(r_z,m)$. 

Another way to encode $|+\ra_L$ in $\PQRM(r_x,r_z,m)$ would be to first encode $|0\ra_L$ in $\PQRM(r_z,r_x,m)$ as in Fig.~\ref{fig:pQRM}(a), then apply a transversal Hadamard gate. Transversal gates preserve fault tolerance, but one may want to avoid as many quantum operations as possible since they are noisy. The direct encoding in Lemma~\ref{lemma:PQRM_plus} and Fig.~\ref{fig:pQRM}(b) is more useful in this sense.

\subsection{Checking for stabilizers}
\label{sec:is_stabilizer}
In later sections~\ref{sec:FT_prep} \& \ref{sec:simulation} where we delve into fault-tolerant encoding, we need to test whether the transversal measurement result (a bit-string) is a stabilizer of the logical state. 
This is a simple task, as we can propagate this string backward through the (classical) encoding circuit to unencode. For example, in Fig.~\ref{fig:polyeval}(a), put the measurement result string on the right side and append an additional zero on the bottom-most wire, then propagate to the left. If all the rows with labels containing $>r$ ones are zero, then the measurement string belongs to $\RM(r,m)^*$. If,  additionally, the propagated result at the bottom-most wire also turns out to be $0$, then this string is in $\oRM(r,m)$, otherwise, it is in $\mathbf{1}+\oRM(r,m)$.

Propagating backward is the same as being encoded by $\bE^{-1}$, but $\bE$ is the inverse of itself, so encoding using the backward or forward version has the same effect.
In later figures, we choose to draw the transversal measurement result passing through $\bE$ in the forward direction, for instance in Fig.~\ref{fig:exp-vs-stim}(a) bottom-right corner.

To save time in our simulations we will use Stim \cite{stim} to post-process if the measurement is a stabilizer, as shown in Fig.~\ref{fig:exp-vs-stim}(b). The noiseless \emph{quantum} CNOT gates are used where $Z$ propagates in the reversed direction. One might think this will cause extra complication, but the acceptance criterion actually becomes simpler. After propagating through the original hypercube encoding circuit, measure in $Z$/$X$ basis at wires initialized to $|0\ra$/$|+\ra$ (initializations are described in Lemma~\ref{lemma:initialization}\&\ref{lemma:PQRM_plus}). If we are testing for $X$/$Z$ type stabilizer, we accept if all $Z$/$X$ basis measurement results are zero.

As an example, Fig.~\ref{fig:exp-vs-stim}(a) tests if the (noisy) transversal $Z$-basis measurement results turn out to be an $X$-type stabilizer. We want to do the same thing with Stim in Fig.~\ref{fig:exp-vs-stim}(b), we add $X$ noise before propagation, then measure noiselessly, and accept if all $Z$-basis measurements are zero. We ignore any $X$-basis measurement result since they cannot be obtained in Fig.~\ref{fig:exp-vs-stim}(a).
The measurement of the third block in Fig.~\ref{fig:fullStim} serves as an example of testing for $Z$-type stabilizer.

\section{Logical gates}
\label{sec:logicalgates}
Now we turn to the question of logical gates for PQRM codes. 

\subsection{Transversal gates}
Being CSS codes, PQRM codes have transversal CNOT gates. Let us discuss what other transversal gates that PQRM codes support. Again we impose $r_x+r_z+1=m$. 

Note that we only consider puncturing one qubit in this work, and hence the lemmas in this section are special cases of \cite{sublogarithmic} that considers puncturing coordinates containing $\leq w$ ones, i.e. $\sum_{i=0}^{w-1}{m\choose i}$ qubits are removed. There, the resulting $\llbracket \sum_{i=w+1}^m {m\choose i}, \sum_{i=0}^w {m\choose i}, \sum_{i=w+1}^{r+1}{r+1\choose i}\rrbracket$ code admits a transversal logical gate at the $\nu$-th level of Cliﬀord hierarchy where $m> \nu r$. Our $\PQRM(r,m-1-r,m)$ code is just the $w=0$ case, and for this $K=1$ code, $\otimes_{j=1}^{2^m-1} \diag(1, \exp(2\pi i/{2^\nu}))$ acting transversally on each qubit leads to a logical action of $\diag(1, \exp(-2\pi i/{2^\nu}))$ on the single logical qubit\footnote{For the $K=\sum_{i=0}^w {m\choose i}$ code ($w\geq 1$) in \cite{sublogarithmic}, such a transversal gate affects \emph{all} logical qubits. It is not yet known how to address the logical qubits individually.}.  

\begin{lemma}
    $\PQRM(\frac{m-1}{2},\frac{m-1}{2},m)$ with odd $m>1$  admits transversal $H$ and $S$ gates.
\end{lemma}
The code has transversal $H$ because $C_X=C_Z=\oRM(\frac{m-1}{2},m)$, and logical $X$ and $Z$ are both $\mathbf{1}$. 

$C_X$ is generated by monomials up to degree $\frac{m-1}{2}$; let us see why it is doubly-even. Since those generators all evaluate to zero at the punctured coordinate $00\dots 0$, the polynomial formalism is not affected by this puncturing.
Generator rows of $C_X$ selected from $\bE$ have weight divisible by $2^{(m+1)/2}$. The overlap between any of two of these rows has even weight, because there is at least one absent variable in their product monomial.

Similarly, $C_X=\oRM(r_x,m)$ with $r_x\leq \frac{m-1}{3}$ is triply-even. Overlap between any three degree $\leq \frac{m-1}{3}$ monomials contain at least one absent variable and is thus even. 
\begin{lemma}
    $\PQRM(r_x,r_z=m-r_x-1,m)$ with $r_x\leq \frac{m-1}{3}$ has transversal $T$ gate. Similarly, the code has transversal $TX=HTH$ gate if $r_z\leq \frac{m-1}{3}$. 
\end{lemma}
As commented in Section~\ref{sec:logical_gates}, one can generalize the conditions for triply-even to weight divisible by $2^\nu$.
\begin{lemma}
    $\PQRM(r_x,m-r_x-1,m)$ with $r_x\leq \frac{m-1}{\nu}$ has transversal $Z^{1/2^{\nu-1}}=\diag(1,\exp(2\pi i/2^\nu))$ gate. Similarly, the code has transversal $X^{1/2^{\nu-1}}$ gate if $r_z\leq \frac{m-1}{\nu}$.
\end{lemma}

In particular, one can see that the $\llbracket 127,1,15\rrbracket$ $\PQRM(3,3,7)$ code admits transversal $H$ and $S$ gate, and the $\llbracket 127,1,7\rrbracket$ $\PQRM(2,4,7)$ code admits transversal $T$ gate.

\subsection{Code switching}
\label{sec:code_switch}

A universal set of quantum gates can be implemented by switching between these two codes, using $\PQRM(3,3,7)$ for Clifford operations and $\PQRM(2,4,7)$ for the $T$ gate. Fortunately, switching between these codes is quite easy by using Steane error correction. Let us first give a little background on the method.

Paetznick and Reichardt suggest implementing the logical Hadamard for the $\PQRM(1,2,4)$ $\llbracket 15,1,3\rrbracket$ code as transversal Hadamard followed by Steane error correction~\cite{15-1-3}. We can interpret their result as follows. After the transversal Hadamard gate, the quantum information is encoded in the $\PQRM(2,1,4)$ code, and to switch the code back to $\PQRM(1,2,4)$, they couple the data block with an ancillary block encoded in $|+\ra_L$ of $\PQRM(1,2,4)$ through transversal CNOT. 

These two codes have many stabilizers in common. In particular, the $X$-stabilizers of the target code (in the ancilla block) are contained in those of the temporary code (in the data block), which lacks some $Z$-type stabilizers present in the target code.
The transversal CNOT copies the unwanted $X$-type stabilizers onto the ancilla and copies the desired $Z$-type stabilizers from the ancilla to the data. Those unwanted $X$-type stabilizers are treated as noise, and thus through transversal $Z$-basis measurements of the ancilla, we know how to correct them. The logical information is not disturbed because the two codes share the same logical operators, and the logical stabilizer of the ancilla does not get copied to the data block.

By the same logic, it is possible to perform a more general transversal code-switching between PQRM codes, as follows.
\begin{theorem}
    Provided $r_x+r_z=r'_x+r'_z=m-1$,  $\PQRM(r_x,r_z,m)$ can be transformed into $\PQRM(r'_x,r'_z,m)$ via Steane error correction, and vice versa. 
    Assuming $r_x>r'_x$, the forward direction is achieved by using the $|+\ra_L$ state of $\PQRM(r'_x,r'_z,m)$ as the ancilla. 
    The reverse direction is achieved by employing $|0\ra_L$ of $\PQRM(r_x,r_z,m)$.
\end{theorem}

\begin{figure}
    \centering
    \includegraphics[width=1.0\linewidth]{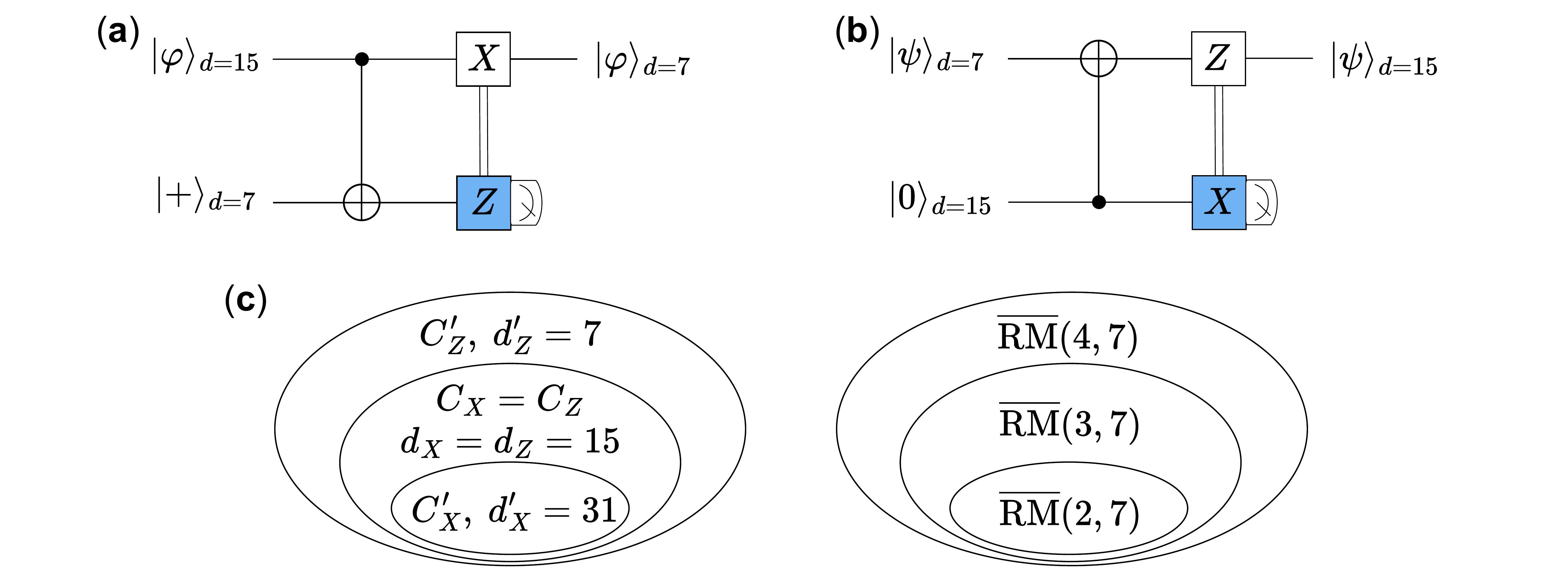}
    \caption{Code switching via Steane error correction (a) from $\PQRM(3,3,7)$ to $\PQRM(2,4,7)$, (b) the reverse direction. These two circuits hold for all PQRM codes. (c) Stabilizer containment relationship. $C_X$/$C_Z$ being the $X$/$Z$-type stabilizers of $\PQRM(3,3,7)$ and $C'_X/C'_Z$ are those of $\PQRM(2,4,7)$. The shown distance, e.g., $d'_X=31$ is the minimum distance between $C'_X=\oRM(2,7)$ and $\mathbf{1}+C'_X$. }
    \label{fig:codeswitch}
\end{figure}

Upon measuring a bit-string $\tbc$, the correction to apply is either $\tbc$ or $\mathbf{1}+\tbc$, depending on whether $\tbc$ is closer to $\oRM(r_x,m)$ or $\mathbf{1}+\oRM(r_x,m)$ in the forward direction; and whether $\tbc$ is closer to $\oRM(r'_z,m)$ or $\mathbf{1}+\oRM(r'_z,m)$ in the reverse direction.

Let us show that the claim holds using the particular example in Fig.~\ref{fig:codeswitch}.
Assume for a moment that no noise is present in Fig.~\ref{fig:codeswitch}(a). 
Without the transversal CNOT, a transversal measurement in the $Z$-basis of $|+\ra_{d=7}$ would yield an $X$-type stabilizer of the state, namely an element of $\RM(2,7)^*$. With the transversal CNOT, the $X$-type stabilizers of $|\varphi\ra_{d=15}$, which is $\oRM(3,7)$ are copied to the ancilla.
The subsequent $Z$-basis measurement will fix the gauge for the data block, i.e., a random $X$-type $\bc\in\oRM(3,7)$ becomes the noise on $|\varphi\ra_{d=7}$. However, the measurement result will be $\bc$ plus a random codeword $\bc'$ of $\RM(2,7)^*$. If we directly undo the noise by $\bc+\bc'$, we will introduce a logical error\footnote{We do not care about the $\oRM(2,7)$ part because they form the $X$-type stabilizers of $\PQRM(2,4,7)$.} if $\bc'\in\mathbf{1}+\oRM(2,7)$.
Therefore, we need to determine if $\bc+\bc'$ is in $\oRM(3,7)$ or $\mathbf{1}+\oRM(3,7)$, this is easy, just check the parity of its weight.


In the presence of noise, when measuring a noisy version $\tbc$ of $\bc+\bc'\in\RM(3,7)^*$, one still applies $\tbc$ or $\mathbf{1}+\tbc$, depending on whether $\tbc$ is closer to $\oRM(3,7)$ or $\mathbf{1}+\oRM(3,7)$. The decoder thus needs to make a binary decision, and the distance of this decoding problem is the distance between $\oRM(3,7)$ and $\mathbf{1}+\oRM(3,7)$, which is $15$.

By a similar argument, for Fig.~\ref{fig:codeswitch}(b), the transversal measurement in $X$-basis yields a noisy version $\tbc$ of $\RM(4,7)^*$. The decoder needs to decide whether it is $\oRM(4,7)$ or $\mathbf{1}+\oRM(4,7)$ that $\tbc$ is closer to. The distance is only $7$, and this switching direction is the most error-prone step in the $T$ gate extended rectangle, i.e., the third block of Fig.~\ref{fig:exRec}(a).

\subsection{Data qubit noise decoding}
\label{sec:data_noise_decode}
Before proceeding to circuit-level simulation, we first investigate how the successive cancellation list (SCL) decoder \cite{Dumer3, SCL-CRC} performs under a plain data qubit noise model.

\begin{figure}[hbt]
    \centering
    \begin{tikzpicture}
{\small

\pgfplotsset{every tick label/.append style={font=\footnotesize}}
\begin{axis}[
width=0.3\textwidth,
log basis y={10},
tick align=inside,
tick pos=left,
x grid style={lightgray211},
xlabel={\small Physical error rate \(\displaystyle p\)},
xmajorgrids,
xmode=log,
xmin=0.001, xmax=0.1,
xtick={0.001, 0.01, 0.1},
xticklabels={
  \(\displaystyle {0.001}\),
  \(\displaystyle {0.01}\),
  \(\displaystyle {0.1}\),
},
minor xtick={0.002,0.005,0.02,0.05},
xminorgrids=true,
scaled x ticks=false,
xtick style={color=black},
y grid style={lightgray211},
ylabel={\small Logical error rate \(\displaystyle P_L\)},
ylabel near ticks,
ymajorgrids,
ymin=1e-08, ymax=1,
ymode=log,
ytick style={color=black},
ytick={1e-8,1e-6,1e-4,0.01,1},
yticklabels={
  \(\displaystyle {10^{-8}}\),
  \(\displaystyle {10^{-6}}\),
  \(\displaystyle {10^{-4}}\),
  \(\displaystyle {0.01}\),
  \(\displaystyle {1}\),
},
minor ytick={1e-7,1e-5,1e-3,0.1},
error bars/y dir=both,
error bars/y explicit,
transpose legend,
legend style={at={(1.05,1.1)},anchor=north west, font=\scriptsize, text opacity=1, draw=none},
legend cell align=left,
legend image post style={xscale=.75}
]


\addplot [darkviolet, mark=*, mark size=0.5, mark options={solid}]
table [x=noise, y=scl, y error plus=error, y error minus=error] {r4.dat};
\addlegendentry{$\llbracket 127,1,7\rrbracket$, phase-flip}

\addplot [darkcyan, mark=*, mark size=0.5, mark options={solid}]
table [x=noise, y=scl, y error plus=error, y error minus=error] {r3.dat};
\addlegendentry{$\llbracket 127,1,15\rrbracket$, bit/phase-flip}

\addplot [blue, mark=*, mark size=0.5, mark options={solid}]
table [x=noise, y=scl, y error plus=error, y error minus=error] {r2.dat};
\addlegendentry{$\llbracket 127,1,7\rrbracket$, bit-flip}

\addplot[lavender, dashed, samples=100, domain=1e-3:1e-1] {413385*x^4};
\addlegendentry{$413385 p^4$}
\addplot[lavender, line width=1pt, dotted, samples=100, domain=1e-3:1e-1] {413385*x^4*(1-x)^123};
\addlegendentry{$413385 p^4 (1-p)^{123}$}

\addplot[limegreen, dashed, samples=100, domain=1e-2:1e-1] {76003785*x^8};
\addlegendentry{$76003785 p^8$}
\addplot[limegreen, forget plot, line width=1pt, dotted, samples=100, domain=1e-2:1e-1] {76003785*x^8*(1-x)^119};

\addplot[lightskyblue, dashed, samples=100, domain=0.03:1e-1] {801540700065*x^16};
\addlegendentry{$801540700065 p^{16}$}

\end{axis}
}
\end{tikzpicture}
    \vspace*{-4mm}
    \caption{Data qubit noise decoding on BSC($p$). List size $8$ for all simulations. The decoder needs to distinguish between whether the added noise is closer to $\oRM(r,m)$ or $\mathbf{1}+\oRM(r,m)$ for $r=4$ (purple), $3$ (green), $2$ (blue).}
    \label{fig:RM_data_qubit_noise}
\end{figure}
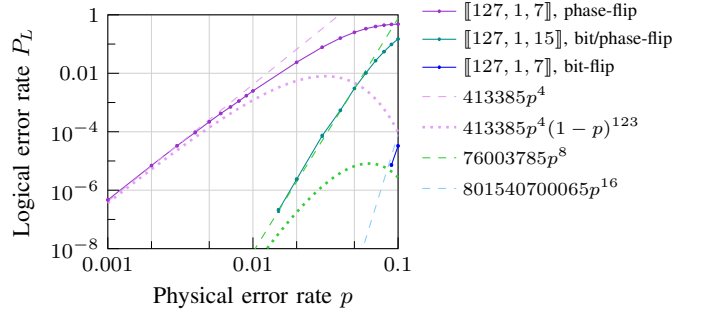

In Fig.~\ref{fig:RM_data_qubit_noise}, we add i.i.d. bit-flip noise with probability $p$ to a codeword $\bc$ from $\RM(r,m)^*$ and let SCL decode the noisy codeword $\tbc$. SCL tries to find the closest codeword $\hbc\in\RM(r,m)^*$ to $\tbc$. We record a logical error if $\hbc$ and $\bc$ belong to different cosets of $\oRM(r,m)$. The purple, green and blue curves correspond to $r=4,3,2$.

In our simulations, we observe that SCL is isotropic, possibly because the distance is odd\footnote{The minimum distance of $\tbc$ to the two cosets must be of different parity. Were there tie-breaking, SCL might be biased due to its implementation.}. The simulation method above has no difference to directly letting SCL decode whether the added noise is closer to $\oRM(r,m)$ or $\mathbf{1}+\oRM(r,m)$, and record a logical error if the latter is decided. In other words, we let $\bc=\mathbf{0}$ in the above simulation. In this picture, we can calculate a few lower bounds in Fig.~\ref{fig:RM_data_qubit_noise} shown in dotted lines, for example, $413385 p^4 (1-p)^{123}$. The coefficients are calculated based on the following results on the number of lowest-weight codewords in classical punctured RM codes. 



\begin{theorem}
\textbf{(Thm. 9 of \cite[Chapter~13]{theoryEC}).} The number of codewords of minimum weight in:\\
(a) $\RM(r,m)^*$ is $A_{2^{m-r}-1}=\prod_{i=0}^{m-r-1}\frac{2^{m-i}-1}{2^{m-r-i}-1}$,\\
(b) $\RM(r,m)$ is $A_{2^{m-r}}=2^r \prod_{i=0}^{m-r-1}\frac{2^{m-i}-1}{2^{m-r-i}-1}$.
\end{theorem}

The coefficient $413385$ is ${7\choose 4}\cdot A_7=35\cdot \frac{2^7-1}{2^3-1}\cdot \frac{2^6-1}{2^2-1}\cdot \frac{2^5-1}{2^1-1}$. The reason is as follows. Take any weight $7$ codeword from $\RM(4,7)^*$, since it is of odd-weight, it lies in $\mathbf{1}+\oRM(4,7)$. Consider an arbitrary weight four error whose support is contained in any weight $7$ codeword from $\mathbf{1}+\oRM(4,7)$, it will lead to a logical error in decoding because it is closer to this coset. Moreover, it cannot be simultaneously contained in two weight-$7$ codewords from $\mathbf{1}+\oRM(4,7)$, since the Hamming distance between any two codewords within this coset code is at least $8$. Similarly, the other two coefficients are calculated as $76003785={15\choose 8}\cdot A_{15}$ and $801540700065={31\choose 16}\cdot A_{31}$.

In the low logical error region, we observe that the simulation results can be fit by dropping the power of $(1-p)$ in the lower bounds, shown in dashed curves in Fig.~\ref{fig:RM_data_qubit_noise}. This plays a role when we estimate the logical error rate in the Clifford extended rectangles later in Table~\ref{table:exRec_simulation}(a).

Let us provide a few more details of the list decoder's usage. The SCL decoder is an adapted version of the aff3ct\cite{Cassagne2019a} package. SCL is a soft-input decoder, for BSC($p$), one initializes the decoder bitwisely as $\ln\frac{1-p}{p}$ for receiving $0$ and $\ln\frac{p}{1-p}$ for $1$. From our experience, SCL is insensitive to the precise value of $p$, therefore, we always use the initialization $+1$ for $0$ and $-1$ for $1$ throughout our simulations. To inform the decoder that the bottom-most bit is punctured, the soft input for that bit is set to $0=\ln\frac{1-0.5}{0.5}$. Besides the closest codeword $\hbc$, SCL also returns an unencoded version of it, i.e., what is on the left side of Fig.~\ref{fig:polyeval}(a). As commented in Section~\ref{sec:is_stabilizer}, if the bottom-most wire is $0$ then $\hbc$ belongs to $\oRM(r,m)$, otherwise $\hbc\in\mathbf{1}+\oRM(r,m)$.

\section{Fault-tolerant state preparation}
\label{sec:FT_prep}
For a fully fault-tolerant computational scheme it remains to construct a fault-tolerant preparation of the $|0\ra_L$ and $|+\ra_L$ states of the two $N=127$, $K=1$ PQRM codes. That is the topic of this section. Moreover, here we are interested in implementing the codes on platforms with long-range connectivity, such as the neutral atom  \cite{Bluvstein_2023} or ion trap architectures. 
In particular, the neutral atom platform has demonstrated the implementation of certain hypercube circuits involving $>200$ qubits \cite{Bluvstein_2023}, hence our designed protocols are tailored to them.

This platform is known for the ability to implement parallel CNOT gates \cite{Evered_2023,Bluvstein_2023}, perfect for the transversal CNOT gates needed in encoding and error detection and correction. Global single qubit gates that apply the \emph{same} operation to each atom are also native to this platform and have a much higher fidelity than two-qubit gates. However, it should be mentioned that the current solution for applying \emph{different} single qubit gates to atoms, e.g., needed in the correction, might involve addressing a single atom at a time.

Atoms can also be coherently transported by a pair of AODs independently controlling the X/Y coordinates, provided the acceleration is not too large. Due to this feature, the native movements implementable with AODs are translations, stretches, and compression of a rectangular grid of atoms \cite{Bluvstein_2023}.

Throughout this section, we assume errors do not spread through permutations, and we also do not take atom loss into account. Classical computations, e.g., decoding, testing if the transversal measurement result, etc., are all assumed to be noiseless.

\subsection{2D layout and encoding circuits}
\label{sec:2d_layout}
We provide a 2D layout of all the atoms in the hypercube in Table~\ref{tab:hypercube_layout}.
The layout is recursively constructed by expanding to the left/above when the dimension of the hypercube grows from even/odd to odd/even. For the PQRM codes, the atom at the all-zero coordinate is punctured. 

The initial (not fault-tolerant) preparation of encoded $|0\ra_L$ and $|+\ra_L$ states follows from Lemmas~\ref{lemma:encode0} and~\ref{lemma:PQRM_plus}. We imagine two patches of atoms being initialized in the $|0\ra$ and $|+\ra$ state, and then rearranged \cite{Barredo} to patterns in Table~\ref{table:init} during loading.
In the hypercube encoding circuit, at the $t^{th}$ time step, parallel CNOT gates are applied to pairs of atoms that have labels differing at the $t^{th}$ bit, as shown in Table~\ref{table:polar}. The atom that would have interacted with the punctured qubit will undergo an identity gate during the global CNOT gate.

\begin{table*}
\setlength\arraycolsep{0.75pt}
\renewcommand{\arraystretch}{1.25}
\[
\begin{array}{|cccccccc|cccccccc|}
\cline{1-16}
\begin{array}{c} 127 \\[-1ex] \scriptstyle \text{1111111} \end{array} & 
\begin{array}{c} 126 \\[-1ex] \scriptstyle \text{1111110} \end{array} & 
\begin{array}{c} 123 \\[-1ex] \scriptstyle \text{1111011} \end{array} & 
\begin{array}{c} 122 \\[-1ex] \scriptstyle \text{1111010} \end{array} & 
\begin{array}{c} 111 \\[-1ex] \scriptstyle \text{1101111} \end{array} & 
\begin{array}{c} 110 \\[-1ex] \scriptstyle \text{1101110} \end{array} & 
\begin{array}{c} 107 \\[-1ex] \scriptstyle \text{1101011} \end{array} & 
\begin{array}{c} 106 \\[-1ex] \scriptstyle \text{1101010} \end{array} & 
\begin{array}{c} 63 \\[-1ex] \scriptstyle \text{0111111} \end{array} & 
\begin{array}{c} 62 \\[-1ex] \scriptstyle \text{0111110} \end{array} & 
\begin{array}{c} 59 \\[-1ex] \scriptstyle \text{0111011} \end{array} & 
\begin{array}{c} 58 \\[-1ex] \scriptstyle \text{0111010} \end{array} & 
\begin{array}{c} 47 \\[-1ex] \scriptstyle \text{0101111} \end{array} & 
\begin{array}{c} 46 \\[-1ex] \scriptstyle \text{0101110} \end{array} & 
\begin{array}{c} 43 \\[-1ex] \scriptstyle \text{0101011} \end{array} & 
\begin{array}{c} 42 \\[-1ex] \scriptstyle \text{0101010} \end{array} \\ 
\begin{array}{c} 125 \\[-1ex] \scriptstyle \text{1111101} \end{array} & 
\begin{array}{c} 124 \\[-1ex] \scriptstyle \text{1111100} \end{array} & 
\begin{array}{c} 121 \\[-1ex] \scriptstyle \text{1111001} \end{array} & 
\begin{array}{c} 120 \\[-1ex] \scriptstyle \text{1111000} \end{array} & 
\begin{array}{c} 109 \\[-1ex] \scriptstyle \text{1101101} \end{array} & 
\begin{array}{c} 108 \\[-1ex] \scriptstyle \text{1101100} \end{array} & 
\begin{array}{c} 105 \\[-1ex] \scriptstyle \text{1101001} \end{array} & 
\begin{array}{c} 104 \\[-1ex] \scriptstyle \text{1101000} \end{array} & 
\begin{array}{c} 61 \\[-1ex] \scriptstyle \text{0111101} \end{array} & 
\begin{array}{c} 60 \\[-1ex] \scriptstyle \text{0111100} \end{array} & 
\begin{array}{c} 57 \\[-1ex] \scriptstyle \text{0111001} \end{array} & 
\begin{array}{c} 56 \\[-1ex] \scriptstyle \text{0111000} \end{array} & 
\begin{array}{c} 45 \\[-1ex] \scriptstyle \text{0101101} \end{array} & 
\begin{array}{c} 44 \\[-1ex] \scriptstyle \text{0101100} \end{array} & 
\begin{array}{c} 41 \\[-1ex] \scriptstyle \text{0101001} \end{array} & 
\begin{array}{c} 40 \\[-1ex] \scriptstyle \text{0101000} \end{array} \\ 
\begin{array}{c} 119 \\[-1ex] \scriptstyle \text{1110111} \end{array} & 
\begin{array}{c} 118 \\[-1ex] \scriptstyle \text{1110110} \end{array} & 
\begin{array}{c} 115 \\[-1ex] \scriptstyle \text{1110011} \end{array} & 
\begin{array}{c} 114 \\[-1ex] \scriptstyle \text{1110010} \end{array} & 
\begin{array}{c} 103 \\[-1ex] \scriptstyle \text{1100111} \end{array} & 
\begin{array}{c} 102 \\[-1ex] \scriptstyle \text{1100110} \end{array} & 
\begin{array}{c} 99 \\[-1ex] \scriptstyle \text{1100011} \end{array} & 
\begin{array}{c} 98 \\[-1ex] \scriptstyle \text{1100010} \end{array} & 
\begin{array}{c} 55 \\[-1ex] \scriptstyle \text{0110111} \end{array} & 
\begin{array}{c} 54 \\[-1ex] \scriptstyle \text{0110110} \end{array} & 
\begin{array}{c} 51 \\[-1ex] \scriptstyle \text{0110011} \end{array} & 
\begin{array}{c} 50 \\[-1ex] \scriptstyle \text{0110010} \end{array} & 
\begin{array}{c} 39 \\[-1ex] \scriptstyle \text{0100111} \end{array} & 
\begin{array}{c} 38 \\[-1ex] \scriptstyle \text{0100110} \end{array} & 
\begin{array}{c} 35 \\[-1ex] \scriptstyle \text{0100011} \end{array} & 
\begin{array}{c} 34 \\[-1ex] \scriptstyle \text{0100010} \end{array} \\ 
\begin{array}{c} 117 \\[-1ex] \scriptstyle \text{1110101} \end{array} & 
\begin{array}{c} 116 \\[-1ex] \scriptstyle \text{1110100} \end{array} & 
\begin{array}{c} 113 \\[-1ex] \scriptstyle \text{1110001} \end{array} & 
\begin{array}{c} 112 \\[-1ex] \scriptstyle \text{1110000} \end{array} & 
\begin{array}{c} 101 \\[-1ex] \scriptstyle \text{1100101} \end{array} & 
\begin{array}{c} 100 \\[-1ex] \scriptstyle \text{1100100} \end{array} & 
\begin{array}{c} 97 \\[-1ex] \scriptstyle \text{1100001} \end{array} & 
\begin{array}{c} 96 \\[-1ex] \scriptstyle \text{1100000} \end{array} & 
\begin{array}{c} 53 \\[-1ex] \scriptstyle \text{0110101} \end{array} & 
\begin{array}{c} 52 \\[-1ex] \scriptstyle \text{0110100} \end{array} & 
\begin{array}{c} 49 \\[-1ex] \scriptstyle \text{0110001} \end{array} & 
\begin{array}{c} 48 \\[-1ex] \scriptstyle \text{0110000} \end{array} & 
\begin{array}{c} 37 \\[-1ex] \scriptstyle \text{0100101} \end{array} & 
\begin{array}{c} 36 \\[-1ex] \scriptstyle \text{0100100} \end{array} & 
\begin{array}{c} 33 \\[-1ex] \scriptstyle \text{0100001} \end{array} & 
\begin{array}{c} 32 \\[-1ex] \scriptstyle \text{0100000} \end{array} \\ 
\cline{9-16}
\begin{array}{c} 95 \\[-1ex] \scriptstyle \text{1011111} \end{array} & 
\begin{array}{c} 94 \\[-1ex] \scriptstyle \text{1011110} \end{array} & 
\begin{array}{c} 91 \\[-1ex] \scriptstyle \text{1011011} \end{array} & 
\begin{array}{c} 90 \\[-1ex] \scriptstyle \text{1011010} \end{array} & 
\begin{array}{c} 79 \\[-1ex] \scriptstyle \text{1001111} \end{array} & 
\begin{array}{c} 78 \\[-1ex] \scriptstyle \text{1001110} \end{array} & 
\begin{array}{c} 75 \\[-1ex] \scriptstyle \text{1001011} \end{array} & 
\begin{array}{c} 74 \\[-1ex] \scriptstyle \text{1001010} \end{array} & 
\begin{array}{c} 31 \\[-1ex] \scriptstyle \text{0011111} \end{array} & 
\begin{array}{c} 30 \\[-1ex] \scriptstyle \text{0011110} \end{array} & 
\begin{array}{c} 27 \\[-1ex] \scriptstyle \text{0011011} \end{array} & 
\begin{array}{c|} 26 \\[-1ex] \scriptstyle \text{0011010} \end{array} & 
\begin{array}{c} 15 \\[-1ex] \scriptstyle \text{0001111} \end{array} & 
\begin{array}{c} 14 \\[-1ex] \scriptstyle \text{0001110} \end{array} & 
\begin{array}{c} 11 \\[-1ex] \scriptstyle \text{0001011} \end{array} & 
\begin{array}{c} 10 \\[-1ex] \scriptstyle \text{0001010} \end{array} \\ 
\begin{array}{c} 93 \\[-1ex] \scriptstyle \text{1011101} \end{array} & 
\begin{array}{c} 92 \\[-1ex] \scriptstyle \text{1011100} \end{array} & 
\begin{array}{c} 89 \\[-1ex] \scriptstyle \text{1011001} \end{array} & 
\begin{array}{c} 88 \\[-1ex] \scriptstyle \text{1011000} \end{array} & 
\begin{array}{c} 77 \\[-1ex] \scriptstyle \text{1001101} \end{array} & 
\begin{array}{c} 76 \\[-1ex] \scriptstyle \text{1001100} \end{array} & 
\begin{array}{c} 73 \\[-1ex] \scriptstyle \text{1001001} \end{array} & 
\begin{array}{c} 72 \\[-1ex] \scriptstyle \text{1001000} \end{array} & 
\begin{array}{c} 29 \\[-1ex] \scriptstyle \text{0011101} \end{array} & 
\begin{array}{c} 28 \\[-1ex] \scriptstyle \text{0011100} \end{array} & 
\begin{array}{c} 25 \\[-1ex] \scriptstyle \text{0011001} \end{array} & 
\begin{array}{c|} 24 \\[-1ex] \scriptstyle \text{0011000} \end{array} & 
\begin{array}{c} 13 \\[-1ex] \scriptstyle \text{0001101} \end{array} & 
\begin{array}{c} 12 \\[-1ex] \scriptstyle \text{0001100} \end{array} & 
\begin{array}{c} 9 \\[-1ex] \scriptstyle \text{0001001} \end{array} & 
\begin{array}{c} 8 \\[-1ex] \scriptstyle \text{0001000} \end{array} \\ 
\cline{13-16}
\begin{array}{c} 87 \\[-1ex] \scriptstyle \text{1010111} \end{array} & 
\begin{array}{c} 86 \\[-1ex] \scriptstyle \text{1010110} \end{array} & 
\begin{array}{c} 83 \\[-1ex] \scriptstyle \text{1010011} \end{array} & 
\begin{array}{c} 82 \\[-1ex] \scriptstyle \text{1010010} \end{array} & 
\begin{array}{c} 71 \\[-1ex] \scriptstyle \text{1000111} \end{array} & 
\begin{array}{c} 70 \\[-1ex] \scriptstyle \text{1000110} \end{array} & 
\begin{array}{c} 67 \\[-1ex] \scriptstyle \text{1000011} \end{array} & 
\begin{array}{c} 66 \\[-1ex] \scriptstyle \text{1000010} \end{array} & 
\begin{array}{c} 23 \\[-1ex] \scriptstyle \text{0010111} \end{array} & 
\begin{array}{c} 22 \\[-1ex] \scriptstyle \text{0010110} \end{array} & 
\begin{array}{c} 19 \\[-1ex] \scriptstyle \text{0010011} \end{array} & 
\begin{array}{c|} 18 \\[-1ex] \scriptstyle \text{0010010} \end{array} & 
\begin{array}{c} 7 \\[-1ex] \scriptstyle \text{0000111} \end{array} & 
\begin{array}{c|} 6 \\[-1ex] \scriptstyle \text{0000110} \end{array} & 
\begin{array}{c} 3 \\[-1ex] \scriptstyle \text{0000011} \end{array} & 
\begin{array}{c} 2 \\[-1ex] \scriptstyle \text{0000010} \end{array} \\ 
\cline{15-16}
\begin{array}{c} 85 \\[-1ex] \scriptstyle \text{1010101} \end{array} & 
\begin{array}{c} 84 \\[-1ex] \scriptstyle \text{1010100} \end{array} & 
\begin{array}{c} 81 \\[-1ex] \scriptstyle \text{1010001} \end{array} & 
\begin{array}{c} 80 \\[-1ex] \scriptstyle \text{1010000} \end{array} & 
\begin{array}{c} 69 \\[-1ex] \scriptstyle \text{1000101} \end{array} & 
\begin{array}{c} 68 \\[-1ex] \scriptstyle \text{1000100} \end{array} & 
\begin{array}{c} 65 \\[-1ex] \scriptstyle \text{1000001} \end{array} & 
\begin{array}{c} 64 \\[-1ex] \scriptstyle \text{1000000} \end{array} & 
\begin{array}{c} 21 \\[-1ex] \scriptstyle \text{0010101} \end{array} & 
\begin{array}{c} 20 \\[-1ex] \scriptstyle \text{0010100} \end{array} & 
\begin{array}{c} 17 \\[-1ex] \scriptstyle \text{0010001} \end{array} & 
\begin{array}{c|} 16 \\[-1ex] \scriptstyle \text{0010000} \end{array} & 
\begin{array}{c} 5 \\[-1ex] \scriptstyle \text{0000101} \end{array} & 
\begin{array}{c|} 4 \\[-1ex] \scriptstyle \text{0000100} \end{array} & 
\begin{array}{c|} 1 \\[-1ex] \scriptstyle \text{0000001} \end{array} & 
\begin{array}{c} 0 \\[-1ex] \scriptstyle \text{0000000} \end{array} \\
\cline{1-16}

\end{array}
\]
\caption{Recursively constructed 2D hypercube layout. Expand to the left/above when the dimension of the hypercube grows from even/odd to odd/even. Coordinates are labeled as $x_mx_{m-1}\cdots x_1$.}
\label{tab:hypercube_layout}
\end{table*}

\begin{table*}[!hbt]
\centering
\begin{tabular}{c|c|c} 
(\textbf{a})\includegraphics[scale = 0.33]{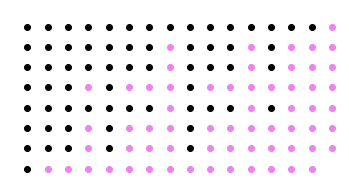} & 
(\textbf{b})\includegraphics[scale = 0.33]{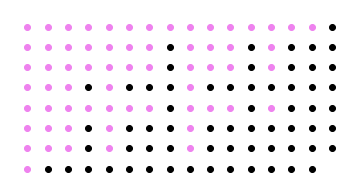} &
(\textbf{c})\includegraphics[scale = 0.33]{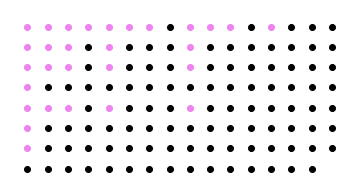}
\end{tabular} 
\caption{Initialization for (\textbf{a}) $|0\ra_{d=15}$ (\textbf{b}) $|+\ra_{d=15}$ (\textbf{c}) $|+\ra_{d=7}$, where violet atoms are initialized to $|+\ra$ and black atoms are initialized to $|0\ra$. } 
\label{table:init} 
\end{table*} 

\begin{table*}[!hbt]
\centering
\begin{tabular}{|c|c|c|c|}
\hline
(\textbf{1})\includegraphics[scale = 0.25]{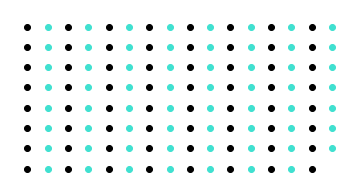} & 
(\textbf{2})\includegraphics[scale = 0.25]{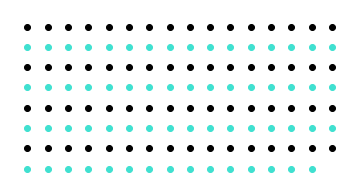} &
(\textbf{3})\includegraphics[scale = 0.25]{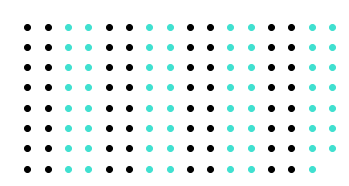} &
(\textbf{4})\includegraphics[scale = 0.25]{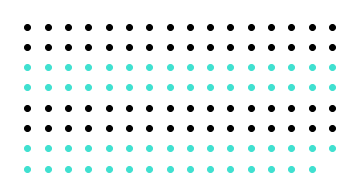}\\\hline
(\textbf{5})\includegraphics[scale = 0.25]{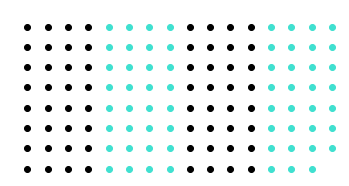} & 
(\textbf{6})\includegraphics[scale = 0.25]{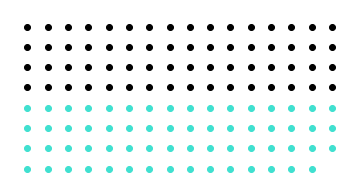} &
(\textbf{7})\includegraphics[scale = 0.25]{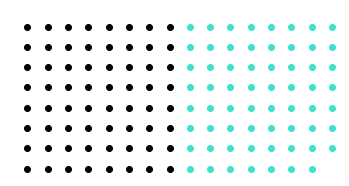} &\\\hline
\end{tabular} 
\caption{The hypercube encoding circuit. Movement is the same for punctured or the original QRM codes, the only difference is whether the bottom right atom (with label $0000000$) is kicked out. At each time step, the atoms labeled in green are translated by AODs in parallel to the neighborhood of the black atoms. For the $|0\ra_L$ preparation, green atoms are the control and black atoms are the target of CNOT gates, cf. Fig.~\ref{fig:pQRM}(a). For the $|+\ra_L$ preparation, black atoms are the control and green atoms are the target, cf. Fig.~\ref{fig:pQRM}(b).} 
\label{table:polar} 
\end{table*}

\subsection{Automorphism-based FT encoding}
\label{sec:auto_FT_encode}

Now let us describe the fault tolerant encoding procedure. To do so, we make use of the following definition from \cite{golay}.
\begin{definition}\textbf{(Strict fault-tolerance \cite{golay})}
An ancilla encoded into a code with distance $d$ is \emph{strictly fault-tolerant} if for all $k\leq\lfloor d/2\rfloor$, faults of probability order $k$ propagates to a residual error of reduced weight at most $k$. 
\end{definition}
The \emph{reduced weight} is the smallest obtainable weight when reducing the residual error by an arbitrary stabilizer of the desired ancilla state, e.g., reduction by a logical $Z$ operator for $|0\ra_L$ is allowed.
Faults of probability \emph{order} $k$ are a combination of $k$ single faults. \emph{A single fault} in the circuit-level noise model can be a flip in the measurement or initialization; one of the Pauli flips $\{X,Y,Z\}$ following a single qubit gate; or one of the $15$ possible errors $\{I,X,Y,Z\}^{\otimes 2}\backslash\{I\otimes I\}$ following a CNOT gate.

\begin{figure}
    \centering
    \includegraphics[width=1.0\linewidth]{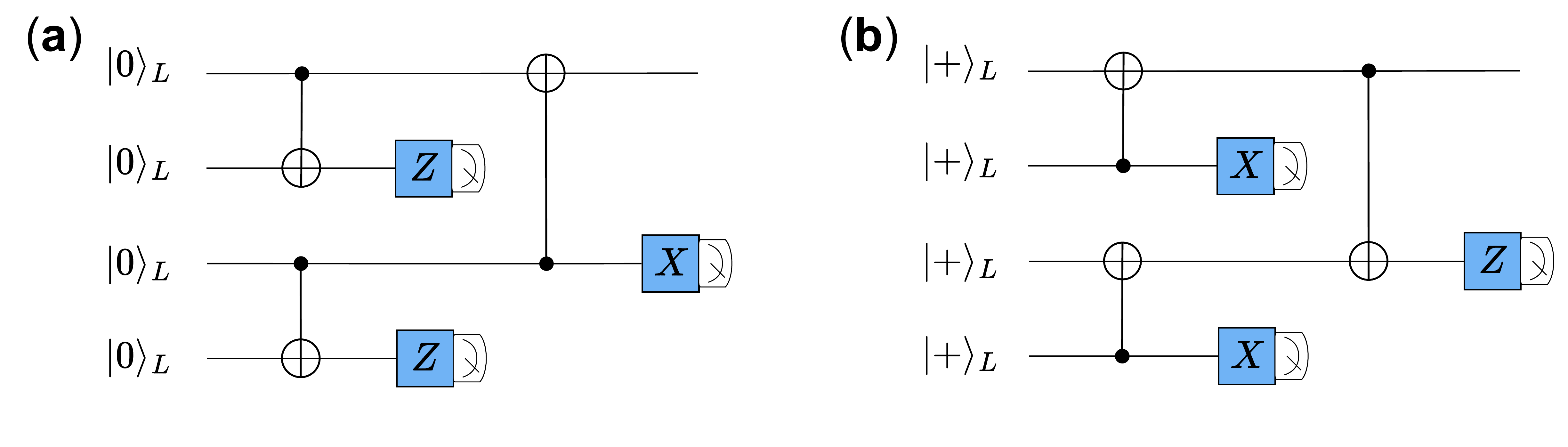}
    \caption{State preparation circuit to create verified $|0\ra_L$ and $|+\ra_L$ state. Only transversal CNOT and measurements are used in these verification circuits. Logical states are still encoded using the hypercube encoding circuit, but each of them undergoes a carefully chosen automorphism permutation before verification circuits to make the preparation fault-tolerant. The output patch is accepted if all transversal measurement results turn out to be stabilizers of the desired state.}
    \label{fig:AutEnc}
\end{figure}

Encoding via the hypercube circuit by itself is clearly not FT, since a single fault on a CNOT gate can lead to a weight-two error. Verification is necessary, we choose to do it after the encoding.
Consider the verification protocol in Fig.~\ref{fig:AutEnc}(a). A verified $|0\ra_L$ state is prepared by checking two pairs of prepared $|0\ra_L$ states against each other for $X$ errors. In the absence of noise, transversal measurements in $Z$-basis yield an $X$-type stabilizer of $|0\ra_L$; this will be our acceptance criterion. We can check whether a bit-string is a stabilizer by Section~\ref{sec:is_stabilizer}.
Then, conditioned on no $X$ errors being detected, the remaining two are checked against each other for $Z$ errors. The output state is accepted if the $X$-basis measurement results turn out to be a $Z$-type stabilizer of $|0\ra_L$.

If the qubits in different patches are not permuted relative to each other, the protocol can only tolerate one fault, but not two. 
A single fault in the encoding circuit cannot break this protocol because upon acceptance, the residual error this fault propagates to will be a stabilizer as well. Two faults can break this protocol. Consider one fault in each patch which happens at the same location in the encoding and propagates to a weight-three error in the output.
The two residual errors cancel at the measurement. We will thus accept this output patch with a weight-three error remaining on it, violating strict FT. Therefore, each patch should be prepared somehow differently. We follow the proposal of \cite{golay} where each patch undergoes a different automorphism permutation before the verification.

\begin{claim}
The permutations in the four columns of Table~\ref{tab:d7_permutations} applied to each of the four patches in Fig.~\ref{fig:AutEnc} gives a strict fault-tolerant preparation protocol of the $|0\ra_L$ and $|+\ra_L$ state of the $\llbracket 127,1,7\rrbracket$ code.
\end{claim}
We verified this claim by exhaustive testing all faults of order up to three. Section~\ref{sec:MITM} details how this can be performed efficiently. 

Exhaustive testing also shows that the permutations in Table~\ref{tab:d15_permutations} applied to $\llbracket 127,1,15\rrbracket$ give a practically fault-tolerant preparation protocol. Concretely, the resulting protocol is strictly FT up to three faults, and all order-four faults except one ($Z$-type) in Fig.~\ref{fig:AutEnc}(a). We also test all faults up to order six and we find in total $732$ order-five $Z$-type faults and $34483$ order-six $Z$-type faults, $88$ order-five $X$-type faults and $5280$ order-six $X$-type faults breaking strict FT for Fig.~\ref{fig:AutEnc}(a). Fig.~\ref{fig:AutEnc}(b) is the dual of Fig.~\ref{fig:AutEnc}(a), so the same number are found, but types are swapped. We call such a protocol \emph{practically fault tolerant} because, in the simulatable regime, preparation failures (violating strict FT) only contribute negligibly to the logical gate extended rectangle failure rate. In fact, in the $>10^{12}$ preparation circuits we simulate for Fig.~\ref{fig:exRec}, strict FT was never violated.

\begin{figure}
    \centering
    \includegraphics[width=1.0\linewidth]{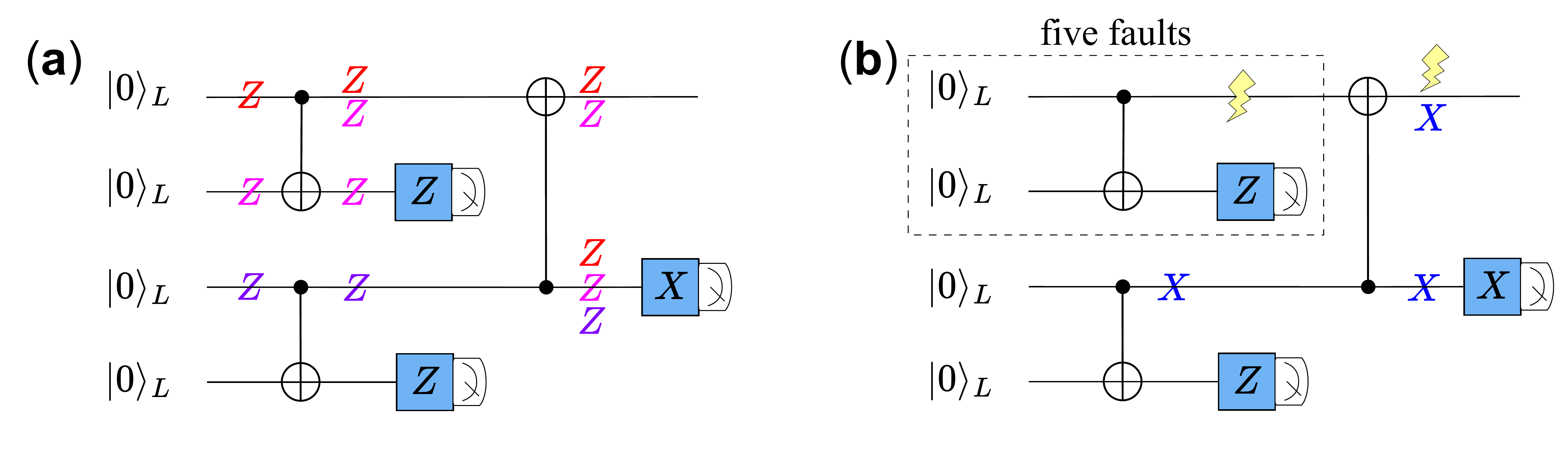}
    \caption{(a) A global malignant event counting across the four patches is necessary for the type of fault tested later, e.g., $Z$-type in this figure. (b) Our counting only assumes faults in the encoding circuit. For example, our reported numbers at the end of Section~\ref{sec:auto_FT_encode} include the order-five fault in the dashed box, but exclude the order-six fault containing the $XI$ error on the copying CNOT in the verification circuit.
    }
    \label{fig:AutEncFailure}
\end{figure}

Let us make a few comments on the above numbers. They are counted by excluding the faults on the verification circuits, see Fig.~\ref{fig:AutEncFailure}(b) for an example. Since verification only consists of transversal CNOT gates and transversal measurements, faults there alone will not break strict FT. One may ask whether some flips in the measurement can cause a false acceptance, thereby violating FT. Indeed, this is possible, but these flips are equivalent to some CNOT faults in the last layer of the hypercube encoding circuit. Therefore, to fulfill strict FT, which is a statement concerning the existence rather than the exact probability of some malignant events, we can restrict attention to faults happening in the encoding circuits.

One may also notice that $Z$-type malignant faults are more common than $X$-type in Fig.~\ref{fig:AutEnc}(a). This is because there are two tests for $X$-flips but only one for $Z$-flips. For $X$-flip tests, we can count order up to six malignant faults separately for the two pairs and then add them up. However, for $Z$-flips, faults from all four patches can conspire to add up to a stabilizer, see Fig.~\ref{fig:AutEncFailure}(a) for an example, leading to violation of strict FT.

Our permutations also guarantee strict FT with \emph{suppression} \cite{golay} for faults of order up to three, which means $s$ faults in the encoding circuits\footnote{Single faults happening on the verification circuits could lead to weight-one residual error on the verified state, e.g., an $XI$ error after the final copying CNOT gate. Therefore, it is impossible to impose suppression when taking them into account.}) lead to residual error weight $\leq s-1$ upon acceptance.

\subsection{Automorphism permutation}
\label{sec:auto_permute}
In Theorem~\ref{thm:automorphsim}, we see that the automorphism group of classical punctured RM codes is $\SL(m,\F_2)$ regardless of the order. Following our discussion at the end of Section~\ref{sec:CSS}, one sees that the PQRM codes inherit this automorphism group from their classical counterparts. Each automorphism permutation of PQRM can be described by a binary invertible matrix $A\in\SL(m,\F_2)$ which permutes the coordinate as $x\mapsto Ax=A(x_1,\dots,x_m)^T$ as in Eq.~\ref{eq:affine}.

For each pair of patches checked against each other, say the first patch undergoes permutation represented by $A$ and the second patch undergoes $B$.
We show in Appendix~\ref{sec:heuristic} that a necessary condition for strict FT (with suppression) is that every column of the relative permutation $A_{\rel}=A^{-1}B$ should contain at least two ones.

Besides this requirement, we would like the $A$ and $B$ permutations to be easy for the AODs to perform. 
We choose $E_{i,j}\in\SL(m,\F_2)$ to be our building block for permutations, where $E_{i,j}$ is the identity matrix with an extra one at the $i^{th}$ row and $j^{th}$ column.
To be consistent with our online source code, we index the row and column starting from zero for $E_{i,j}$, and use the convention of $A(x_m,\dots,x_1)^T$ in this section and Table~\ref{tab:d7_permutations}\&\ref{tab:d15_permutations}.

On one hand, $\SL(m,\F_2)$ is generated\footnote{This is because $E_{i,j}$'s correspond to elementary row operations when multiplied on the left, through which every binary invertible matrix can be turned into the identity matrix (Gaussian elimination).} by all the $E_{i,j}$'s and multiplying $E_{i,j}$ on the right is equivalent to adding column $i$ to column $j$, suitable for creating $A_{\rel}$ with every column having at least two ones.
On the other hand, the $E_{i,j}$ are naturally implemented by AODs. 
The effect of $E_{i,j}$ on the atom array with the layout in Table~\ref{tab:hypercube_layout} is that the atoms with labels satisfying $x_{m-j}=1\wedge x_{m-i}=0$ are swapped with those $x_{m-j}=1\wedge x_{m-i}=1$. We term this a \emph{sub-hypercube swap}. 
One can see that, in our recursively constructed 2D hypercube layout, all sub-hypercubes, specified by some coordinate bits being fixed to some values, always form a rectangular grid.

\begin{table*}[hbt]
    \centering
    \setlength{\tabcolsep}{2pt} 
    \renewcommand{\arraystretch}{1} 
    \begin{tabular}{|m{3.7cm}|m{3.7cm}|m{3.7cm}|m{3.7cm}|}
        \hline
        $E_{1,4}$\includegraphics[width=3.2cm, height=1.6cm]{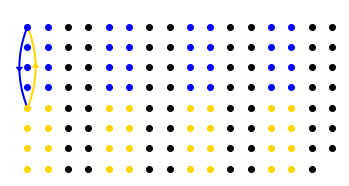} & $E_{1,4}$\includegraphics[width=3.2cm, height=1.6cm]{Eij/E14.png} & $E_{3,6}$\includegraphics[width=3.2cm, height=1.6cm]{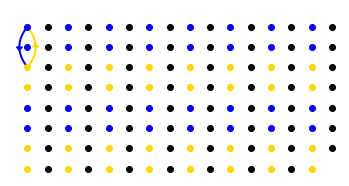} & $E_{3,6}$\includegraphics[width=3.2cm, height=1.6cm]{Eij/E36.png} \\ \hline
        $E_{0,3}$\includegraphics[width=3.2cm, height=1.6cm]{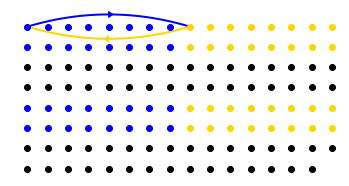} & $E_{0,3}$\includegraphics[width=3.2cm, height=1.6cm]{Eij/E03.png} & $E_{6,5}$\includegraphics[width=3.2cm, height=1.6cm]{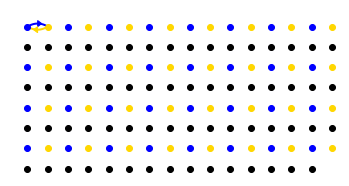} & $E_{6,5}$\includegraphics[width=3.2cm, height=1.6cm]{Eij/E65.png} \\ \hline
        $E_{5,4}$\includegraphics[width=3.2cm, height=1.6cm]{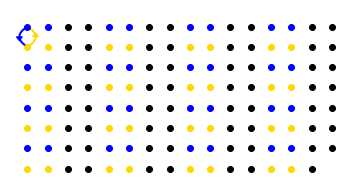} & $E_{4,2}$\includegraphics[width=3.2cm, height=1.6cm]{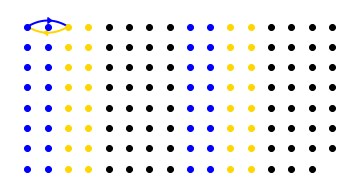} & $E_{5,0}$\includegraphics[width=3.2cm, height=1.6cm]{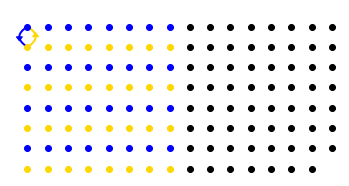} & $E_{4,0}$\includegraphics[width=3.2cm, height=1.6cm]{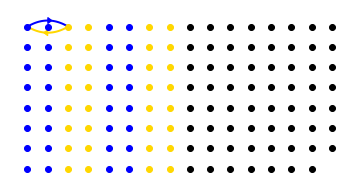} \\ \hline
        $E_{4,3}$\includegraphics[width=3.2cm, height=1.6cm]{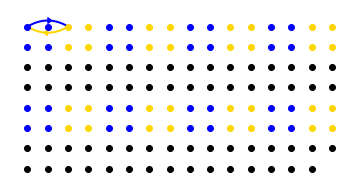} & $E_{0,5}$\includegraphics[width=3.2cm, height=1.6cm]{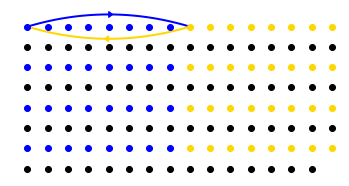} & $E_{6,4}$\includegraphics[width=3.2cm, height=1.6cm]{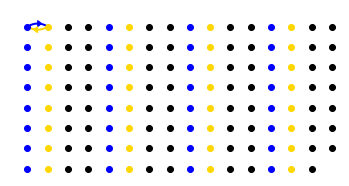} & $E_{2,5}$\includegraphics[width=3.2cm, height=1.6cm]{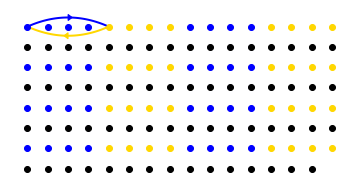} \\ \hline
        $E_{3,2}$\includegraphics[width=3.2cm, height=1.6cm]{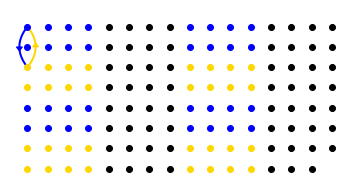} & $E_{6,0}$\includegraphics[width=3.2cm, height=1.6cm]{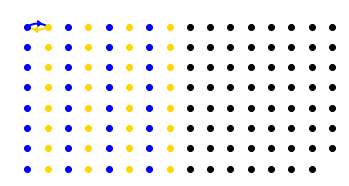} & $E_{2,6}$\includegraphics[width=3.2cm, height=1.6cm]{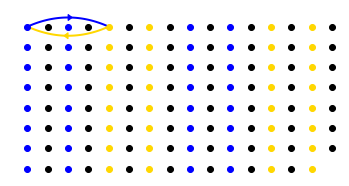} & $E_{1,2}$\includegraphics[width=3.2cm, height=1.6cm]{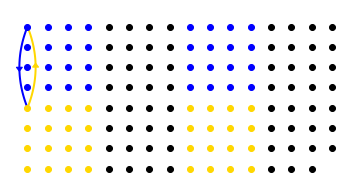} \\ \hline
        $E_{2,1}$\includegraphics[width=3.2cm, height=1.6cm]{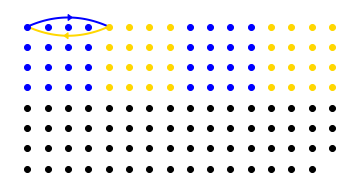} & $E_{5,1}$\includegraphics[width=3.2cm, height=1.6cm]{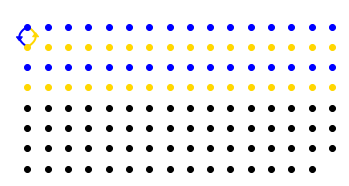} & $E_{0,2}$\includegraphics[width=3.2cm, height=1.6cm]{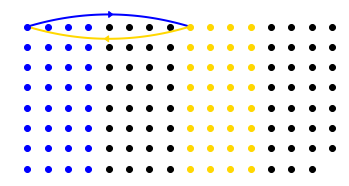} & $E_{6,1}$\includegraphics[width=3.2cm, height=1.6cm]{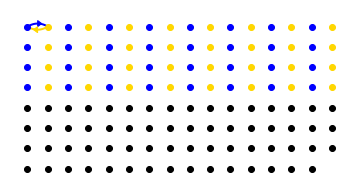} \\ \hline
        $E_{1,0}$\includegraphics[width=3.2cm, height=1.6cm]{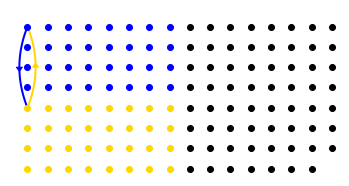} & $E_{2,6}$\includegraphics[width=3.2cm, height=1.6cm]{Eij/E26.png} & $E_{3,1}$\includegraphics[width=3.2cm, height=1.6cm]{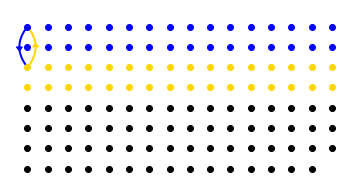} & $E_{5,3}$\includegraphics[width=3.2cm, height=1.6cm]{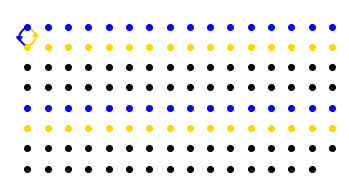} \\ \hline
    \end{tabular}
    \caption{Permutations for fault-tolerant encoding of the $\llbracket 127,1,7\rrbracket$ code, applicable to both $|0\ra_{d=7}$ and $|+\ra_{d=7}$, only the latter is used in the simulations of this work.
    From top to bottom in time, Each column consists of the sub-hypercube swaps to be applied to one of the four patches of $\llbracket 127,1,7\rrbracket$ in Fig.~\ref{fig:AutEnc} before the verification. Atoms in the same color are collectively translated by AODs following the trajectory in the same color. 
}
    \label{tab:d7_permutations}
\end{table*}

We detail how the $E_{i,j}$'s in Table~\ref{tab:d7_permutations}\&\ref{tab:d15_permutations} are found in Appendix~\ref{sec:heuristic}. 
On a high level, the procedure is to first check strict FT up to order three ($d=7$) and four ($d=15$) for \emph{pairs} of permutations. Going from $d=7$ to $d=15$, we inherit the permutations in the second and third columns of Table~\ref{tab:d7_permutations}. The exhaustive counting across the four patches is conducted after fixing those permutations. The fast testing and counting are enabled by the following meet-in-the-middle technique.

\begin{table*}[hbt]
    \centering
    \setlength{\tabcolsep}{2pt} 
    \renewcommand{\arraystretch}{1} 
    \begin{tabular}{|m{3.7cm}|m{3.7cm}|m{3.7cm}|m{3.7cm}|}
        \hline
        $E_{1,6}$\includegraphics[width=3.2cm, height=1.6cm]{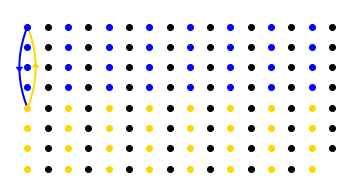} & $E_{1,4}$\includegraphics[width=3.2cm, height=1.6cm]{Eij/E14.png} & $E_{3,6}$\includegraphics[width=3.2cm, height=1.6cm]{Eij/E36.png} & $E_{4,5}$\includegraphics[width=3.2cm, height=1.6cm]{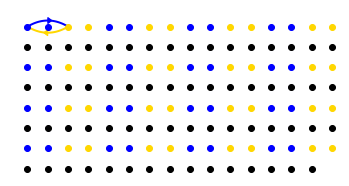} \\ \hline
        $E_{2,3}$\includegraphics[width=3.2cm, height=1.6cm]{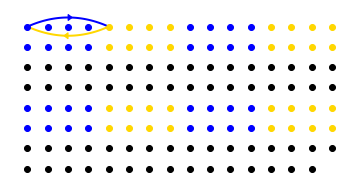} & $E_{0,3}$\includegraphics[width=3.2cm, height=1.6cm]{Eij/E03.png} & $E_{6,5}$\includegraphics[width=3.2cm, height=1.6cm]{Eij/E65.png} & $E_{3,4}$\includegraphics[width=3.2cm, height=1.6cm]{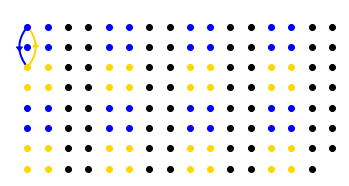} \\ \hline
        $E_{0,1}$\includegraphics[width=3.2cm, height=1.6cm]{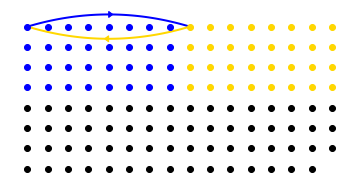} & $E_{4,2}$\includegraphics[width=3.2cm, height=1.6cm]{Eij/E42.png} & $E_{5,0}$\includegraphics[width=3.2cm, height=1.6cm]{Eij/E50.png} & $E_{4,0}$\includegraphics[width=3.2cm, height=1.6cm]{Eij/E40.png} \\ \hline
        $E_{3,6}$\includegraphics[width=3.2cm, height=1.6cm]{Eij/E36.png} & $E_{0,5}$\includegraphics[width=3.2cm, height=1.6cm]{Eij/E05.png} & $E_{6,4}$\includegraphics[width=3.2cm, height=1.6cm]{Eij/E64.png} & $E_{2,5}$\includegraphics[width=3.2cm, height=1.6cm]{Eij/E25.png} \\ \hline
        $E_{4,3}$\includegraphics[width=3.2cm, height=1.6cm]{Eij/E43.png} & $E_{6,0}$\includegraphics[width=3.2cm, height=1.6cm]{Eij/E60.png} & $E_{2,6}$\includegraphics[width=3.2cm, height=1.6cm]{Eij/E26.png} & $E_{1,2}$\includegraphics[width=3.2cm, height=1.6cm]{Eij/E12.png} \\ \hline
        $E_{6,0}$\includegraphics[width=3.2cm, height=1.6cm]{Eij/E60.png} & $E_{5,1}$\includegraphics[width=3.2cm, height=1.6cm]{Eij/E51.png} & $E_{0,2}$\includegraphics[width=3.2cm, height=1.6cm]{Eij/E02.png} & $E_{6,1}$\includegraphics[width=3.2cm, height=1.6cm]{Eij/E61.png} \\ \hline
        $E_{1,2}$\includegraphics[width=3.2cm, height=1.6cm]{Eij/E12.png} & $E_{2,6}$\includegraphics[width=3.2cm, height=1.6cm]{Eij/E26.png} & $E_{3,1}$\includegraphics[width=3.2cm, height=1.6cm]{Eij/E31.png} & $E_{5,3}$\includegraphics[width=3.2cm, height=1.6cm]{Eij/E53.png} \\ \hline
    \end{tabular}
    \caption{Permutations for fault-tolerant encoding of the $\llbracket 127,1,15\rrbracket$, applicable to both $|0\ra_{d=15}$ and $|+\ra_{d=15}$.}
    \label{tab:d15_permutations}
\end{table*}

\subsection{Meet-in-the-middle}
\label{sec:MITM}
Let us illustrate the idea of meet-in-the-middle (MITM) using the simplest example, where we are testing a pair of permutations to see whether it is strict FT for order-two X-type faults, e.g., the top two patches of Fig.~\ref{fig:AutEnc}(a). If both faults happen in the same patch within the hypercube encoding circuit, then upon acceptance, the residual error has reduced weight zero. We therefore only need to consider one fault on each patch. One could exhaustively test by taking a fault $f_1/f_2$ in patch one/two, and calculate its residual error $e_1 / e_2$ before verification. $e_1$ is copied to patch two by the transversal CNOT gates and the subsequent transversal measurement yields (up to stabilizers) $e_1+e_2$. If $e_1+e_2$ is a stabilizer (when we accept), we look at whether the residual error $e_1$ has reduced weight larger than two. Strict FT is broken if such pairs of faults exist. As we described in Section~\ref{sec:is_stabilizer}, to test if a bit-string $e$ is a stabilizer, we pass it through the hypercube encoding circuit again, and see whether the propagated results at certain positions (denoted by syndrome $s_e$) are all zero. Therefore, $s_{e_1}=s_{e_2}$ if $e_1+e_2$ is a stabilizer.

To trade space for time, MITM instead creates an inverse dictionary (hash table) for the second patch, with key-value\footnote{The value does not matter here since only $e_1$ contributes to the output residual error.} pairs $(s_{e_2},e_2)$. We iterate through all single fault $f_1$ of the first patch, each time we consult the dictionary for the presence of $s_{e_1}$. If present, indicating there is a fault $f_2$ on the second patch such that $s_{e_1}+s_{e_2}=0$, we look at if the reduced weight of $e_1$ is larger than two.

Say there are $F$ different faults on each patch. MITM improves the time complexity from $\Theta(F^2)$ to $\Theta(F)$. This might not seem significant for two faults, but for three faults, MITM improves the testing time of a pair from ten minutes to a few seconds. This suddenly renders our heuristic search for $E_{i,j}$'s viable.

For four faults, MITM improves time complexity from $\Theta(F^4)$ to $\Theta(F^2)$ as follows. Again if all four faults happen on one patch, then the residual error is a stabilizer. We only need to consider (3,1), (2,2), (1,3) faults distributed on the two patches. We construct three inverse dictionaries for order-two faults: both purely on patch one, both purely on patch two, and one on each patch. For the last case, when iterating through fault $f_1$ on patch one and $f_2$ on patch two, we store the key-value pair $(s_{e_1}+s_{e_2},e_1)$ into the dictionary\footnote{The actual value kept in the dictionary is a list containing all the $e_1$'s with the same key.}. The value $e_1$ is the residual error of this order-two fault after the transversal CNOT. For faults $f_1,f_2$ purely on patch one, we store $(s_{e_1}+s_{e_2},e_1+e_2)$. For faults $f_1,f_2$ purely on patch two, we store $(s_{e_1}+s_{e_2},0)$. Therefore, for distribution (3,1), we can split into (1,1) and (2,0) and use the two dictionaries to test MITM. If a key is present in both dictionaries, their corresponding values are added and checked for reduced weight.

MITM works similarly when counting across four patches in Fig.~\ref{fig:AutEnc}. For example, an order-six fault distributed as, e.g., (4,0,1,1) can be tested MITM using (2,0,0,1) and (2,0,1,0).
Constructing dictionaries for order-three faults suffices. Faults might be counted multiple times, but the factor can be easily determined, e.g., ${4\choose 2}$ in this case.

\begin{figure*}
    \centering
    \includegraphics[width=1.0\linewidth]{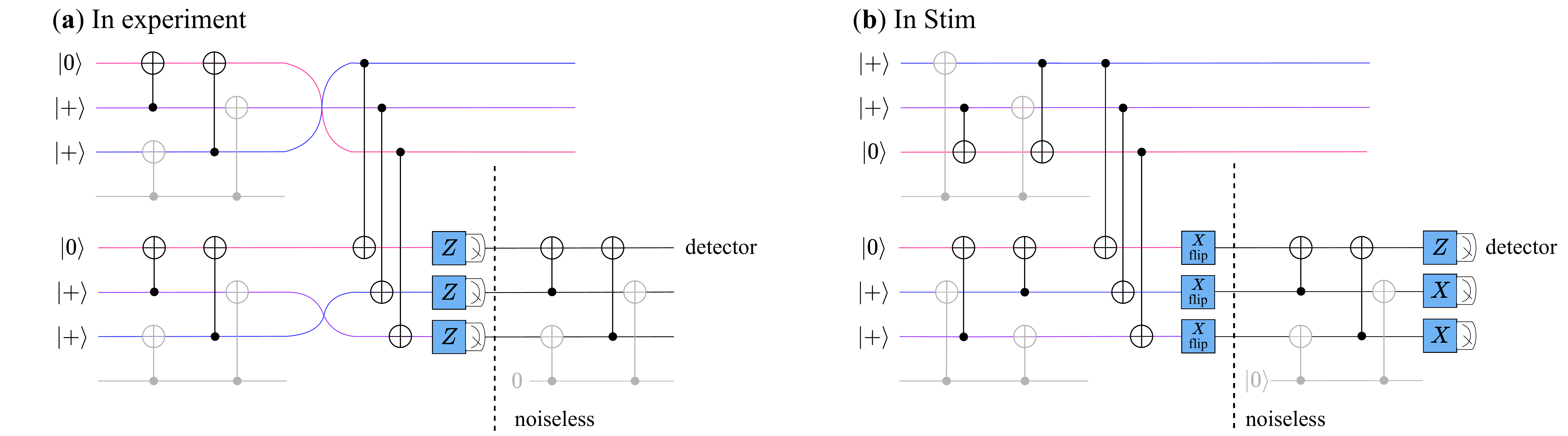}
    \caption{How an automorphism X-flip verification is implemented in (a) experiments and in (b) Stim. (a) In experiments, each logical block is encoded using the canonical hypercube encoding circuit in Table~\ref{table:polar}, and then permute qubits within each block. The $X$-flips are then copied from the data block to the ancillary block via transversal CNOT gates. Each qubit within the ancillary block is measured individually in the $Z$ basis. Feed the measurement result through the hypercube encoding circuit on a noiseless classical computer. If zeros are obtained on all the wires where $|0\ra$'s are assigned initially, then the data block is accepted. (b) In Stim, we directly implement the permuted assignment and hypercube encoding circuit. After the error-copying CNOT gates, a noiseless canonical hypercube encoding circuit is applied. Then each qubit is measured in $Z$ or $X$ basis if the canonical input assignment is $|0\ra$ or $|+\ra$. The data block is accepted if all the $Z$-basis measurements turn out to be zero, cf.  Section~\ref{sec:is_stabilizer}.}
    \label{fig:exp-vs-stim}
\end{figure*}

Let us explain how Stim \cite{stim} enables us to calculate the residual error $e$ (before verification) and the syndrome $s_e$. Say we want to test for $X$-type faults. Given a permutation $x\mapsto Ax$, we directly implement the initialization and hypercube encoding circuit on the permuted qubits. If $x$ is initialized to $|0\ra$ and a CNOT gate is applied to $y,z$ in the canonical settings, then in Stim, initialize $Ax$ to $|0\ra$ and apply a CNOT gate to $Ay,Az$.
We add noise to this permuted circuit, and then we add in a noiseless canonical hypercube encoding circuit, followed by noiseless $Z$/$X$ basis measurements at wires canonically initialized to $|0\ra$/$|+\ra$, cf. Section~\ref{sec:is_stabilizer}. The second block of Fig.~\ref{fig:exp-vs-stim} gives an example (ignore the transversal CNOT gates). If everything were noiseless, all measurements would be zero because of the automorphism. However, since we are testing for $X$-type faults, we only add detectors to $Z$ basis measurements. We compile a detector error model (DEM) herein. DEM merges the equivalent faults and tells us what detectors each fault triggers, this is precisely the $s_e$. The residual error $e$ is obtained by extracting the location/type of the fault and propagating it to the end of the permuted encoding circuit.

\begin{figure*}[h]
    \centering
    \includegraphics[width=0.8\linewidth]{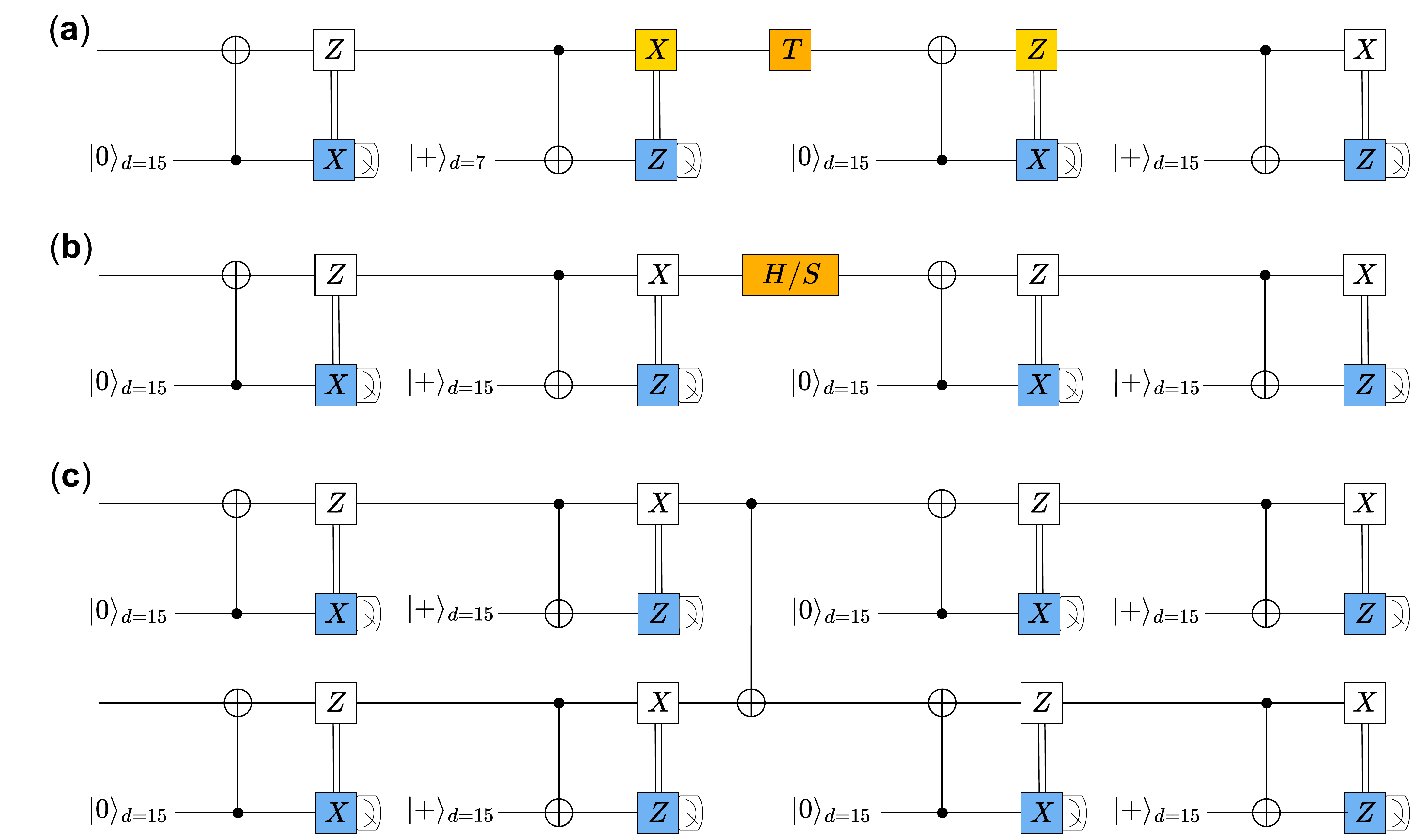}
    \caption{Three types of extended rectangles (exRec) involving $\llbracket 127,1,15\rrbracket$. (a) code switching to $\llbracket 127,1,7\rrbracket$ and back for performing a logical $T$ gate on $\llbracket 127,1,15\rrbracket$. (b) transversal logical $H$ or $S$ gate. (c) transversal CNOT gate. In Steane error correction, CNOTs and measurements are noisy, and the ancillary blocks have residual errors from preparation. Logical transversal CNOT and single-qubit gates $T/H/S$ are also noisy. 
    The correction operation is either noisy (indicated as yellow in $T$ exRec) or noiseless (done in software, i.e. Pauli frame tracking\cite{pauli-frame-tracking}) in Clifford exRec.}
    \label{fig:exRec}
\end{figure*}

\begin{table*}[h]
\centering
\begin{tabular}{c c} 
\begin{tikzpicture}
{


\pgfplotsset{every tick label/.append style={font=\normalem}}
\begin{axis}[
width=0.4\textwidth,
log basis y={10},
tick align=inside,
tick pos=left,
x grid style={lightgray211},
xlabel={\small Physical error rate \(\displaystyle p_{\CNOT}\)},
xmajorgrids,
xmode=log,
xmin=0.0009, xmax=0.0045,
xtick={0.0005,0.001, 0.002, 0.003, 0.004},
xticklabels={
  \(\displaystyle {0.0005}\),
  \(\displaystyle {0.001}\),
  \(\displaystyle {0.002}\),
  \(\displaystyle {0.003}\),
  \(\displaystyle {0.004}\),
},
minor xtick={0.0008, 0.0012, 0.0015, 0.003},
xminorgrids=true,
scaled x ticks=false,
xtick style={color=black},
y grid style={lightgray211},
ylabel={\small Logical error rate \(\displaystyle P_L\)},
ylabel near ticks,
ymajorgrids,
ymin=1e-9, ymax=2e-4,
ymode=log,
ytick style={color=black},
ytick={1e-10,1e-8,1e-6,1e-4,1e-2},
yticklabels={
  \(\displaystyle {10^{-10}}\),
  \(\displaystyle {10^{-8}}\),
  \(\displaystyle {10^{-6}}\),
  \(\displaystyle {10^{-4}}\),
  \(\displaystyle {0.01}\),
},
minor ytick={1e-9,1e-7,1e-5,1e-3},
error bars/y dir=both,
error bars/y explicit,
transpose legend,
legend style={at={(0.05,0.95)},anchor=north west, font=\normalem, text opacity=1.0, draw=none, fill opacity=0},
legend cell align=left,
legend image post style={xscale=.75}
]

\addlegendentry{$p_{\corr}=0,\ p_{\SPAM}=p_{\CNOT}$}
\addlegendimage{empty legend}

\addplot [darkviolet, mark=*, mark size=0.5, mark options={solid}]
table [x=pCNOT, y=pL, y error plus=error, y error minus=error] {CNOT_s2_c0.dat};
\addlegendentry{CNOT simulation}

\addplot [darkviolet, dashed, mark=*, mark size=0.5, mark options={solid}]
table [x=pCNOT, y=pSum] {CNOT_s2_c0_estimate.dat};
\addlegendentry{CNOT estimation}

\addplot [lavender, dashed, mark=*, mark size=0.5, mark options={solid}]
table [x=pCNOT, y=pSum] {Hadamard_s2_c0_estimate.dat};
\addlegendentry{$H$ estimation}

\addplot [lightteal, dashed, mark=*, mark size=0.5, mark options={solid}]
table [x=pCNOT, y=pSum] {S_s2_c0_estimate.dat};
\addlegendentry{$S$ estimation}

\end{axis}
}
\end{tikzpicture} &
\begin{tikzpicture}
{

\pgfplotsset{every tick label/.append style={font=\normalem}}
\begin{axis}[
width=0.4\textwidth,
log basis y={10},
tick align=inside,
tick pos=left,
x grid style={lightgray211},
xlabel={\small Physical error rate \(\displaystyle p_{\CNOT}\)},
xmajorgrids,
xmode=log,
xmin=0.00015, xmax=0.005,
xtick={0.0002,0.0005,0.001, 0.002, 0.004},
xticklabels={
  \(\displaystyle {0.0002}\),
  \(\displaystyle {0.0005}\),
  \(\displaystyle {0.001}\),
  \(\displaystyle {0.002}\),
  \(\displaystyle {0.004}\),
},
minor xtick={0.0008, 0.003},
xminorgrids=true,
scaled x ticks=false,
xtick style={color=black},
y grid style={lightgray211},
ylabel={\small Logical error rate \(\displaystyle P_L\)},
ylabel near ticks,
ymajorgrids,
ymin=1e-7, ymax=1,
ymode=log,
ytick style={color=black},
ytick={1e-6,1e-5,1e-4,1e-3,1e-2,1e-1},
yticklabels={
  \(\displaystyle {10^{-6}}\),
  \(\displaystyle {10^{-5}}\),
  \(\displaystyle {10^{-4}}\),
  \(\displaystyle {10^{-3}}\),
  \(\displaystyle {0.01}\),
  \(\displaystyle {0.1}\),
},
minor ytick={1e-9,1e-7,1e-5,1e-3},
error bars/y dir=both,
error bars/y explicit,
transpose legend,
legend style={at={(0.05,0.95)},anchor=north west, font=\normalem, text opacity=1, draw=none, fill opacity=0},
legend cell align=left,
legend image post style={xscale=.75}
]

\addlegendentry{$p_{\corr}=p_{\CNOT}$, $T$ simulations}
\addlegendimage{empty legend}

\addplot [darkviolet, mark=*, mark size=0.5, mark options={solid}]
table [x=pCNOT, y=pL, y error plus=error, y error minus=error] {T_s2_c10.dat};
\addlegendentry{$p_{\SPAM}=p_{\CNOT}$}

\addplot [blue, mark=*, mark size=0.5, mark options={solid}]
table [x=pCNOT, y=pL, y error plus=error, y error minus=error] {T_s2_c10_SPAM_half_CNOT.dat};
\addlegendentry{$p_{\SPAM}=p_{\CNOT}/2$}


\end{axis}
}
\end{tikzpicture}
\end{tabular} 
\caption{Logical failure rate of the extended rectangles shown in Fig.~\ref{fig:exRec}. Assuming $p_{\single}=0.2\cdot p_{\CNOT}$ (orange squares in Fig.~\ref{fig:exRec}) for all the simulations. (a) Clifford (CNOT, $H$, $S$) extended rectangles where only $p_{\SPAM}=p_{\CNOT}$ is assumed. The corrections before the transversal logical gates can be done in software (Pauli frame tracking) and thus are assumed to be noiseless, i.e., $p_{\corr}=0$. (b) Code-switching rectangle for logical $T$ gate. Assume $p_{\corr}=p_{\CNOT}$ for correction operations to apply at the yellow squares in Fig.~\ref{fig:exRec}. We additionally simulate $p_{\SPAM}=p_{\CNOT}/2$ to see how much the error rate can improve.} 
\label{table:exRec_simulation} 
\end{table*}

\section{Performance}
\label{sec:simulation}

The analysis of fault-tolerant procedures in \cite{AGP} proceeds by compiling a universal noisy quantum circuit into overlapping extended rectangles (exRec). In each exRec, a logical gate is preceded by a leading error correction (LEC) and followed by a trailing error correction (TEC). Let us discuss the purpose of LEC. The fault-tolerant Steane error correction (EC)\cite{SteaneEC} can correct most of the errors, but will still introduce some errors onto the data block. This is inevitable because the transversal CNOT gates used to copy data errors onto the ancilla, the transversal measurements, the ancilla per se, and the corrections before a non-Clifford gate are all noisy. Therefore, there will be some residual errors before applying the logical gates, and incorporating the LEC in the simulation accounts for them.

Define the exRec to be correct, if all the noise in the circuit does not cause the TECs to make the wrong decisions. For example, an exRec fails when faults in LEC and the transversal gate accumulate to a weight-eight uncorrectable error for the $d=15$ code so that the decoder at TEC gives a correction that leads to a logical error.

We implement all the simulations in Stim and we use the following circuit-level noise model. CNOT gates are followed by $\{I,X,Y,Z\}^{\times 2}\backslash\{I\otimes I\}$ each with probability $p_{\CNOT}/15$. State preparation and measurement (SPAM) noise is affected by a bit-flip\footnote{Other works \cite{heußen2024codeswitching,cross2009comparative} sometimes model SPAM by single-qubit depolarizing noise, i.e., affected by $X,Y,Z$ error with probability $p/3$ each, this is equivalent to using $\frac{2}{3}\cdot p_{\SPAM}$ in our bit-flip noise model.} with probability $p_{\SPAM}$. Throughout, we consider two cases $p_{\SPAM}=p_{\CNOT}$ and $p_{\SPAM}=p_{\CNOT}/2$. Our ancilla preparation simulation does not concern single qubit gates, they are only relevant in the exRec simulations in Fig.~\ref{fig:exRec}. There, we assume single qubit depolarizing noise with strength $p_{\single}=p_{\CNOT}/5$ for the transversal $H/S/T$ gates marked in orange. For the correction around the transversal $T$ gate for code-switching (yellow), all qubits are assumed to undergo single qubit depolarizing noise $p_{\corr}=p_{\CNOT}$. This is because the correction might be of high weight\footnote{Recall for code-switching, we apply either $\tbc$ or $\mathbf{1}+\tbc$ for correction, where $\tbc$ is the measurement result. One can reduce $\tbc$ by stabilizers of the target code before applying it, but the weight might still be large.} and the current solution in the neutral atom platform might involve applying Pauli correction sequentially for all relevant qubits. For Clifford exRec, no correction is physically applied, as they can be recorded in Pauli-frame tracking. We did not model idling errors, though they are present in applying the permutations, waiting for the measurement results, etc. However, we want to emphasize that this does not invalidate our fault-tolerance claim, as these idling errors can be absorbed into errors on some preceding or subsequent CNOT gates.

We did not perform full stabilizer simulation for exRec, instead, we only keep track of the Pauli errors through the circuit via \texttt{stim.FlipSimulator}. 
The simulation is not end-to-end either, because the ancilla acceptance rate is too low, cf. Table~\ref{table:acceptance_rate}. To solve this problem, we simulate all ancilla preparations in advance, e.g., as in Fig.~\ref{fig:fullStim}. Especially, we gather the residual error at the red bar upon acceptance (when all the detectors put on patches 2,3,4 turn out to be zero). When doing the exRec simulation, these pre-stored ancilla residual errors are loaded into their corresponding Steane EC blocks via \texttt{stim.FlipSimulator.broadcast\_pauli\_errors}. Upon seeing the noise\footnote{The noise is a combination of the data block error prior to this Steane EC block, the error of the ancilla, the copying CNOTs' error on the ancilla, and the measurement error.}, the decoders at TECs need to decide whether it is closer to $\oRM(3,7)$ or $\mathbf{1}+\oRM(3,7)$ (the only exception is the third block of Fig.~\ref{fig:exRec}(a), where the decoder needs to decide between $\oRM(4,7)$ and $\mathbf{1}+\oRM(4,7)$, cf. Section~\ref{sec:code_switch}), record a logical error if the latter is decided. As we commented in Section~\ref{sec:data_noise_decode}, SCL is isotropic and it will give the same decision as if in the full-stabilizer simulation of Steane EC, where the noise plus a random codeword of $\RM(3,7)^*$  (or $\RM(2,7)^*$ in code-switching) is observed.

There is one important thing related to this simulation method. When loading the ancilla residual error, one can reduce the error by any stabilizer of the \emph{state} entitled by degeneracy. Especially, one can reduce the error by any logical stabilizer, and it is absolutely necessary to do so when the noise is closer to the non-trivial logical coset.
For examples, suppose the residual error when preparing the $|0\ra_{d=15}$ ancilla is a $Z$-logical from $\mathbf{1}+\oRM(3,7)$.
One can reduce this error to zero, however, had one not done so, the decoder for Steane EC will decide the noise is closer to $\mathbf{1}+\oRM(3,7)$ (assume no faults happen elsewhere) and by our criterion, a logical $Z$ error will be recorded. Therefore, when loading the ancilla, we always use another decoder dedicated to finding the closest state stabilizer to the residual error and subsequently reduce it by that.

\begin{table}[hbt]
\scriptsize
\centering
\begin{tabular}{|c|c|c|c|c|}
\hline
& \multicolumn{4}{c|}{Acceptance rate}\\\hline
$p_{\CNOT}$ & \multicolumn{2}{c|}{$p_{\CNOT}=p_{\SPAM}$} & \multicolumn{2}{c|}{$p_{\CNOT}=2p_{\SPAM}$}\\\hline
& $|0/+\ra_{d=15}$ & $|+\ra_{d=7}$ & $|0/+\ra_{d=15}$ & $|+\ra_{d=7}$\\\hline
$0.005$ & $6.95\times 10^{-7}$ & $8.46\times 10^{-7}$ & $6.72\times 10^{-6}$ & $7.7\times 10^{-6}$\\\hline
$0.004$ & $1.23\times 10^{-5}$ & $1.37\times 10^{-5}$ & $7.2\times 10^{-5}$ & $8.06\times 10^{-5}$\\\hline
$0.003$ & $2.07\times 10^{-4}$ & $2.24\times 10^{-4}$ & $7.87\times 10^{-4}$ & $8.5\times 10^{-4}$\\\hline
$0.002$ & $3.5\times 10^{-3}$ & $3.69\times 10^{-3}$ & $8.53\times 10^{-3}$ & $8.98\times 10^{-3}$\\\hline
$0.001$ & $5.92\%$ & $6.08\%$ & $9.24\%$ & $9.48\%$\\\hline
$0.0008$ & $10.42\%$ & $10.64\%$ & $14.87\%$ & $15.18\%$\\\hline
$0.0005$ & $24.3\%$ & $24.65\%$ & $30.39\%$ & $30.79\%$\\\hline
$0.0002$ & $56.82\%$ & $57.12\%$ & $62.1\%$ & $62.43\%$\\\hline
$0.0001$ & $75.38\%$ & $75.58\%$ & $78.81\%$ & $79.01\%$\\\hline
\end{tabular} 
\caption{Acceptance rate for two sets of parameter choices, $p_{\CNOT}=p_{\SPAM}$ and $p_{\CNOT}=2p_{\SPAM}$. Being accepted means that auxiliary ancilla measurements in the verification circuit are stabilizers of the desired state, i.e., all detectors in patch $2,3,4$ in Fig.~\ref{fig:fullStim} are zero.
} 
\label{table:acceptance_rate} 
\end{table}

To simulate the logical $T$ gate exRec, following \cite{105-1-9}, we twirl the non-Pauli errors obtained by propagating Pauli errors across the transversal $T^{\dagger}$ gates back to Pauli. The propagation rule is that $Z$ error remains $Z$ error, and $X$ error transforms as $T^{\dagger}XT=X(I-iZ)/\sqrt{2}$. We implement the twirling by keeping the Pauli error as before and adding additional $Z$ errors with a 50\% chance at locations where $X$ errors are present.

Following \cite{cross2009comparative,golay}, we set the input noise to LECs to zero, this is because the residual errors just before the logical gates are independent of this input, if the corrections there are successful\footnote{Consider a naive circuit consisting of two logical gates. By the chain rule, the probability of both gates being successful is $\Pr[\text{gate 1 succeeds}] \cdot \Pr[\text{gate 2 succeeds} | \text{gate 1 succeeds}]$. One can view the second term as the success rate of the gate 2 exRec. Gate 1 succeeds if its TEC, which is also the LEC of gate 2 succeeds. In our Clifford simulations, by setting the input noise to LECs to zero, our decoders at LEC never encounter logical errors.}.
Let us be more concrete about this independence statement. 
The $Z$ component of the residual error after the first $Z$-type error Steane EC block consists of those from the $|0\ra_{d=15}$ ancilla, the transversal $X$-basis measurement errors, the CNOT error on both the ancilla and data. After the $X$-type error Steane EC block, the $Z$ component gets contribution from the CNOT and the $|+\ra_{d=15}$ ancilla.
The $X$ component comes solely from the $X$ faults (ancilla, CNOT, measurement) on this Steane EC block.

That being said, we want to compare our $T$ exRec simulation to \cite[Fig.~1]{heußen2024codeswitching}. They propose to implement the $T$ gate by transversal code-switching between a 2D and 3D color code. The numerics they show (Fig.~6) for $d=5$ code are 
very similar to our $d=7$ code. There are a few reasons for this besides their codes having a much smaller blocklength ($17$ and $49$).
The major one is that their LEC (Fig.~1) has one less Steance EC block compared to ours. Their $|0\ra_T$ state is noisy and so is our first ancilla in the Steane EC. Therefore, they still lack the noise contribution from one transversal CNOT and one transversal measurement. There is also the difference that they model SPAM as single-qubit depolarizing noise (equivalent to our $p_{\SPAM}=\frac{2}{3}p_{\CNOT}$) and we assume a high correction noise $p_{\corr}=p_{\CNOT}$.

We want to give a few comments on the numerics shown in Table~\ref{table:exRec_simulation}. 
For every datapoint in the CNOT exRec simulation, we simulate $2.5\cdot 10^{11}$ preparation protocols for $|0\ra_L$ and $|+\ra_L$ each (or in total for $p_{\CNOT}=0.003,0.004$).
We describe in Appendix~\ref{sec:simulation} how we efficiently handle the storage and loading of such a large amount of the ancilla (residual errors).
Strict FT is verified to never be violated in the preparation simulations. 
This is also the reason for observing a $p^8$ scaling at this physical error rate regime. However, as $p_{\CNOT}$ further decreases, since there is one order-four fault that could break strict FT for the $d=15$ code (cf. end of Section~\ref{sec:auto_FT_encode}), the scaling will eventually become $A\cdot p^4$ for some small $A$. 

In our CNOT simulations, we observe that $ZI$ is the most common logical error, followed by $IX$. This is because we correct $Z$ errors first and this result is in agreement with \cite[Fig.~10]{golay}. 
The logical error rates for single qubit Clifford gate ($H$ and $S$) are too small to determine, therefore we develop an estimation method that coincides quite well with the CNOT simulation. The estimation uses the average number of errors that each decoder in TECs sees and calculates the logical error using the green dashed curve in Fig.~\ref{fig:RM_data_qubit_noise}. For example, if the decoders in the third and fourth Steane EC blocks of Fig.~\ref{fig:exRec}(b) see $\overline{w}_Z$ $Z$-flips and $\overline{w}_X$ $X$-flips on average, then the logical error rate is estimated as $76003785\cdot\left(\left(\frac{\overline{w}_Z}{127}\right)^8+\left(\frac{\overline{w}_X}{127}\right)^8\right)$.

\section{High-rate codes}
\label{sec:high_rate}
\renewcommand{\arraystretch}{0.9}
\setlength\arraycolsep{2pt}

Fault-tolerant implementation of high-rate codes has gained much attention in recent years. 
Such codes serve as a much more efficient memory for storing quantum information, yet how to compute directly on that information is not so clear. 
In particular, logical qubits are encoded together in the same block, making them difficult to address individually. 

Current solutions include teleporting to an ancilla block, performing computation there, and teleporting back \cite{Bravyi_2024, xu2023constantoverhead}; using code deformations to perform Clifford gates via measurement \cite{cohen2022, cross2024ancilla}; and homomorphic CNOT \cite{huang2022homomorphic} with a masked ancilla code patch \cite{xu2024fastparallelizable}, etc. The ancilla system used in all these methods is not encoded in the same code as the data block, creating the extra challenge of how to prepare and decode the ancillas. 

Here we show how to directly perform all Clifford operations on a high-rate family of QRM codes with the help of another ancilla block encoded in the \emph{same} code. 
In particular, we consider the $\llbracket 2^{2r}, {2r\choose r},2^r\rrbracket$ $\QRM(r-1,r-1,2r)$ codes, which includes the $\llbracket 4,2,2\rrbracket$ code \cite{Vaidman1996}, the $\llbracket 16,6,4\rrbracket$ tesseract code \cite{tesseract}. Two further codes that can be decoded reasonably well by SCL are $\llbracket 64,20,8\rrbracket$ and $\llbracket 256,70,16\rrbracket$\footnote{One can obtain the data qubit noise decoding performance using \cite{gongaa-pw-qpc_2023} and a command like \texttt{./build/apps/program -N 256 -Kx 163 -Kz 163 -l 16 -px 0.03 -n 10000 -con RM -seed 42}}.

Let $k={2r\choose r}$ be the number of logical qubits encoded in $\QRM(r-1,r-1,2r)$. 
Logical Pauli operators can be implemented transversally, of course, so we need only concern ourselves with the logical Clifford operations modulo the logical Pauli operations. 
Their action on the logical Pauli operators forms a representation of the symplectic group $\Sp(2k,\F_2)$. 
In particular, we need not concern ourselves with phase changes to the logical operators that the Clifford operations may cause, since these can be implemented by logical Pauli operations. 

\begin{figure*}
    \centering
    \includegraphics[width=0.8\linewidth]{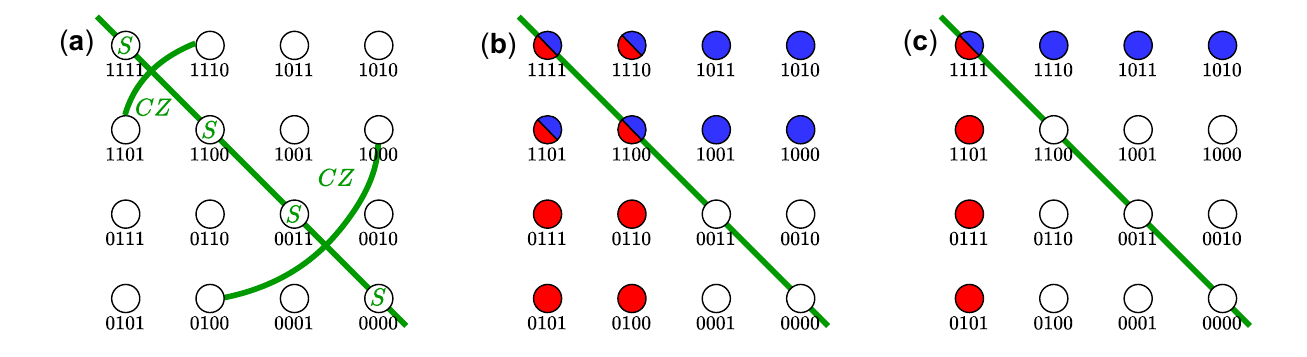}
    \caption{Fold-transversal gate on $\QRM(r-1,r-1,2r)$. Coordinates are labeled as $x_{2r}x_{2r-1}\cdots x_2x_1$. Apply $S$ gates to the qubits on the diagonal, i.e., whose labels satisfy $x_{2i}=x_{2i-1},\forall i\in [1,r]$. Apply CZ gates to all pairs reflected along the diagonal. (b) The $X$-type stabilizer $x_4$ (blue) gets mapped to the product of itself and the $Z$-type stabilizer $x_3$ (red). (c) The $X$-type logical operator $x_4x_2$ (blue) gets mapped to the product of itself and its pairing $Z$ logical $x_3x_1$ (red), up to some phase. This figure intends to mimic \cite[Fig.~3]{eberhardt2024logical}.}
    \label{fig:fold-transversal}
\end{figure*}

The stabilizers and logical operators of $\QRM(r-1,r-1,2r)$ are generated by the degree $\leq r-1$ and degree $r$ monomials. A stabilizer/logical is (represented by) a monomial means that it is supported on all the coordinates that this monomial evaluates to one. We show an example in Fig.~\ref{fig:fold-transversal}, where qubits are labeled by and placed at coordinates $0000$ to $1111$. 
The $X$-type stabilizer $x_4$ shown in Fig.~\ref{fig:fold-transversal}(b) evaluates to one at the eight coordinates colored blue (those having $x_4=1$ while $x_3,x_2,x_1$ taking on arbitrary values).


The $k={2r\choose r}$ logical pairs can be obtained by propagating 
$X$ or $Z$ placed on each row with label (input coordinate) containing exactly $r$ ones, following Lemma~\ref{lemma:initialization}. 
The propagation result can be described in general as follows. Assume we assign $X$/$Z$ on input coordinate $c_{m}\cdots c_{1}$, then $X$ propagates to the monomial which is the product of all $x_i$ where $c_i=1$. $Z$ propagates to the product of all $(1-x_i)$ where $c_i=0$. We can reduce this $Z$-type logical by any $Z$ stabilizer (degree $\leq r-1$ monomials), and the result is neatly the product of all $x_i$ where $c_i=0$. One can see that an $X$/$Z$ logical operator pair is just two \emph{complementing} monomials, i.e., two degree $r$ monomials whose product is $x_{2r}\cdots x_1$. As a sanity check, the pair indeed anticommutes because their product only evaluates to one at one coordinate.


The $X$- and $Z$-type logical operators can then each be separated into two groups, as follows. 
Let $\cL$ contain all the degree $r$ monomial where $x_1$ is present, and choose an arbitrary order within $\cL$. 
Then define $\overline{\cL}$ so that the $i^{th}$ element forms the complementing monomial of the $i^{th}$ element from $\cL$. Observe that each set contains $k/2$ monomials. 
For example, consider $r=2$. There we may partition the six degree-two monomials into $\cL=\{x_1x_2,x_1x_3,x_1x_4\}$ and $\overline{\cL}=\{x_3x_4,x_2x_4,x_2x_3\}$. 
The $Z$-type operators from monomials in $\overline{\cL}$ are paired with (anticommute with) the corresponding $X$-type monomials in $\cL$, and vice versa. For concreteness, we order the logical operators for symplectic matrix calculation as $X$-type logical operators from $\cL$, $X$-type from $\overline{\cL}$, then $Z$-type from $\overline{\cL}$, and finally $Z$-type from $\cL$.

The symplectic group $\Sp(2k,\F_2)$ can be generated by three types of generator matrices acting on the logical space (see \cite{Sp-gen} and Appendix~\ref{sec:Sp-gen}):
CNOT-type $\left(\begin{array}{c|c} U & 0\\ \hline 0 & U^{-T} \end{array}\right)$, where $U\in\SL(k,\F_2)$ and $U^{-T}:=(U^T)^{-1}=(U^{-1})^T$; phase-type $\left(\begin{array}{c|c} I & S\\ \hline 0 & I \end{array}\right)$, where $S$ is an arbitrary symmetric $k\times k$ matrix (not necessarily invertible); and Hadamard-type $\left(\begin{array}{c|c} 0 & I\\ \hline I & 0 \end{array}\right)$.

We will show how to effect all three kinds of transformations. 
First let us consider the effect of automorphism permutations of QRM codes.
Since the permutation is a code automorphism, stabilizers are mapped to stabilizers. 
Only the logical operators are affected, and moreover $X$-type logicals remain $X$-type and the same for $Z$-type.
Observe that the affine part of the automorphism group has no effect on the logical operators. A transformation of the form $x\mapsto x+b$ for some $b\in \F^{m}$ simply results in additional terms of lower degree. These correspond to stabilizers and can therefore be removed. 
Then the action of an RM automorphism $A\in\SL(m,\F_2)$ can be represented by a $k\times k$ matrix $\phi_A$ describing the transformation of the $X$-type logical operators. 
By the anticommutation relations, the $Z$-type logical operators automatically transform as $(\phi_A^T)^{-1}=(\phi_A^{-1})^T$. 
In Appendix~\ref{sec:homomorphism}, we prove that the map $\phi:\SL(2r,\F_2)\to \SL(k,\F_2)$, $A\mapsto \phi_A$ is a group representation, i.e., $\phi$ is a group homomorphism satisfying $\phi_A\phi_B=\phi_{AB}$.

Let us give an example, again for $r=2$. 
Suppose $A$ has the following effect: $x_1\mapsto x_1$, $x_2\mapsto x_2+x_4$, $x_3\mapsto x_3$, $x_4\mapsto x_1+x_4$. Therefore, the six degree-two monomials are transformed into linear combinations of degree-two monomials, as follows: 
$x_1 x_2\mapsto x_1x_2+x_1x_4$, $x_1x_3\mapsto x_1x_3$, $x_1x_4\mapsto x_1(x_1+x_4)=x_1^2+x_1x_4=x_1+x_1x_4\equiv x_1x_4$ (again the degree-one monomials are stabilizers, and can therefore be removed), $x_3x_4\mapsto x_3x_1+x_3x_4$, $x_2x_4\mapsto (x_2+x_4)(x_1+x_4)\equiv x_2x_1+x_2x_4+x_4x_1$, $x_2x_3\mapsto x_2x_3+x_4x_3$. The matrix form of $A$ and $\phi_A$ are thus given as
\begin{equation}
\label{eq:phi}
A=\begin{psmallmatrix}
    1&0&0&0\\0&1&0&1\\0&0&1&0\\1&0&0&1
\end{psmallmatrix},\quad \phi_A=\begin{psmallmatrix}
    1&0&1&0&0&0\\
    0&1&0&0&0&0\\
    0&0&1&0&0&0\\
    0&1&0&1&0&0\\
    1&0&1&0&1&0\\
    0&0&0&1&0&1
\end{psmallmatrix}.
\end{equation}

Now to the three kinds of $\Sp(2k,\F_2)$ generators. 
We first show that we can achieve the CNOT-type gates through automorphism permutation and transversal CNOT with an ancillary block. This is a technique introduced in \cite{Grassl_2013}. This ancilla block is the \emph{same} code as our data block. For QRM codes, we need to have such an ancilla block ready for Steane error correction, so it does not introduce additional technical difficulties. This technique is extremely useful because it greatly enlarges the possible CNOT-type gates compared to using permutations alone. As an example, we already know that the automorphism of QRM is $\GA(m,\F_2)$ but only $\SL(m,\F_2)$ is needed for the family we are considering. Therefore, through permutations alone, one expects to do at most $|\SL(m,\F_2)|$ different CNOT-type gates. However, this technique can enlarge the number of operations to $|\SL(k={m\choose m/2},\F_2)|$. 
In Appendix \ref{sec:CNOT_type}, we review this technique and give the proof that $\QRM(r-1,r-1,2r)$ can achieve all $\SL(k,\F_2)$ CNOT-type gates on the data block. As an overview, by making use of \cite{Grassl_2013}, we only need to prove that the $\F_2$-algebra of $H=\text{im }\phi$ can generate any binary $k\times k$ matrix, i.e., any matrix can be written as a sum of certain matrices from $H$.

Next we proceed to the Hadamard-type gate. 
A transversal Hadamard gate swaps $X$- and $Z$-type stabilizers and logical operators. 
Since the stabilizers are both monomials of degree no larger than $r-1$, the code is preserved. 
The effect on the logical operators is represented by the following matrix in $\Sp(2k,\F_2)$:
$$\tilde{H}=\tiny\left(\begin   {array}{c|c} 0\ 0 & 0\ I\\ 0\ 0 & I\ 0 \\ \hline 0\ I & 0\ 0\\ I\ 0 & 0\ 0 \end{array}\right)\,.$$
Here the identity matrices are of shape $k/2\times k/2$. We obtain the Hadamard-type gate $\left(\begin{array}{c|c} 0 & I\\ \hline I & 0 \end{array}\right)$ by multiplying a CNOT-type gate $\tiny\left(\begin{array}{c|c} 0\ I & 0\ 0\\ I\ 0 & 0\ 0 \\ \hline 0\ 0 & 0\ I\\ 0\ 0 & I\ 0 \end{array}\right)$ with $\tilde{H}$. Note that this operation interchanges $\cL$ and $\overline{\cL}$. 

Finally we come to the phase-type gates. 
To this end, first consider the action of transversal $S$. 
It preserves the code since $SXS^{\dagger}=iXZ$ and $SZS^{\dagger}=Z$, for each $X$-type stabilizer there is a $Z$-type stabilizer supported on the same qubits, and no extra phase is introduced because stabilizers have weight divisible by four ($RM(r-1,2r)$ is doubly-even).
The corresponding matrix is
$$\tilde{S}=\tiny\left(\begin{array}{c|c} I\ 0 & 0\ I\\ 0\ I & I\ 0 \\ \hline 0\ 0 & I\ 0\\ 0\ 0 & 0\ I \end{array}\right).$$
However, $\tilde{S}$ cannot complete the group to the full Clifford group because the upper right block of $\tilde{S}$, which is $\begin{psmallmatrix}0&I\\I&0\end{psmallmatrix}$, cannot be written as $AA^T$ for any matrix $A$ over $\F_2$. 

This problem can be remedied by the use of a fold-transversal gate, defined below. Its matrix representative is 
\begin{equation}
\label{eq:CZ}
    \widetilde{CZ}=\left(\begin{array}{c|c} I & T_{CZ}\\ \hline 0&I\end{array}\right),
\end{equation}
where $T_{CZ}$ is a $k\times k$ symmetric permutation matrix, and most importantly, contains at least one $1$ on its main diagonal. In Appendix~\ref{sec:phase_type} we show that such a $T_{CZ}$ can be written as $UU^T$ for some $U\in\SL(k,\F_2)$. Thus, using two CNOT-type gates we can obtain
$$\tilde{S}'=\left(\begin{array}{c|c} I & I\\ \hline 0 & I \end{array}\right)=\left(\begin{array}{c|c} U^{-1} & 0\\ \hline 0 & U^T \end{array}\right) \left(\begin{array}{c|c} I & T_{CZ}\\ \hline 0 & I \end{array}\right) \left(\begin{array}{c|c} U & 0\\ \hline 0 & U^{-T} \end{array}\right).$$
With the help of $\tilde{S}'$, we can then achieve any phase-type gate; the proof is given in Appendix~\ref{sec:phase_type}.

We now describe the fold-transversal gate shown in Fig.~\ref{fig:fold-transversal}. The 2D hypercube layout follows from Table.~\ref{tab:hypercube_layout}. We draw a diagonal between qubit $0$ and $2^{2r}-1$. One can verify that this diagonal passes through all qubits with labels satisfying $x_1=x_2,\ x_3=x_4,\dots ,x_{2r-1}=x_{2r}$, no matter how large $r$ is. The fold-transversal gate involves applying $S$ gates to all qubits on the diagonal and CZ gates to pairs reflected along the diagonal. Clearly, $S$ gates and $CZ$ gates both leave $Z$-type stabilizers and logical operators invariant, and so does the fold-transversal gate. 

To understand the effect on $X$-type operators, define $\tau$ to be a map on polynomials which substitutes every $x_{2i}$ with $x_{2i-1}$ and every $x_{2i-1}$ with $x_{2i}$, $i\in [r]$. For example, $\tau(x_4x_2)=x_3x_1$. This $\tau$ describes how a polynomial transforms when being reflected along the diagonal, i.e., all the coordinates where polynomial $p(x_1,\dots, x_{2r})$ evaluates to one are mapped to those where $\tau(p(x_1,\dots, x_{2r}))$ evaluates to one. Therefore, an $X$-type stabilizer or logical operator, represented by certain polynomial $p$, transforms under the fold-transversal gate to the product of itself and the $Z$-type monomial $\tau(p)$ up to some phase; see Fig.~\ref{fig:fold-transversal}(b,c). We can also let $\tau$ act on a coordinate since a coordinate can also be represented by a polynomial. For example, $1010$ is the only place $x_4 (1-x_3) x_2(1-x_1)$ evaluates to one. The reflection and thus $\tau$ swaps every two neighboring bits of the coordinate. For example, qubit $1010$ is reflected to $0101$.

As in \cite{fold_transversal,eberhardt2024logical}, we need to verify that this fold-transversal gate does not introduce an extra phase for the stabilizers. 
It suffices to do this for the generators, i.e., $X$-type stabilizer monomials. 
For each qubit $j$ not on the diagonal but in the support of a stabilizer monomial $m_S$, we distinguish between the following two cases. If qubit $j$ does not lie in (the support of) $\tau(m_S)$, then it gains no phase because the reflected qubit $\tau(j)$ is not in $m_S$ and thus $CZ_{j,\tau(j)}X_j I_{\tau(j)} CZ^{\dagger}_{j,\tau(j)}=X_j Z_{\tau(j)}$. If qubit $j$ lies in $\tau(m_S)$, then qubit $\tau(j)$ lies within $m_S$. From $CZ_{j,\tau(j)}X_j X_{\tau(j)} CZ^{\dagger}_{j,\tau(j)}=-(X_jX_{\tau(j)}) (Z_jZ_{\tau(j)})$, qubit $j$ and $\tau(j)$ can be seen as both gaining a phase $+i$. On the other hand, if a qubit lies both on the diagonal and in $m_S$, then it transforms as $SXS^{\dagger}=iXZ$, gaining a phase of $+i$ as well. Therefore, the total phase accumulated through the transform is $+i$ to the power of $\wt(m_S\wedge \tau(m_S))$. The term inside the bracket is the overlap between monomials $m_S$ and $\tau(m_S)$, which is just their product monomial. 
Since stabilizer monomials are of degree $\leq r-1$, there must be at least two absent variables in this product monomial, and thus the overlap is divisible by four and the total gained phase is $1$.

We now verify that this fold-transversal gate has an action on the logical space represented by the matrix in Eq.~\ref{eq:CZ}. Under this gate, an $X$-type logical representative, i.e., a degree $r$ monomial $m_L$, transforms to the product of itself and $Z$-type monomial $\tau(m_L)$. It is clear the $\tau$ preserves degree, so $\tau(m_L)$ is a logical operator. Moreover, $\tau$ simply permutes the variables, so the monomials are also permuted and it follows that $T_{CZ}$ is a permutation matrix. 
Further, $T_{CZ}$ is symmetric\footnote{Recall that the rows and columns of $T_{CZ}$ are the $X$- and $Z$-type logical monomials (degree-$r$) paired up in complement.}, i.e., $\tau(\overline{\tau(m_L)})=\overline{m_L}$, because $\tau(\tau(m_L))=m_L$ and $\overline{\tau(m_L)}=\tau(\overline{m_L})$ where the overline denotes taking the complementing monomial. Finally, $T_{CZ}$ contains at least one $1$ on its main diagonal implies there is at least an $m_L$ such that $\tau(m_L)=x_{2r}\cdots x_1/m_L=\overline{m_L}$. Taking $m_L=x_{2r-1}x_{2r-3}\cdots x_3 x_1$ clearly satisfies this.

We want to give a few comments for the $\llbracket 4,2,2\rrbracket$ code. In this case, the transversal CNOT trick is not needed, as affine automorphism alone can implement all the $|\SL(2,\F_2)|$ CNOT-type gates. However, besides $\tilde{S}'$, $\tilde{S}$ is also needed for the full Clifford group, the reason becomes manifest in our proof in Appendix~\ref{sec:phase_type}. For $r>2$, $\tilde{S}$ can be recovered using $\tilde{S}'$ (though one may not want to do so for FT purposes).

For PQRM codes, since affine automorphism preserves polynomial degree, logical operators represented by higher degree monomials cannot interact with the all-one logical, implying that the full Clifford group including the all-one logical is impossible. This has already been discussed in \cite{Grassl_2013} saying that $\llbracket 15,7,3\rrbracket$ $\PQRM(1,1,4)$ can achieve all CNOT-type gates when excluding the all-one logical operator.

One may be interested in $\QRM(r-1,r-1,2r+1)$ as well. 
In particular, this case could be relevant for the $[[127,1,7]]$ and $[[127,1,15]]$ codes discussed in the previous section, because they can be seen as gauge fixing the $[[127,71,7]]$ $\PQRM(2,2,7)$ codes. For $\QRM(r-1,r-1,2r+1)$ codes the logical operators are represented by degree $r$ and $r+1$ monomials. Using affine automorphism, transitions from degree $r+1$ to $r$ are allowed but the reverse is prohibited. Therefore, for the same reason, the full Clifford group when considering both degrees is impossible when using only the affine automorphism. We only briefly investigated this case, as it is harder to prove $\phi$ is a homomorphism when two degrees are involved. However, based on our MAGMA \cite{MAGMA} calculation for $\QRM(0,0,3)$ (where the homomorphism was verified by hand to hold), the possible CNOT-type gates on the data block using \cite{Grassl_2013} form a group $\SL(3,\F_2)\ltimes (\SL(3,\F_2)\ltimes \Z_2^9)$. This means that when we order the basis such that $\cL$ and $\overline{\cL}$ contain respectively all the three degree $1$ and $2$ monomials, all the invertible block lower triangular matrices (two $3\times 3$ blocks) are achievable, which is a consequence of $\F_2H$ achieving any binary matrices whose upper right block is all zero. For larger $r$, we expect $\phi$ to still be a homomorphism and the CNOT-type gates $\SL(l,\F_2)\ltimes (\SL(l,\F_2)\ltimes \Z_2^{l^2})$, $l={2r+1\choose r}$ to be achievable using \cite{Grassl_2013}. 

\section{Conclusion and outlook}
In this work, we show that code automorphisms can greatly facilitate quantum computation and FT encoding of logical states. The quantum Reed-Muller code family we explore is versatile in gate set and appears to fit quite well in the neutral atom platform. We only focus on the transversal $T$ gate for these codes, but they admit other non-Clifford gates, e.g., CCZ or multi-control Z as well \cite{15-1-3,Haah2018, barg2024RM}. There are several other aspects that future work could explore and improve upon.

First, the acceptance rate of state preparation in our protocol is quite low at $p_{\CNOT}=0.001$. To increase the acceptance rate it may be possible to exploit the recursive structure of the RM code family ($\RM(r,m)$ follows from the Plotkin $(u+v|v)$ construction, where $u\in\RM(r,m-1)$, $v\in\RM(r-1,m-1)$), so that a factory-based FT encoding scheme \cite{Q1-factory} can be devised.

In fact, \cite{Q1-factory} does this for a code closely related to RM, called Q1. Q1 and RM share the same encoding circuit but differ at input assignments (code constructions), and they form the two extremes of the polarization weight (PW) construction \cite{beta_expansion,gong2023list} of polar codes \cite{polar}. One type of Q1 stabilizers are of weight two, contrary to RM codes being non-degenerate, i.e., weight of stabilizers at least the distance. Q1 is also easier to decode, SCL with list size one is already the maximum-likelihood (ML) decoder (though the logical error rate is not particularly low). However, for Q1 to sustain the $\sqrt{N}$ distance, only one logical qubit can be encoded, while RM and the entire PW quantum polar code family \cite{gong2023list} have a more flexible rate-distance tradeoff. The classical PW polar codes are known to have the block lower-triangular affine automorphism \cite{BLTA_complete,geiselhart2021BLTA,LTA}, so we expect some of the techniques in this work to be transferrable to their quantum counterparts as well.

Secondly, the logical error rate of the code-switching $T$ gate is not low enough at the physical error rate $10^{-3}$ for implementing practically relevant quantum algorithms. However, as already mentioned in the introduction, such a "noisy" $T$ gate can be used in \cite[Fig.~8]{Beverland2021}, \cite{Haah2018} to distill a high-fidelity $|T\ra$ state for the $d=15$ code. We want to give some more remarks regarding the future study of the distillation protocols.
First, there is no need to implement the entire code-switching  $T$ gate for the ancilla blocks. After switching to $d=7$ and performing a transversal $T^{\dagger}$ gate, there is no need to switch back to $d=15$. Instead, directly measuring these blocks transversally in $X$ basis suffices.
Second, future work should incorporate the injection stage besides distillation to estimate the $T$ \emph{gate} fidelity. In particular, the residual errors on the data block should be taken care of by leading error corrections, and the residual errors of the $T$ state should be considered as well.
Third, given the failure rate of our code-switching $T$ gate, we expect one round of distillation to suffice in bringing the $T$ state infidelity down to a similar level to the Clifford operations. Nevertheless, the protocol will still be very challenging to simulate because of the large number of ancilla and their preparation simulations (maybe results can be reused after random shuffle), and difficulties in simulating rare logical failures. 
Using distillation protocols based on QRM \cite{Haah2018}, one can still benefit from the hypercube layout and native AOD manipulations.
Given the recursive instruction of the QRM codes, it may be interesting to explore the recent magic state cultivation protocol \cite{cultivation} as well.

There is also a lot to explore from the coding theory perspective. To have PQRM codes admitting a transversal $T$ gate while having a higher distance than the $\llbracket 127,1,7\rrbracket$ code, one can resort to the $\PQRM(3,6,10)$ $\llbracket 1023,1,15\rrbracket$ code. Though the state preparation success rate will likely be too low with our method herein.  
Any polar-related code, obtained by choosing some rows from $\bE$ punctured by \emph{one} column, cannot fill in the intermediate distances either.
One could either use other cyclic codes than PQRM \cite{grassl1999BCH,cyclic-CSS-T} or use codes based on doubling transform \cite{MSDlowoverhead,bravyi2015doubled,95-1-7} to remedy this and be more efficient in blocklength. For example, the $\llbracket 95,1,7\rrbracket$ code \cite{95-1-7}  saves $32$ qubits compared to $\llbracket 127,1,7\rrbracket$. There are further higher distance codes based on doubling a quadratic residue code \cite{jain2024transversalT}\footnote{The codes in \cite{jain2024transversalT} are not triply-even in our definition. However, it is possible to find a bi-partition of qubits into two sets $A$ and $B$, such that for any $X$-stabilizers, the difference in the support sizes of it intersecting $A$ and $B$ is divisible by eight\cite{bravyi2015doubled}, thereby $T$ on $A$ and $T^{\dagger}$ on $B$ implements a logical $T$ gate. To construct codes that have weights of $X$-stabilizers divisible by $8$, one may use the $\llbracket 49,1,9 \rrbracket$ self-dual doubly-even BCH code from \cite{grassl1999BCH} and then the $\llbracket 47,1,11\rrbracket$ QR code from \cite{Steane_1999}. The resulting codes are $\llbracket 193,1,9\rrbracket$ and $\llbracket 287,1,11\rrbracket$ and transversal $T^{\dagger}$ implements a logical $T$ gate.}. However, efficient encoding and decoding are not yet known for these codes. Whether the rich automorphism of QR codes \cite[Chapter~16]{theoryEC} can be inherited for these doubled codes for FT encoding protocols, and whether they can be easily implementable on hardware platforms needs further research. 

We should also comment on the decoding perspective of RM codes. RM codes have been proven to achieve capacity under ML decoding for binary memoryless symmetric channels \cite{reeves,abbe_proof_2023-1}. However, practical decoders like SCL have a noticeable gap to ML. In fact, RM codes are notoriously hard to decode as the blocklength grows. Nonetheless, there are some caveats to this statement. It is probably only necessary to consider RM codes with blocklength up to $1024$ in the near term, because with that we can already have PQRM codes of distance $15$ admitting transversal $T$, and QRM codes encoding $252$ qubits of distance $32$. Moreover, the final working regime for a quantum computer will be at a very low physical error rate (or after sufficient concatenation), so that SCL or a bounded distance decoder like Reed's majority logic decoder \cite{RM-Reed} could still excel. The situation for quantum computation is unlike the classical communication scenario, where performance near channel capacity is critical.
With all these arguments, we believe it is foreseeable that the ASICs \cite{asic1,asic2,asic3} designed for SCL decoding of 5G polar codes could be reused to some extent for RM. These works may not be directly applicable to RM because the decoders are optimized to some specific code constructions, but they do have the SC (list size one) component built-in, which could be reused for the parallelizable SC-based automorphsim ensemble decoding \cite{geiselhart2021ensemble,geiselhart2021BLTA}.

There is a lot to explore for the high rate $\QRM(r-1,r-1,2r)$ codes as well.
The fold-transversal gate is not FT in our definition because of the two-qubit gates involved, and its  performance degradation needs further investigation. Though our proof is constructive, the sequence of permutations and (fold) transversal gates involved in exerting arbitrary Clifford gates could be very long. It will be interesting to study how the circuit depth can be systematically minimized. 
We also did not discuss how to fault-tolerantly prepare these codes, though we expect similar automorphism-based verification protocols to work. There the linear part ($+b$) in Eq.~\ref{eq:affine} may be exploited. They are also easy to achieve through AODs: just swap the green and black atoms in Table~\ref{table:polar} if some bit of $b$ is one.

Finally, how to construct fault-tolerant non-Clifford gates on high-rate QRM codes is also an interesting topic. One can either explore other high-rate QRM codes admitting transversal CCZ \cite{barg2024RM}, or consider teleporting from, e.g., $\llbracket 256,70,16\rrbracket$ to $\llbracket 127,1,7\rrbracket$. Teleportation may be the most straightforward method, but the question again is fault tolerance. Perhaps one can avoid the high-weight bare-ancilla stabilizer measurement by doing transversal CNOT with a specially prepared ancilla state, similar to code switching in our protocol.

\section*{Acknowledgment}
We thank Wenchao Xu and Zhanchuan Zhang for discussing what operations are currently native to the neutral atom array platform. We thank John Preskill for encouraging us to explore high-rate codes.
Numerical simulations were performed on the ETH Zürich Euler cluster.
\bibliographystyle{IEEEtran}
\bibliography{main}
\appendices
\section{CNOT-type gates}
\label{sec:CNOT_type}
We prove that for $\QRM(r-1,r-1,2r)$, by interleaving permutations with transversal CNOT gates with an ancillary block, all CNOT-type gates are achievable on the data block encoding $k={2r\choose r}$ logical qubits.
In fact, all CNOT-type gates on the $2k$ logical qubits (data plus ancilla) are also achievable, but in the end, we will restrict ourselves to the block-diagonal subgroup that acts trivially on the ancillary block. In other words, the technique \cite{Grassl_2013} will allow us to prove the upper-left $2k\times 2k$ $X$-type logical block of $\Sp(4k,\F_2)$ can take on $\SL(2k,\F_2)$. The paired-up $Z$-type logical operators (bottom-right block) will automatically transform as inverse transpose, so we do not need to deal with them explicitly. 

All operations are over $\F_2$ throughout these sections, and we will use the following notations. $E_{i,j}$ is the elementary matrix that has an all-one diagonal and a one at row $i$ column $j$, call this entry $(i,j)$. $F_{i,j}$ is zero everywhere except at entry $(i,j)$, i.e., $E_{i,j}=I+F_{i,j}$. $\SWAP_{i,j}$ has a diagonal full of ones except at $(i,i)$ and $(j,j)$, and additionally, ones at $(i,j)$ and $(j,i)$. 
Multiplying an arbitrary matrix $M$ by $E_{i,j}$ on the left has the effect of adding the $j$th row of $M$ to the $i$th row, and leaving everything else as it was. 
Multiplying $M$ by $\SWAP_{i,j}$ on the left has the effect of exchanging row $i$ and row $j$. It is easy to verify that $\SWAP_{i,j}=E_{i,j}E_{j,i}E_{i,j}$.

\subsection{Generators of $\Sp(2k,\F_2)$}
\label{sec:Sp-gen}
It is written in \cite{Sp-gen} that the symplectic group $\Sp(2k,\F_2)$ is generated by translations, rotations, and semi-involutions. Translations and rotations are just our phase-type $\left(\begin{array}{c|c} I & S\\ \hline 0 & I \end{array}\right)$ and CNOT-type $\left(\begin{array}{c|c} U & 0\\ \hline 0 & U^{-T} \end{array}\right)$ gates, where $S=S^T$, $U\in\SL(k,\F_2)$, $U^{-T}:=(U^T)^{-1}=(U^{-1})^T$. However, we only declare one Hadamard-type gate $\left(\begin{array}{c|c} 0 & I\\ \hline I & 0 \end{array}\right)$ in Section~\ref{sec:high_rate}, while their semi-involutions are in a more general form $\left(\begin{array}{c|c} Q & I-Q\\ \hline I-Q & Q \end{array}\right)$, where $Q$ is a diagonal matrix. We thus need to prove that semi-involutions can be generated by the three types of generators we give.

Since $Q$ is a diagonal matrix of size $k\times k$, assume it has zeros at entries $(i_1,i_1),\dots,(i_t,i_t)$, then $I-Q$ has ones at these entries. Therefore, $Q$ can be obtained from $\left(\begin{array}{c|c} I & 0\\ \hline 0 & I \end{array}\right)$ by swapping row $i_1$ with row $i_1+k$, and swapping row $i_2$ with row $i_2+k$, etc. In other words, $\left(\begin{array}{c|c} Q & I-Q\\ \hline I-Q & Q \end{array}\right)=\SWAP_{i_t,i_t+k}\cdots \SWAP_{i_1,i_1+k}$. Now we make use of the identity $\SWAP_{i,i+k}=E_{i,i+k}E_{i+k,i}E_{i,i+k}$. Note that $E_{i,i+k}=\left(\begin{array}{c|c} I & F_{i,i}\\ \hline 0 & I \end{array}\right)$ is a phase-type gate, $E_{i,i+k}$ is of shape $2k\times 2k$ while $F_{i,i}$ has shape $k\times k$ and is indeed symmetric. Moreover, $E_{i+k,i}=\left(\begin{array}{c|c} I & 0\\ \hline F_{i,i} & I \end{array}\right)=\left(\begin{array}{c|c} 0 & I\\ \hline I & 0 \end{array}\right) \left(\begin{array}{c|c} I & F_{i,i}\\ \hline 0 & I \end{array}\right)\left(\begin{array}{c|c} 0 & I\\ \hline I & 0 \end{array}\right)$, finishing the proof.

\subsection{Interleaving permutations with transversal CNOT}
In this section we review the techniques in \cite{Grassl_2013} that are relevant to us, while adapting their notations to ours.

In Section~\ref{sec:high_rate}, we give an example of the following mapping $\phi:\SL(2r,\F_2)\to \SL(k,\F_2)$, $A\mapsto \phi_A$. $A$ is an affine transform of the variables $x_{2r},\dots,x_{1}$, and $\phi_A$ is an affine transform of the $k={2r\choose r}$ degree $r$ monomials.
When performing a permutation of qubits (coordinates) by $A$, we effectively get a CNOT-type gate $\phi_A$ on the $X$-type logical space.

Now we introduce an ancillary block that employs exactly the same code as the data block. Since QRM codes are CSS codes, they admit transversal CNOT gates. The action of a transversal CNOT on the $2k$ (data plus ancilla) $X$-type logical space, depending on the direction, can either be
$$\left( \begin{array}{c|c} I & 0\\ \hline I & I \end{array}\right)\text{ or } \left( \begin{array}{c|c} I & I\\ \hline 0 & I \end{array}\right).$$
Consider the following group generated by automorphism permutations on data or ancilla block and transversal CNOT gates between them:
$$G_{12}=\left\langle \left( \begin{array}{c|c} I & 0\\ \hline I & I \end{array}\right), \left( \begin{array}{c|c} I & I\\ \hline 0 & I \end{array}\right), \left( \begin{array}{c|c} \phi_A & 0\\ \hline 0 & \phi_B \end{array}\right),\forall A,B\in\SL(2r,\F_2) \right\rangle.$$

Generators of this group (permutations, transversal CNOT) can be implemented fault-tolerantly. Moreover, \cite[Thm. 4]{Grassl_2013} says that if the $\F_2$-algebra of $\text{im } \phi$ can achieve any binary $k\times k$ matrix, then $G_{12}\cong \SL(2k,\F_2)$. Let us summarize their proof in our notation.

First, note that the following matrices are in $G_{12}$:
$$\left( \begin{array}{c|c} \phi_A & 0\\ \hline 0 & I \end{array}\right) \left( \begin{array}{c|c} I & I\\ \hline 0 & I \end{array}\right) \left( \begin{array}{c|c} \phi_A & 0\\ \hline 0 & I \end{array}\right)^{-1}=\left( \begin{array}{c|c} I & \phi_A\\ \hline 0 & I \end{array}\right),$$
and hence so do the following matrices
$$\left( \begin{array}{c|c} I & T\\ \hline 0 & I \end{array}\right)\text{ and } \left( \begin{array}{c|c} I & 0\\ \hline T & I \end{array}\right),\quad T=\sum_{A\in\SL(2r,\F_2)}\alpha_A \phi_A\text{, }\alpha_A\in \F_2$$ because they form an abelian group
\begin{equation}
\label{eq:abelian}
\left( \begin{array}{c|c} I & T_1\\ \hline 0 & I \end{array}\right) \left( \begin{array}{c|c} I & T_2\\ \hline 0 & I \end{array}\right)=\left( \begin{array}{c|c} I & T_1+T_2\\ \hline 0 & I \end{array}\right).
\end{equation}

The above assumption says that $T$ can be any binary $k\times k$ matrix. Therefore, $M_1=\left( \begin{array}{c|c} I & F_{i,j}\\ \hline 0 & I \end{array}\right)$ and $M_2=\left( \begin{array}{c|c} I & 0\\ \hline F_{j,l} & I \end{array}\right)$ belong to $G_{12}$. So does the commutator of $M_1$ and $M_2$ (if $i\neq k$) $$M_2^{-1}M_1M_2M_1^{-1}=\left( \begin{array}{c|c} I+F_{i,l} & 0\\ \hline 0 & I \end{array}\right).$$
The above $F_{i,j}$'s are for matrices of shape $k\times k$.
Therefore, $G_{12}$ contains all the possible $E_{i,j}$'s for shape $2k\times 2k$ matrices ($\left( \begin{array}{c|c} I & 0\\ \hline 0 & I+F_{i,l} \end{array}\right)$ can be similarly obtained) and $\SL(2k,\F_2)$ can be generated herein.

Had $G_{12}\cong \SL(2k,\F_2)$ been true, when restricting ourselves to the block-diagonal subgroup of $G_{12}$ that acts trivially on the ancillary code block, we can then achieve all the $\SL(k,\F_2)$ operations on the data block.

Therefore, we are left to prove that $\sum_{A\in\SL(2r,\F_2)}\alpha_A \phi_A$, $\alpha_{A}\in \F_2$, $\phi_A\in\SL(k,\F_2)$ includes any binary $k\times k$ matrix. We will first prove that $\phi$ is a group homomorphism $\phi_A\phi_B=\phi_{AB}$ so that $H:=\text{im }\phi$ is a group as well. Then $\F_2H$ will be a ring because given $T_1=\sum_{A\in\SL(2r,\F_2)}\alpha_A \phi_A$ and $T_2=\sum_{B\in\SL(2r,\F_2)}\beta_B \phi_B$ in $\F_2H$, we have $T_1+T_2\in\F_2H$ and also $T_1T_2=\sum_A\sum_B \alpha_A\beta_B\phi_{AB}\in\F_2H$ (in fact, we only need $H$ to be a magma for the inclusion to hold). 

\subsection{Homomorphism}
\label{sec:homomorphism}
We prove $\phi_A\phi_B=\phi_{AB}$ by reducing the problem to the Cauchy-Binet formula. One first notices the following fact:

When giving $A=(a_{ij})$, consider the entry in $\phi_A$ that lies in row $x_{s_1}x_{s_2}\cdots x_{s_r}$ and column $x_{u_1}x_{u_2}\cdots x_{u_r}$. It is the coefficient of $x_{u_1}x_{u_2}\cdots x_{u_r}$ in affine-transformed $x_{s_1}x_{s_2}\cdots x_{s_r}$, i.e., in $\prod_{i=1}^r (\sum_{j=1}^m a_{s_i,j}x_j)$.
This coefficient can be written as
$\sum_{\sigma}a_{s_1,\sigma(u_1)}a_{s_2,\sigma(u_2)}\cdots a_{x_r,\sigma(u_r)}$, where $\sigma$ iterates over all the permutations on $r$ elements. Since addition is over $\F_2$ (permanent is the same as determinant), one recognizes the above as the minor $\det (A_{S,U})$, i.e., the determinant of the submatrix $A_{S,U}$, where only rows corresponding to $x_i$, $i\in S=\{s_1,\dots,s_r\}$ and columns corresponding to $x_j$, $j\in U=\{u_1,\dots,u_r\}$ of $A$ are kept.

The entry in $\phi_A\phi_B$ corresponding to row $x_{s_1}\cdots x_{s_r}$ and column $x_{t_1}\cdots x_{t_r}$ is thus $\sum_U \det (A_{S,U}) \det (B_{U,T})$, where $T=\{t_1,\dots,t_r\}$ and $U$ iterates over all size $r$ subset of $[m]$. On the other hand, the same entry in $\phi_{AB}$ is $\det (AB_{S,T})$. The two things being equal is a consequence of the Cauchy-Binet formula.

\subsection{Constructive proof}
\label{sec:constructive_proof}

Let us first prove that if there is an $F_{i,j}\in\F_2 H$, then all the other $F_{i',j'}\in\F_2 H$. The idea is to sandwich $F_{i,j}$ between the images of some permutation matrices $P_L, P_R$ in $\SL(2r,\F_2)$ under $\phi$. 
If ${(\phi_{P_L})}_{i',i}=1$ and ${(\phi_{P_R})}_{j,j'}=1$, then $F_{i',j'}=\phi_{P_L}F_{i,j}\phi_{P_R}\in\F_2 H$. This is because $\phi_{P_L}, \phi_{P_R}$ are also permutation matrices, as monomials remain monomials, not polynomials under variable permutation. There is plenty of freedom in the choices of $P_L, P_R$. We illustrate the idea by continuing our $r=2$ example. The ordered basis is still $\{x_1x_2,x_1x_3,x_1x_4,x_3x_4,x_2x_4,x_2x_3\}$.

Say we already have $F_{3,4}\in\F_2 H$, i.e., the right-hand side of Eq.~\ref{eq:phi1r+1}, and we want to find $P_L$ and $P_R$ such that $F_{1,1}=\phi_{P_L}F_{3,4}\phi_{P_R}\in\F_2 H$. ${(\phi_{P_L})}_{3,1}=1$ means $x_1x_4\mapsto x_1x_2$, so we can choose $P_L$ to have the effect of swapping $x_4$ and $x_2$ while leaving other variables invariant. The general idea is to pair up variables that only occur in one of the $i^{th}$ and the $i'^{th}$ monomial and let $P_L$ swap each pair (this is always possible because the two monomials both have degree $r$). Similarly, for ${(\phi_{P_R})}_{1,4}=1$, $x_1x_2\mapsto x_3x_4$, we can let $P_R$ swap $(x_1,x_3)$ and $(x_2,x_4)$. We can verify $\phi_{P_L}F_{3,4}\phi_{P_R}=F_{1,1}$ as 
\begin{equation}
\label{eq:swapFij}
\scriptsize
\begin{psmallmatrix}
    0&0&1&0&0&0\\
    0&1&0&0&0&0\\
    1&0&0&0&0&0\\
    0&0&0&0&0&1\\
    0&0&0&0&1&0\\
    0&0&0&1&0&0\\
\end{psmallmatrix}\begin{psmallmatrix}
    0&0&0&0&0&0\\
    0&0&0&0&0&0\\
    0&0&0&1&0&0\\
    0&0&0&0&0&0\\
    0&0&0&0&0&0\\
    0&0&0&0&0&0
\end{psmallmatrix}\begin{psmallmatrix}
    0&0&0&1&0&0\\
    0&1&0&0&0&0\\
    0&0&0&0&0&1\\
    1&0&0&0&0&0\\
    0&0&0&0&1&0\\
    0&0&1&0&0&0\\
\end{psmallmatrix}=\begin{psmallmatrix}
    1&0&0&0&0&0\\
    0&0&0&0&0&0\\
    0&0&0&0&0&0\\
    0&0&0&0&0&0\\
    0&0&0&0&0&0\\
    0&0&0&0&0&0
\end{psmallmatrix}.
\end{equation}

Next, we construct a sequence of addition and multiplication of elements in $\F_2H$ to yield an $F_{i,j}$. Our building block is $\phi_{E_{i,j}}+\phi_{E_{j,i}}+\phi_{\SWAP_{i,j}}$ and we denote this sum as $\phi_{i\Leftrightarrow j}$.
Notice that each of the three matrices only concerns the transformation of two variables $x_i$ and $x_j$ while leaving the other variables invariant. We claim that $\phi_{i\Leftrightarrow j}$ is diagonal and the ones are the on row/columns corresponding to degree-$r$ monomials that either contain both $x_i,x_j$ or contain neither of $x_i,x_j$. We again illustrate this using the $r=2$ example.

Consider the three matrices that only change $x_1$ or $x_2$, while $x_3$ and $x_4$ remain themselves. $E_{1,2}$ has the effect of $x_1\mapsto x_1+x_2$ and $x_2\mapsto x_2$, $E_{2,1}$ has the effect of $x_1\mapsto x_1$ and $x_2\mapsto x_2+x_1$, and $\SWAP_{1,2}$ has the effect of swapping $x_1$ and $x_2$. One can compute that $\phi_{E_{1,2}}+\phi_{E_{2,1}}+\phi_{\SWAP_{1,2}}=$
\begin{equation}
\scriptsize
\label{eq:phi12}
\begin{psmallmatrix}
    1&0&0&0&0&0\\
    0&1&0&0&0&1\\
    0&0&1&0&1&0\\
    0&0&0&1&0&0\\
    0&0&0&0&1&0\\
    0&0&0&0&0&1\\
\end{psmallmatrix}+\begin{psmallmatrix}
    1&0&0&0&0&0\\
    0&1&0&0&0&0\\
    0&0&1&0&0&0\\
    0&0&0&1&0&0\\
    0&0&1&0&1&0\\
    0&1&0&0&0&1\\
\end{psmallmatrix}+\begin{psmallmatrix}
    1&0&0&0&0&0\\
    0&0&0&0&0&1\\
    0&0&0&0&1&0\\
    0&0&0&1&0&0\\
    0&0&1&0&0&0\\
    0&1&0&0&0&0\\
\end{psmallmatrix}=\begin{psmallmatrix}
    1&0&0&0&0&0\\
    0&0&0&0&0&0\\
    0&0&0&0&0&0\\
    0&0&0&1&0&0\\
    0&0&0&0&0&0\\
    0&0&0&0&0&0
\end{psmallmatrix}
\end{equation}
is indeed diagonal, and has ones at monomial row/column $x_1x_2$, $x_3x_4$.

This is true for any $r$. Besides the diagonal, $\phi_{E_{i,j}}$ has additional ones at $(m,m\cdot\frac{x_j}{x_i})$ where monomials $m$ contains $x_i$ but not $x_j$; similarly, $\phi_{E_{j,i}}$ has additional ones besides diagonal at $(m,m\cdot\frac{x_i}{x_j})$ where monomials $m$ contains $x_j$ but not $x_i$. $\phi_{\SWAP_{i,j}}$ does not have ones on diagonal on the previous two types of rows, instead, it has both types of additional ones. Terms added twice are canceled, and things left are precisely those $(m,m)$'s where either $x_ix_j|m$ or $x_i\nmid m \wedge x_j\nmid m$.

Knowing that $\phi_{i\Leftrightarrow j}$ belongs to $\F_2H$, the product of several such diagonal matrices will also be in $\F_2H$. One observes that multiplying binary diagonal matrices is just taking intersections of their ones on the diagonal. Therefore, the product $\phi_{1,\dots, r}:=\phi_{1\Leftrightarrow 2}\phi_{2\Leftrightarrow 3}\cdots\phi_{r-1\Leftrightarrow r}$ only has two ones on the diagonal. More specifically, the ones are at the row/column corresponding to $x_1x_2\cdots x_r$ and $x_{r+1}x_{r+2}\cdots x_{2r}$, since they are the only two monomials that contain either all of $\{x_1,x_2,\dots,x_r\}$ or none of them. 

Finally, $(\phi_{E_{1,r+1}}\phi_{1,\dots,r})+\phi_{1,\dots,r}$ contains a single one located at row $x_1x_{r+2}\cdots x_{2r}$ and column $x_{r+1}x_{r+2}\cdots x_{2r}$. Continuing our $r=2$ example, $\phi_{1,...,2}$ is just $\phi_{1\Leftrightarrow 2}$, i.e., the right-hand side of Eq.~\ref{eq:phi12}. $(\phi_{E_{1,3}}\phi_{1,\dots,r})+\phi_{1,\dots,r}$ calculated below indeed contains a single one at row $x_1x_4$ and column $x_3x_4$.
\begin{equation}
\scriptsize
\label{eq:phi1r+1}
\begin{psmallmatrix}
    1&0&0&0&0&1\\
    0&1&0&0&0&0\\
    0&0&1&1&0&0\\
    0&0&0&1&0&0\\
    0&0&0&0&1&0\\
    0&0&0&0&0&1\\
\end{psmallmatrix}\begin{psmallmatrix}
    1&0&0&0&0&0\\
    0&0&0&0&0&0\\
    0&0&0&0&0&0\\
    0&0&0&1&0&0\\
    0&0&0&0&0&0\\
    0&0&0&0&0&0
\end{psmallmatrix}+\begin{psmallmatrix}
    1&0&0&0&0&0\\
    0&0&0&0&0&0\\
    0&0&0&0&0&0\\
    0&0&0&1&0&0\\
    0&0&0&0&0&0\\
    0&0&0&0&0&0
\end{psmallmatrix}=\begin{psmallmatrix}
    0&0&0&0&0&0\\
    0&0&0&0&0&0\\
    0&0&0&1&0&0\\
    0&0&0&0&0&0\\
    0&0&0&0&0&0\\
    0&0&0&0&0&0
\end{psmallmatrix}.
\end{equation}

Let us see why this is true in general. Multiplying a matrix $A$ with a diagonal matrix $B$ having only two ones on row/column $i,j$ is just replacing the $i^{th},j^{th}$ column of $B$ with the $i^{th},j^{th}$ column of $A$. Therefore, we only need to look at column $x_1x_2\cdots x_r$ and $x_{r+1}x_{r+2}\cdots x_{2r}$ of $\phi_{E_{1,r+1}}$. No monomial other than $x_1x_2\cdots x_r$ itself will be mapped to a polynomial that contains the term $x_1x_2\cdots x_r$ under the map $x_1\mapsto x_1+x_{r+1}$. There are two possibilities to obtain the term $x_{r+1}\cdots x_{2r}$, one is from itself, and the other from $x_1x_{r+2}\cdots x_{2r}\mapsto x_1x_{r+2}\cdots x_{2r}+x_{r+1}x_{r+2}\cdots x_{2r}$. Thereby completing the proof.

To summarize, 
\begin{equation}
\label{eq:Fij}
(\phi_{E_{1,r+1}}+\phi_I)\prod\limits_{i=1}^{r-1}(\phi_{E_{i,i+1}}+\phi_{E_{i+1,i}}+\phi_{\SWAP_{i,i+1}})
\end{equation}
gives a matrix containing a single one. By sandwiching it between images of certain permutation matrices, we can obtain any other matrices with a single one. Therefore, $\F_2H$ includes arbitrary binary $k\times k$ matrices, and by \cite{Grassl_2013}, we can achieve all CNOT-type gates.

\section{Phase-type gate}
\label{sec:phase_type}
We prove the following lemma that, if $T_{CZ}$ is a $k\times k$ symmetric permutation matrix, and contains at least a one on its main diagonal, then there exists $U\in\SL(k,\F_2)$, such that $T_{CZ}=UU^T$. Conjugating $T_{CZ}$ by $\SWAP_{i,j}$ (symmetric) has the effect of exchanging row $i$ and $j$, meanwhile exchanging column $i$ and $j$. A symmetric matrix remains symmetric after conjugation. A permutation matrix stays as a permutation matrix, having exactly a single one in each row or column. Moreover, a diagonal entry remains on the diagonal after conjugation. Therefore, it is possible to find a sequence of $\SWAP_{i,j}$'s, such that conjugating $T_{CZ}$ by them we obtain the following
$$\begin{psmallmatrix}
    0&1& & & & & & &\\
    1&0& & & & & & &\\
     & &0&1& & & & &\\
     & &1&0& & & & &\\
     & & & &\sddots& & \\
     & & & & &1& & &\\
     & & & & & &\sddots&\\
     & & & & & & & &1
\end{psmallmatrix}$$
where we assume there are $l$ preceding blocks of $\begin{psmallmatrix}0&1\\1&0\end{psmallmatrix}$ and $k-2l\geq 1$ ones on the diagonal. 

We only need to prove that this matrix can be written as $UU^T$. We give the following construction of $U$. For $1\leq i \leq l$, the $(2i-1)^{th}$ row of $U$, denoted by $\mathbf{u}_{2i-1}$, has ones at columns $\{1,\cdots,2i-1,2i\}$ and row $\mathbf{u}_{2i}$ has ones at columns $\{1,\cdots,2i-1,2i+1\}$. Row $\mathbf{u}_{2l+1}$ has ones at columns $\{1,\cdots,2l+1\}$. The rest of the rows each have a single one on the diagonal. This works because $(UU^T)_{i,j}$ is the parity of the overlap size between $\mathbf{u}_i$ and $\mathbf{u}_j$.

The final missing piece is to prove that with the help of $\tilde{S}'=\left(\begin{array}{c|c} I & I\\ \hline 0 & I \end{array}\right)$, we can then achieve any phase-type gate.
Using Eq.~\ref{eq:abelian} and the identity
$$\left( \begin{array}{c|c} U & 0\\ \hline 0 & U^{-T} \end{array}\right) \left( \begin{array}{c|c} I & I\\ \hline 0 & I \end{array}\right) \left( \begin{array}{c|c} U^{-1} & 0\\ \hline 0 & U^T \end{array}\right)=\left( \begin{array}{c|c} I & UU^T\\ \hline 0 & I \end{array}\right),$$
we only need to prove that any binary symmetric $k\times k$ matrices can be written as a sum of $UU^T$ for some $U$'s in $\SL(k,\F_2)$.

Let us start with $k=2$, i.e., for the $\llbracket 4,2,2\rrbracket$ code, as mentioned in the main text, in this case we will need the help of $\tilde{S}=\tiny\left(\begin{array}{c|c} 1\ 0 & 0\ 1\\ 0\ 1 & 1\ 0 \\ \hline 0\ 0 & 1\ 0\\ 0\ 0 & 0\ 1 \end{array}\right)$. With its upper right block $\tilde{S}=\begin{psmallmatrix}0&1\\1&0\end{psmallmatrix}$ and the upper right block of $\tilde{S}'$ which is $\begin{psmallmatrix}1&0\\0&1\end{psmallmatrix}$, and together with $E_{1,2}E_{1,2}^T=\begin{psmallmatrix}0&1\\1&1\end{psmallmatrix}$ and  $E_{2,1}E_{2,1}^T=\begin{psmallmatrix}1&1\\1&0\end{psmallmatrix}$, we indeed get all the $2\times 2$ binary symmetric matrices.

For $k>2$, by making use of the above lemma we know that $\SWAP_{i,j}$ can be written as $UU^T$ for some $U$, since it is a symmetric permutation matrix and has $k-2\geq 1$ ones on the diagonal. Therefore, we obtain $F_{i,i}=E_{i,j}E_{i,j}^T+\SWAP_{i,j}$ and $F_{j,j}=E_{j,i}E_{j,i}^T+\SWAP_{i,j}$. Furthermore, all the other bases for binary symmetric matrices in the form of $F_{i,j}+F_{j,i}$ are obtainable by adding $F_{l,l}$'s to $\SWAP_{i,j}$, $l\neq i,j$. This finishes the proof that any phase-type gate is achievable.

\section{Simulation methods}
\label{sec:simulation_detail}
\begin{figure}
    \centering
    \includegraphics[width=1.0\linewidth]{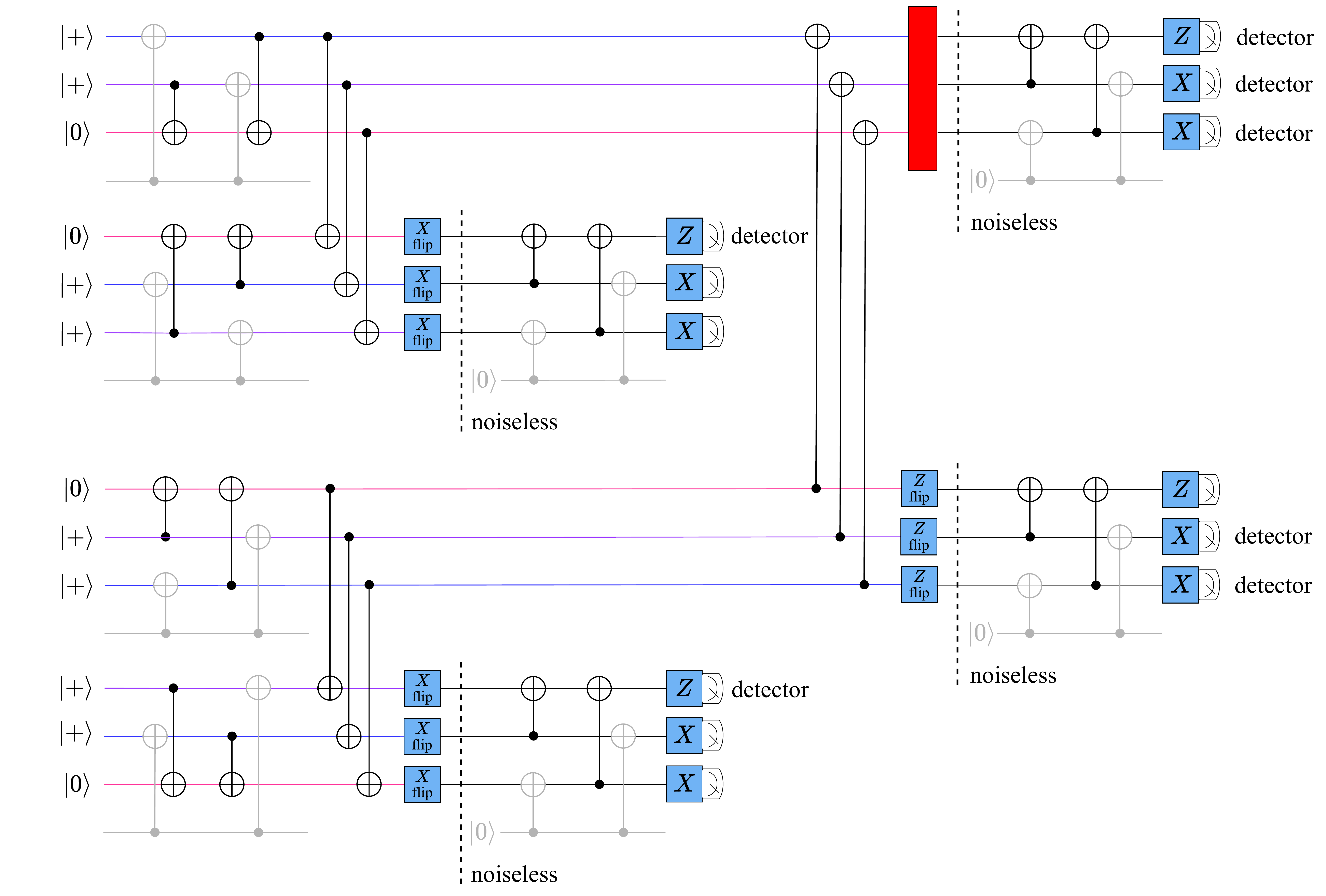}
    \caption{Ancilla preparation in Stim for gathering residual error. Effectively implementing Fig~\ref{fig:AutEnc}(a). An ancilla passes the test if all detectors put on ancilla 2,3,4 are zero. Upon acceptance, we store the faults (column numbers) that happened as returned by DEM. A propagation dictionary is calculated in advance with each key being a single fault in this circuit, and the value being the residual error this fault creates at the red bar.
    When an exRec loads this ancilla, the residual error is obtained by multiplying those of each occurred fault.}
    \label{fig:fullStim}
\end{figure}

We use Fig.~\ref{fig:fullStim} to describe how the ancilla preparation is simulated using the Stim detector error model (DEM). Given the four permutations, we directly implement the permuted circuit for each logical patch, followed by the transversal CNOT gates for copying errors. 
Every operation has added noise, further, we add flips to mimic the measurement errors before the noiseless post-processing. The post-processing consists of the canonical hypercube encoding circuit and $Z$/$X$ measurements at wires initialized to $|0\ra$/$|+\ra$. For checking against $X$/$Z$ faults at patch $2,3,4$, the detectors are only placed at $Z$/$X$ basis measurements. Please note that it is a good practice to measure every qubit at the end, otherwise, Stim might unexpectedly optimize out certain faults that have no effect on measurements. It is the same for detectors, therefore, we place detectors on every measurement of the output patch. This way, any fault that affects the output patch will be captured by DEM. 

We pre-compute a propagation dictionary for the DEM. Each key corresponds to a column of DEM, which is a single fault within the whole circuit. Its value is the residual error at the red bar in Fig.~\ref{fig:fullStim} obtained by propagating this fault.
To facilitate propagation, we insert \texttt{TICK} between layers of CNOTs in the permuted encoding circuit and verification, with the entire noiseless post-processing circuit happening in the same TICK as the preceding copying CNOT gates. We also create a noiseless version of the permuted encoding circuit and the three copying CNOT gates (the post-processing circuits and measurements are not involved). Specifically, it is chopped into a list of circuits, with the $i^{th}$ entry being the layer of CNOTs separated by TICK $i$ and TICK $i+1$.

We then let the noisy circuit explain DEM errors column by column.
The following information about a single fault is extracted: at which TICK and what Pauli flip(s) (for CNOT gate, it might be two flips for a single fault) happen. We propagate the flip(s) through all layers of CNOTs after that TICK using \texttt{stim.PauliString.propagate}.

The idea behind handling a huge amount of ancilla preparation simulations is to only store what faults (the column numbers of DEM) happened, and reassemble the residual errors using the propagation dictionary during load. 


We divide simulations into smaller chunks, each file stores the results of $2.5\cdot 10^8$ preparation simulations.
The file format is as follows. If rejected, we save nothing. If DEM tells us no fault happens, we increase the counter for no fault by one.
If $s\geq 2$ faults happen, we store the corresponding column numbers.
If only one fault happens, there are only $351=3\cdot 127$ columns of DEM that can make this happen. Since strict FT implies the residual error will have weight one\footnote{It could have weight zero, which is some fault that propagates to a stabilizer before verification. However, these kinds of faults are optimized out by DEM, since they trigger no detectors.}. They are faults equivalent to a single qubit $X,Y,Z$ flip after all the gates. Therefore, we keep track of another counter for the occurrences of the $351$ columns.

Each ancilla in the exRec simulation loads from a different file.
To sample and reconstruct ancilla residual errors, we construct a fault dictionary whose keys are sorted tuples of column numbers, and values are the number of occurrences.
For example, there is a key \texttt{none} whose value is the number of no-fault simulations. There is a key for each of the $351$ columns and a key for every $\geq 2$ fault event that ever happened in the simulations. With \texttt{random.choice}, we can perform a weighted random sampling of keys (the weights are specified by their values). 

When loading an order-$s$ fault, the keys tell us the $s$ column numbers. We take their residual errors from the propagation dictionary and multiply them together. The whole procedure finishes with a dedicated decoder reducing the combined error by stabilizers of the state. 

\section{Heuristic search}
\label{sec:heuristic}
We discuss how the permutations in Table~\ref{tab:d7_permutations}\&\ref{tab:d15_permutations} are found. For the protocol in Fig.~\ref{fig:AutEnc}(a) to be strictly FT up to $s$ faults ($s=3$ for $d=7$, $s=4$ for $d=15$), the pair $1$\&$2$ tested each other for $X$ flips should be strictly FT up to $s$ $X$-type faults, so should pair $3$\&$4$. Similarly, pair $1$\&$3$, $1$\&$4$, $2$\&$3$, $2$\&$4$ should be strictly FT up to $s$ $Z$-type faults.
Since we want the permutations to be applicable to the $|+\ra_L$ state preparation in Fig.~\ref{fig:AutEnc}(b) as well, we need to test every pair of blocks (six possibilities) against both $X$ and $Z$-type faults. This is only a necessary (but not sufficient) condition for the entire preparation protocol to be FT. Nevertheless, when we search for the permutations, we only test strict FT between each pair since it is more time-efficient and allows for simple heuristics to develop. They turn out to also guarantee strict FT up to $s$ faults on the entire preparation protocol, the only exception is a single $Z$-type order-four fault for $d=15$.

Let us only focus on a pair of patches checked against each other, say the first patch undergoes permutation represented by $A$ and the second patch undergoes $B$.
The analysis is the same as no permutation on the first patch, and permutation $A^{-1}B$ is applied to the second patch. 
We show that every column of the relative permutation $A_{\rel}=A^{-1}B$ should contain at least two ones. Otherwise, we can construct an order-two fault, one in each patch, and both propagate to the same weight-three error.

We can describe the propagation of a single $Z$ fault in the plain hypercube encoding circuit as follows. For example, $\overleftarrow{0101}000$ denotes the fault that happens on the qubit labeled with $0101000$ and specifically between timestep $3$ and $4$, this fault will propagate to qubits with labels $\{0000000, 0001000,0100000,0101000\}$ at the end. $0000000$ is punctured, hence the residual error has weight three. This shorthand notation allows one to read out what the fault propagates to easily -- turn each one covered by the overline arrow to zero freely.  

Now suppose $A_{\rel}$ has a single one in the second and fourth column, say in row six and seven respectively. Assume the above fault happened in the second patch, after the permutation $A_{\rel}$, the residual error gets mapped to\footnote{We use the convention of $A(x_m,\dots,x_1)^T$ here and index row and column starting from zero for $E_{i,j}$.} $A_{\rel}(0001000)^T$, $A_{\rel}(0100000)^T$, $A_{\rel}(0101000)^T$, which are $\{0000001,0000010,0000011\}$. They cancel the fault $\overleftarrow{0000011}$ on the first patch exactly. One can also see that, in this case, adding the fourth column to the second, or the reverse, will not help with making $A_{\rel}$ FT.

One can also see that, if we want suppression, i.e., order-two faults lead to residual error $\leq 1$ upon acceptance, then $A_{\rel}$ cannot have a single column that contains a single one. It is not necessary to be this stringent with two faults, but this pre-sieving makes the search easier when we proceed to permutations robust against order-three faults.

$A_{\rel}$ containing at least two ones in each column implies that $A_{\rel}$ consists of a product of at least seven $E_{i,j}$'s. Starting from the identity matrix, reading the product from left to right corresponds to adding a column to another each time. Since no column contains a single one, $j$ must take on all values between $0$ and $6$. However, the shortest $A_{\rel}$ we find against three faults has length nine. We can split $A_{\rel}$ into two halves for $A$ and $B$, thus initially we hope to find four permutations containing five $E_{i,j}$ each.

Five is not enough. Call the permutations in the four columns $A$,$B$,$C$,$D$. We also need to make sure $A^{-1}B$, $C^{-1}D$, $A^{-1}C$, $A^{-1}D$, $B^{-1}C$, $B^{-1}D$ are strictly FT. If the four permutations contain five $E_{i,j}$ each, $j$ does not take on (at least) two values in each of $A,B,C,D$, there must be a pair where $j$ does not take on seven values.

Nevertheless, we first find a pair of $A$ and $B$ each containing five $E_{i,j}$'s while making $A^{-1}B$ robust to three faults, then bootstrap from them. They are the last five rows of the first two columns of Table~\ref{tab:d7_permutations}. Namely $A=E_{5,4}E_{4,3}E_{3,2}E_{2,1}E_{1,0}$ (note that the earlier $E_{i,j}$ to apply in time is on the left of the product) and $B=E_{4,2}E_{0,5}E_{6,0}E_{5,1}E_{2,6}$. We bootstrap them by extending the same $E_{i,j}$'s in the first two rows. Since $E_{i,j}$ is the inverse of itself, the relative permutation after bootstrapping remains the same.

$A$ is written down by hand, and $B$ is found by random search as follows. Since by our choice of $A$, $j$ did not take on value $5,6$, we require them to be present in $B$. Furthermore, we also impose $3,4$ not in $B$. Writing $B$ as $E_{i_1,j_1}E_{i_2,j_2}E_{i_3,j_3}E_{i_4,j_4}E_{i_5,j_5}$, we let $j_1,\dots,j_5$ be a random permutation of $0,1,2,5,6$ then choose each $i_k$ randomly from $\{0,1,2,3,4,5,6\}\backslash\{j_k\}$. 

When searching for $C$ and $D$ (the last five rows in column three and four), we require $j$ to lack the value $3,5$ and $4,6$ respectively. $C$ and $D$ will also be bootstrapped by some other $E_{i,j}$'s in the first two rows. The total bootstrap term $E$ consists of four $E_{i,j}$'s (two from $A,B$ and two from $C,D$), and we require the $j$'s to take on values $3,4,5,6$  
($E$ finally ends up being $E_{0,3}E_{1,4}E_{3,6}E_{6,5}$). 
We randomly search for $C,D,E$ such that the relative permutations $C^{-1}D$, $A^{-1}EC$,$A^{-1}ED$, $B^{-1}EC$, $B^{-1}ED$ are all robust against three faults.

Another useful trick is that, since there are $29$ X-type stabilizers and $98$ Z-type stabilizers of $|+\ra_{d=7}$, testing the robustness of a permutation against $Z$-type faults is much faster. We use this to sieve the permutations to be tested for FT against $X$-type faults. It is worth mentioning that, also because of this stabilizer number imbalance, the automorphism-based encoding protocol we use for $|+\ra_{d=7}$ is Fig.~\ref{fig:AutEnc}(a). This way, more detectors can be added, forming a more stringent acceptance condition than using Fig.~\ref{fig:AutEnc}(b).

With the above heuristic, it only took us a few hours to find Table~\ref{tab:d7_permutations} on a MAC Air.

Proceed to $\llbracket 127,1,15\rrbracket$, we reuse the middle two columns of Table~\ref{tab:d7_permutations} since this pair is already strict FT against four faults. We search for the fourth column on a supercomputer by running many processes in parallel (each one runs for four hours), the input to each process is what values $j$ should take on. For every valid fourth column, we search for its corresponding first column. For example, in Table~\ref{tab:d15_permutations}, one can see that $j$ takes on all values in the second column while lacking $3$ in the third and $6$ in the fourth. We thus impose $j$ to take on $3$ and $6$ twice in search of the first column using multiple processes. We gather around ten possible permutation quadruples in a day and we finally choose to present the permutations in Table~\ref{tab:d15_permutations} because they have the least number of order-four faults violating strict FT across the four patches.

We believe the permutation depth can be improved. Besides developing better heuristics and doing more extensive searches, future work could also explore the following ways. First, other native operations of AODs like squeezing or stretching may be incorporated. Second, if the unnecessary CNOT gates (control on $|0\ra$ or target on $|+\ra$) in the hypercube encoding circuit could be removed physically\footnote{We do not consider this optimization because of the rectangular grid constraint, but this may be possible on other platforms.}, then finding permutations will become easier. Finally, one could consider merging some of the permutations into the encoding.
For example, a permutation matrix factor $A_p$ of a qubit-permutation\footnote{Any $A\in\SL(m,\F_2)$ is a permutation of qubits, by $x\mapsto Ax$. We are now concerned with $A$ being written as some other invertible matrices times a permutation matrix $A_p$.} $A$ can be absorbed into the hypercube encoding circuit in Table~\ref{table:polar}, rather than doing it afterward and causing idling errors. $A_p$ permuting the variables $x_m,\dots,x_1$ corresponds to implementing the $m$ layers of CNOTs in a different order. The initialization in Table~\ref{table:init} stays the same, because the assignment only concerns how many ones each label has, which is a quantity invariant under permuting $x_m,\dots,x_1$.

\end{document}